\documentclass[review]{elsarticle} 

\usepackage{hyperref}
\usepackage{changepage}
\journal{Electrochimica Acta}

\usepackage{lineno}

\usepackage{amsmath} 
\usepackage{amssymb}
\usepackage{upgreek}
\usepackage{bm} 
\usepackage{mathtools, nccmath} 
\biboptions{sort&compress}
\usepackage{makecell}  
\usepackage{float} 
\usepackage{tablefootnote}
\usepackage{subcaption} 
\usepackage{algorithm} 
\usepackage{algpseudocode}
\usepackage[margin=2cm]{geometry}
\usepackage[capitalise, nameinlink]{cleveref}

\usepackage{placeins}

\usepackage{todonotes} 
\let\oldtodo\todo
\renewcommand{\todo}[1]{\oldtodo[inline]{#1}}

\usepackage[normalem]{ulem}

\begin{document}

\begin{frontmatter}
\title{Corrosion of metal reinforcements within concrete and localisation of supporting reactions under natural conditions}

\author[1]{Tim Hageman\corref{A1}}
\cortext[A1]{Corresponding authors}
\ead{tim.hageman@eng.ox.ac.uk}

\author[2]{Carmen Andrade}

\author[1]{Emilio Martínez-Pañeda\corref{A1}}
\ead{emilio.martinez-paneda@eng.ox.ac.uk}

\address[1]{Department of Engineering Science, University of Oxford, Oxford OX1 3PJ, UK}
\address[2]{International Center of Numerical Methods in Engineering (CIMNE), Madrid 28010, Spain}

\begin{abstract} 
Corrosion in concrete prevents in-situ observation, necessitating models to provide insight into the local reaction currents. We present a computational method for predicting corrosion rates of reinforcements within concrete under natural conditions, i.e. requiring the corrosion current to be supported by equal cathodic currents. In contrast to typical corrosion models, where these two counteracting currents are required to be co-located, we allow these currents to be separated such that pitting corrosion can be supported by cathodic reactions over a much larger area. Pitting corrosion is investigated, elucidating the effects of the concrete porosity and water saturation, the presence of dissolved oxygen, and chlorine concentration within the pore solution. The presented model is capable of capturing the dynamic growth of acidic regions around corrosion pits, showing the limited region over which the hydrogen evolution reaction occurs and how this region evolves over time. The ability of oxygen to diffuse towards the metal surface due to increased porosity is seen to have a major effect on the corrosion rate, whereas changes in the chlorine concentration (and thus changes in the conductivity of the pore solution) play a secondary role. Furthermore, external oxygen is seen to enhance corrosion but is not required to initialise and sustain acidic corrosion pits.\\
\end{abstract}



\end{frontmatter}

\section{Introduction}
\label{sec:intro}

While reinforcements within concrete play a crucial role, they are prone to corrosion, in turn leading to internal stresses \citep{Page1982, Arya1996}, local fractures \citep{Khan2014, Korec2023, Korec2024b}, and de-bonding between the reinforcement and the surrounding concrete \citep{Fang2004, Lee2002, Zhao2011, Michel2013}. Initially, the solution contained within the pores of the concrete is alkaline, creating passivation layers on the steel and preventing corrosion \citep{Page1982, Angst2011}. However, under the influence of chlorine ions, this passivation layer can rupture, thereby enabling pitting corrosion \citep{Wang2021, Angst2019, Macdonald2021, Wang2013}. Once a pit forms, the local environment becomes acidic due to corrosion product hydrolysis, preventing repassivation even if the original source of the depassivation is removed \citep{Pourbaix1975, Taylor2018a, Li2021a, Turnbull1983, Pickering1972}. The cathodic reactions occurring on the surrounding surface consume the currents supplied by corrosion, balancing anodic and cathodic reactions and altering the overall electric potential of the metal \citep{Hoar1967, Galvele1981, Frankel2017, Frankel1990, Nguyen2021}. Cathodic and anodic reactions occur over different regions, with the latter being highly localised. Over time, the continued growth of corrosion pits leads to significant loss of reinforcement cross-section, reducing structural integrity and increasing the risk of failure in concrete structures. The precipitation of corrosion products further induces tensile stresses in the surrounding concrete, promoting cracking, spalling, and reducing the durability and service life of reinforced concrete structures \citep{korec2024phase}. Therefore, it is crucial to establish the changes in the local environment, the range over which these changes are induced, and how long these changes are being sustained by the presence of a corrosion pit under realistic circumstances.

There has been significant interest in bringing insight into these complex electro-chemo-mechanical phenomena through both theoretical and experimental work. The basic equations controlling mass transport have been utilised to enhance our understanding of these processes \citep{Sharland1987, Jafarzadeh2019, Newman1962, Newman2021a} but there remain notable experimental hurdles which hinder validation and quantitative understanding. The difficulty of conducting laboratory tests on small pits goes beyond the small scales involved, as these pits evolve continuously, with experiments being unable to capture intermediate processes and the complexity of a multispecies electrolyte. Of particular relevance to this work, laboratory tests capturing pitting corrosion in detail are impossible in concrete where the physical cover prevents the direct observation of where a pit is generated, or how it grows \citep{Ali2024, Angst2018, Feliu2005}.

Experimentally, corrosion experiments are performed under natural or accelerated conditions \citep{Frankenthal1972, Elsener2005}. Under natural conditions, the corrosion rate is slow, with experiments taking months to provide usable corrosion rate data \citep{Andrade1978}. Other techniques frequently applied are based on imposing variations in the current source and measuring the system response through impedance-based techniques \citep{Darowicki2004, Duarte2014, Liu2025, Zhao2007}. Although these experiments are not fully realistic as the whole bar is polarized, which is not the case in natural conditions, they provide information about corrosion rates in a reasonable time. Accelerated corrosion experiments allow the data to be obtained on a practical time scale by either providing a set corrosion current or imposing an electric potential to the metal, such that the rate of corrosion and the material dissolution become measurable within the typical time span of experiments \citep{Lin2010, Nguyen2021, Sola2019, Hay2020}. Simplifications are also made when numerically simulating corrosion in concrete: Most models assume the dominant reaction step in the corrosion process \textit{a priori}, neglecting the role of the other reactions, and assume this dominant reaction rate to be diffusion driven \citep{Galvele1976, Galvele1981, Nguyen2021, Sun2019}, activation controlled \citep{Sun2021a, Khatami2021}, or impose a constant rate matching that of experimentally observed corrosion currents \citep{Korec2023}, usually corrected for environmental conditions such as mechanical stresses \citep{Kovacevic2023, Cui2021}. In reality, however, the actual dominant mechanism  depends on the local environment, and can differ between different parts of the domain and change over time, making it relevant to develop models that dynamically capture reaction rates. These simplifications prevent existing models from capturing realistic conditions.

Another important unknown is the space over which anodic and cathodic reactions occur. Reinforcements within concrete can be isolated from other current sources, except for what is conducted by the water within the concrete pores and through the rebar itself. This poses the requirement that the corrosion is sustained by cathodic reactions, consuming all electrons produced by the corrosion reactions. This condition is built into some Pourbaix diagrams, enforcing local reaction currents to be conserved by directly including both anodic and cathodic reactions within a single reaction process \citep{Pourbaix1974, Pourbaix1990}. These cathodic and anodic reactions do not need to occur over the same area of the metal, however, allowing pitting corrosion in a localised area to be sustained by hydrogen and oxygen evolution reactions over the much larger rebar surface. It has been argued that these supporting reactions are still localised around the corrosion pits, with the majority of the rebar not partaking in any reactions \citep{Pickering1972, Chen1997}. Observing this localisation and determining if reaction currents dependent on the properties of the concrete the metal rebar is contained within through experiments alone is unfeasible. Experiments have shown that the conductivity of the concrete is a major indicator of its corrosion ability \citep{Morris2002, Andrade2018, Andrade2018a, Shevtsov2021}. Similarly, pencil electrode tests demonstrated that increases in ion concentrations are associated with increased corrosion rates, although part of this effect was allocated to chlorine ions causing the passivation layer to dissolve \citep{Woldemedhin2015}. These experimentally observed dependences have then been confirmed through numerical simulations of pencil electrodes, providing a detailed look into each individual reaction current and the location of these currents \citep{Makuch2024, Duddu2016, Mai2018}. The non-local current conservation condition, allowing for anodic reactions to be supported by cathodic reactions over a much wider area, has also recently been integrated in a computational model, showing the role of this coupling and the interactions between reaction areas for pencil electrodes \citep{Hageman2023a}. However, these studies are yet to be extended to concrete, separating increases in conductivity due to ion concentrations, porosity, and saturation changes and their overall effect on the corrosion rate. 

Key questions remain thus to be answered. To what extent does the area surrounding pits partake in sustaining corrosion, and does this depend on the species reacting to provide the cathodic currents? Furthermore, how are this localisation and these reaction currents dependent on the properties of the concrete within which the metal rebar is contained? Here, we present a new computational model to estimate the corrosion of reinforcements within concrete, under the requirement that the corrosion current is fully supported by hydrogen and oxygen evolution reactions. This requirement emulates natural corrosion, where the rebar is fully isolated from any external current and potential sources. As this creates a direct link between the corrosion rate and the cathodic reaction rates, this model allows for in-depth analysis of the processes causing corrosion in natural conditions. We furthermore do not impose any specific reaction areas, rather letting the computational model resolve these by itself, to study which areas of the rebar partake in the electrochemical reactions. These two important and novel model features allow us to gain new physical insight and answer the pressing questions formulated above. The governing equations and reactions included within the model are provided in \cref{sec:modelOverview}, after which in \cref{sec:Results} this model is used to investigate the processes surrounding a single corrosion pit. The results demonstrate that (part of) the supporting cathodic reactions are highly localised around corrosion pits. The effects of oxygen diffusing into the concrete are further elucidated, and through parametric studies, the roles of the concrete porosity, saturation, and electrolyte conductivity are investigated. Concluding remarks end the manuscript in Section \ref{Sec:Conclusions}. 

\section{Model overview}
\label{sec:modelOverview}
\begin{figure}
    \centering
    \includegraphics{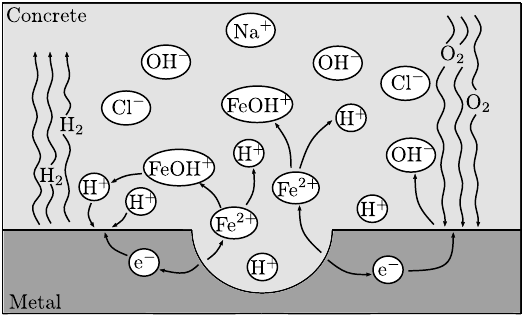}
    \caption{Schematic overview of the electro-chemical system modelled, consisting of porous concrete saturated with an electrolyte and metal reinforcements. The electrolyte contains ions and dissolved oxygen, causing electro-chemical reactions with the metal restricted by the charge conservation across reactions.}
    \label{fig:overview}
\end{figure}

To study localised corrosion phenomena, we consider a metal rebar embedded within concrete, considering the processes schematically shown in \cref{fig:overview}. The concrete is porous, with porosity $\phi$, and the pore saturation with water is indicated by $S_\text{w}$. If this saturation is $S_\text{w}=1$, then the concrete pores are fully filled with electrolyte (water), whereas the pores are considered partially air-filled as this saturation approaches zero. Low concentrations of oxygen are considered, such that the oxygen is fully dissolved within the water, characterising this dissolved oxygen by its concentration $C_{\text{O}_2}$. For unsaturated conditions, this concentration of oxygen is also used to indicate the amount of oxygen in the air contained within pores, assuming equal concentrations of oxygen between the water and air phases due to the large contact area between the two. Within the electrolyte other ionic species are present, here considering $\text{H}^+$ and $\text{OH}^-$ ions to represent water, $\text{Fe}^{2+}$ and $\text{FeOH}^+$ to capture the corrosion products, and $\text{Na}^+$ and $\text{Cl}^-$ as ``filler" ions that do not directly partake in any reactions but do contribute to the overall conductivity of the electrolyte. Concentrations of these species are indicated by $C_\pi,\;\pi=\text{H}^+,\text{OH}^+, \text{Fe}^{2+}, \text{FeOH}^+, \text{Na}^+, \text{Cl}^-$. The rebar has a single corrosion pit, allowing for localised corrosion solely on this pitted area. The remainder of the rebar is assumed to have intact corrosion protection (e.g. passivation layers, coating), such that only the oxygen and hydrogen evolution reactions occur at its surface. It should be noted that we do not model the breakdown of this passivation layer (e.g. due to interactions with chlorine or mechanical loading), and the changes in porosity due to a build-up of corrosion products. This allows our simulations to assume a constant geometry that does not change over time, allowing steady-state reaction currents and their corresponding reaction areas to be obtained. \\

A note regarding concentrations and units used throughout this paper: Concentrations of ions are given relative to the volume of the electrolyte, not the total volume of the concrete, air-filled pores, and electrolyte combined. Where unsaturated conditions are considered, the concentration of oxygen is relative to the total volume of the pores, as this oxygen is contained in both the air and the electrolyte. SI units are consistently used for all quantities throughout this paper, including concentrations. As such, concentrations are given in mol of ionic species per $\text{m}^3$ electrolyte, $\text{mol}/\text{m}^3$. 

\subsection{(Electro-) chemical reactions}
As anodic reaction, the corrosion of iron on the surface of the corrosion pit is considered:
\begin{equation}
    \text{Fe}^{2+}+2\text{e}^-\xrightleftharpoons[k_{\text{c}}']{k_{\text{c}}}\text{Fe}
\end{equation}
and as cathodic reactions, the hydrogen and oxygen evolution reactions are considered, and allowed to occur on the complete metal surface:
\begin{align}
    \text{O}_2+2\text{H}_2\text{O} +4\text{e}^-&\xrightleftharpoons[k_{\text{o}}']{k_{\text{o}}} 4 \text{OH}^- \\
    2\text{H}^+ + 2\text{e}^-&\xrightharpoonup{k_{\text{h}}} \text{H}_2 
\end{align}
In these reactions, the rates are governed by the forward reaction constants $k$ and backward reaction constants $k'$. The hydrogen evolution reaction is assumed to be uni-directional, with the produced hydrogen gas diffusing out of the simulated domain fast enough to not impact the reaction rates. As we include both the forward and backward reactions for the corrosion and oxygen evolution, the diffusion-controlled and rate-controlled mechanisms are both included. If diffusion of corrosion products is limiting, the corrosion reaction will be in near equilibrium, solely producing iron ions to compensate for the diffusion. In contrast, if the electric potential of the metal is negative, the corrosion rate itself will directly govern the amount of iron ions produced. Including both of these mechanisms within a single scheme allows for capturing a large range of conditions, without selecting the limiting reaction mechanism a priori. Furthermore, as no assumptions are made regarding the direction of the reactions, the oxygen and corrosion reactions can be imposed on a surface without requiring knowledge of whether these reactions actually occur on this surface. This allows the location of reactions to be studied, with the reaction rate reducing, becoming zero, or even becoming negative in areas where, for instance, the electric overpotential is not sufficiently high to make the oxygen evolution reaction occur. 

The reaction rates $\nu_\pi$ and reaction currents $i_\pi$ for the corrosion, oxygen, and hydrogen surface reactions are respectively given by the Butler-Volmer equation as:
\begin{align}
    \begin{split}
    \nu_\text{c} = \frac{i_\text{c}}{2F} = k_\text{c} &\frac{C_{\text{Fe}^{2+}}}{C_\text{ref}} \exp\left(-\alpha_\text{c} \frac{\eta F}{RT}\right) \\ &-  k_\text{c}'  \exp\left((1-\alpha_\text{c}) \frac{\eta F}{RT}\right) \end{split} \label{eq:surf_c}\\
    \begin{split}
    \nu_\text{o} = \frac{i_\text{o}}{4F} = k_\text{o} & \frac{C_{\text{O}_2}}{C_\text{ref}} \exp\left(-\alpha_\text{o} \frac{\eta F}{RT}\right) \\ &- k_\text{o}' \frac{C_{\text{OH}^{-}}}{C_\text{ref}} \exp\left((1-\alpha_\text{o}) \frac{\eta F}{RT}\right) \end{split} \label{eq:surf_o}\\
    \nu_\text{h} = \frac{i_\text{h}}{2F} = k_\text{h} & \frac{C_{\text{H}^+}}{C_\text{ref}} \exp\left(-\alpha_\text{h} \frac{\eta F}{RT}\right) \label{eq:surf_h}  
\end{align}
using the charge transfer coefficients $\alpha_\pi$, Faraday constant $F$, reference temperature $T$, and gas constant $R$. To allow for consistent units between the reaction rate constants, a reference concentration $C_\text{ref}=1000\;\text{mol}/\text{m}^3$ is included, such that all reaction rate constants are given in $\text{mol}/\text{m}^2/\text{s}$. These reaction rate constants are related to typical current densities reported in literature via $k_\pi=i_{0,\pi}/|z_\pi|F$. The electric overpotential $\eta=E_\text{m}-\varphi-E_{\text{eq},\pi}$ is based on the potential difference between the metal and electrolyte, $E_\text{m}-\varphi$, and includes an offset for the reaction equilibrium potential $E_{\text{eq},\pi}$ under reference conditions. \footnote{Note that the reference equilibrium potential used here is assumed to be a constant material parameter (taken as the equilibrium potential at pH 0). While this equilibrium potential is typically dependent on the pH, with the pH providing an offset, this is already taken into account explicitly by the multiplication of the reaction rates with the concentrations. For instance, for the hydrogen evolution reaction, this could be equally written as:
\begin{equation*}
\begin{split}
    \nu_\text{h} &= k_\text{h} \frac{C_{\text{H}^+}}{C_\text{ref}} \exp\left(-\alpha_\text{h} \frac{\eta F}{RT}\right) \\
    &= k_\text{h} \text{exp}\left( -\alpha_\text{h}\frac{F}{RT}\left(E_\text{m}-\varphi - E_{\text{eq},\pi} - \frac{RT\text{ln}(10)}{\alpha_\text{h}F}\text{pH} \right) \right)
\end{split} \label{eq:surf_NOTUSED}
\end{equation*}
with the factor in front of the pH being $\approx 0.0592$, the typical factor used to represent the pH dependence in the Nernst equation. While this equation could equally be used instead of \cref{eq:surf_h}, it is chosen here to use \cref{eq:surf_h} due to its explicit (and linear) dependence on the concentrations, facilitating the implementation of numerical solution schemes. \label{footnote:Eeq}}

The metal potential is considered an independent variable, being allowed to change throughout the duration of the simulation. This allows an additional constraint to be introduced: the charge conservation condition requires all electro-chemical reactions that occur on the metal surface to be supported by other reactions on the metal. This is enforced as requiring the total reaction currents, the local currents integrated over the metal area, to fulfil \citep{Hageman2023a}:
\begin{equation}
\begin{split}
    I_\text{c}+I_\text{o}+I_\text{h}+I_\text{ext}=\int_\Gamma i_\text{c}+i_\text{o}+i_\text{h}\;\text{d}\Gamma +I_\text{ext} \\ = \int_\Gamma 2F\nu_\text{c} + 4F\nu_\text{o} + 2F\nu_\text{h} \;\text{d}\Gamma +I_\text{ext} = 0 \end{split}
    \label{eq:current_conservation}
\end{equation}
Here, we will consider the case where there is no external current source, $I_\text{ext}=0$. This is equivalent to considering the metal to be isolated from any current or potential sources, instead of considering a case in which only the naturally occurring reaction currents are included. Any currents resulting from corrosion within the pit need to be consumed by hydrogen or oxygen reactions somewhere on the metal surface. As we assume the metal to be well-conducting, this balance does not need to hold locally, only globally, allowing anodic and supporting cathodic reactions to be spatially separated. As a result of this current conservation, the electric potential of the metal $E_\text{m}$ will change, altering the overpotential $\eta$ and thus adapting the reaction rates to preserve reaction currents. It should be noted that this significantly differs from accelerated corrosion experiments/simulations, where the majority of reaction currents are enforced through an external current source to accelerate the corrosion process, where either the corrosion current, $I_\text{c}$, or the electric potential of the metal, $E_\text{m}$, are directly imposed. 

In addition to the surface reactions, the water auto-ionisation reaction and hydrolysis of the corrosion products occur within the electrolyte:
\begin{align}
    \text{H}_2\text{O} &\xrightleftharpoons[k_{\text{w}}']{k_{\text{w}}} \text{H}^+ + \text{OH}^-\\
    \text{Fe}^{2+} + \text{H}_2\text{O} &\xrightleftharpoons[k_{\text{fe}}']{k_{\text{fe}}} \text{FeOH}^+ + \text{H}^+ \label{React:FeOH}  \\
    \text{FeOH}^{+} + \text{H}_2\text{O} &\xrightarrow{k_{feoh}} \text{Fe(OH)}_2 + \text{H}^+ \label{React:FeOH2}
\end{align}
where the last reaction is assumed to be unidirectional, such that the produced $\text{Fe}(\text{OH})_2$ does not further partake in any reactions, and thus does not need to be tracked within the model. The water auto-ionisation is always in a near-equilibrium, enforced through a penalty factor $k_\text{eq}$, which allows the volume reactions for the $\text{OH}^-$ ions and contribution towards the $\text{H}^+$ reactions to be obtained as:
\begin{equation}
    \begin{split}
    R_{\text{H}^+1}=R_{\text{OH}^-} = k_{\text{w}}\frac{C_{\text{H}_2\text{O}}}{C_\text{ref}} - k_{\text{w}}'\frac{C_{\text{H}^+}C_{\text{OH}^-}}{C_\text{ref}^2}  \\ = k_{\text{eq}} \left(K_{\text{w}}-\frac{C_{\text{H}^+} C_{\text{OH}^-}}{C_\text{ref}^2} \right) \end{split}\label{eq:water_react}
\end{equation}
using equilibrium constant $K_\text{w}=10^{-14}$. The iron hydrolysis reactions, in contrast, occur at a slower rate determined by the volume reaction constants $k_\text{fe}$, $k_\text{fe}'$ and $k_\text{feoh}$ (all with units $\text{mol}/\text{m}^3/\text{s})$. These reaction constants then dictate the reaction rates based on the local concentrations as:
\begin{align} \label{Re:Iron1}
    R_{\text{Fe}^{2+}}&=-k_{\text{fe}}\frac{C_{\text{Fe}^{2+}}}{C_\text{ref}}+k_{\text{fe}}'\frac{C_{\text{H}^+}C_{\text{FeOH}^+}}{C_\text{ref}^2} \\
    R_{\text{FeOH}^+}&=k_{\text{fe}}\frac{C_{\text{Fe}^{2+}}}{C_\text{ref}}-k_{\text{fe}}'\frac{C_{\text{H}^+}C_{\text{FeOH}^+}}{C_\text{ref}^2} - k_\text{feoh} \frac{C_{\text{FeOH}^+}}{C_\text{ref}}
    \label{Re:Iron2}\\
    R_{\text{H}^+2}&=k_{\text{fe}}\frac{C_{\text{Fe}^{2+}}}{C_\text{ref}}- k_{\text{fe}}'\frac{C_{\text{H}^+}C_{\text{FeOH}^+}}{C_\text{ref}^2} + k_\text{feoh} \frac{C_{\text{FeOH}^+}}{C_\text{ref}}
\end{align}
where the contributions of the hydrolysis and auto-ionisation to the $\text{H}^+$ reaction rate are combined, $R_{\text{H}^+}=R_{\text{H}^+1}+R_{\text{H}^+2}$. A result of these hydrolysis reactions is that any $\text{Fe}^{2+}$ ion that is produced from corrosion reactions (producing two electrons) can react to generate up to two $\text{H}^+$ ions in turn. These two $\text{H}^+$ ions can then react with the metal surface, consuming two electrons. This theoretically makes the corrosion reactions self-sustaining once initialised, even in the absence of oxygen. However, when iron ions are allowed to diffuse out of the concrete, and when they do not fully react to form $\text{Fe(OH)}_2$, the corrosion process will need a secondary sink for electrons (e.g. the oxygen evolution reaction). Similarly, as concrete pore solutions are typically basic, corrosion reactions will not be able to initiate by themselves without this secondary sink for electrons being present.

\subsection{Diffusion and species transport}
Within the electrolyte contained within the pores of the concrete, the movement of all ionic species and dissolved oxygen is governed by the Nernst-Planck equation:
\begin{equation}
    S^*_\pi \phi \dot{C}_\pi +\bm{\nabla}\cdot\left(-D_\pi^{\text{eff}} \bm{\nabla}C_\pi\right)+\frac{z_\pi F}{RT}\bm{\nabla}\cdot\left(-D_\pi^{\text{eff}} C_\pi \bm{\nabla}\varphi\right) + \phi S^* R_\pi = 0 \label{eq:massconserv}
\end{equation}
which considers electro-migration for species with charge $z_\pi$ due to gradients in electrolyte potential $\varphi$ and concentration-based diffusion, while assuming negligible fluid flow to occur within the pores of the concrete. The factor $S^*$ corrects the capacity of the pores in the case of non-saturated conditions, for ionic species being equal to the saturation $S^*=S_\text{w}$, whereas for oxygen always being equal to one $S^*_\text{o}=1$. 

Both diffusion and electro-migration involve the effective diffusivity of the ionic species, $D_\pi^\text{eff}$. This effective diffusivity is the ``scaled'' diffusivity which takes into account that the ions can not diffuse through a free electrolyte, but through an unsaturated porous material. This effective diffusivity is given by \citep{Tartakovsky2019, Ghanbarian2013}:
\begin{equation}
    D_\pi^{\text{eff}} = \frac{D_\pi \phi}{\tau} S_\text{eff}^2 = \phi^{3/2} D_\pi \left(\frac{S_\text{w}-S_\text{irr}}{1-S_\text{irr}}\right)^2 \label{eq:Deff}
\end{equation}
where $D_\pi$ is the diffusivity of the ionic species $\pi$ in the electrolyte (not contained within concrete pores), and $\tau$ is the tortuosity factor, taken as a function of the concrete porosity via $\tau=\phi^{-1/2}$ \citep{Tartakovsky2019}. The effects of saturation are included through the term $S_\text{eff}$, which takes into account a saturation limit below which the electrolyte within the pores becomes disconnected via $S_\text{irr}=0.2$ \citep{Bentz1991, Liu2022}. This saturation dependence is used in the definition of the effective diffusivity for all ionic species, whereas $S_\text{eff}=1$ is assumed to obtain the effective diffusivity of oxygen. 

Finally, to model the electric current flowing through the porous concrete we consider the concrete skeleton itself to be non-conductive, with the conductivity solely determined by the water within the concrete. Within the model, this conductivity is then enforced through the electroneutrality condition:
\begin{equation}
    \sum_\pi \phi z_\pi C_\pi = 0 \label{eq:electroneutrality}
\end{equation}
By requiring each individual point within the electrolyte to remain at neutral charge, divergence-free electric currents within the electrolyte are enforced implicitly \citep{Feldberg2000}. 

Combined, \cref{eq:electroneutrality,eq:massconserv} model the movement of ions and the electric potential required to conserve electric currents. As such, we do not need to prescribe an ``effective conductivity'' for the electric field, with this electric field instead following directly from the model using the concentrations of all species present to determine the conductivity. Nevertheless, while not used within the simulations, we define an effective conductivity purely for post-processing purposes. The effective diffusivity is related to the resistivity $\rho$ of the concrete (the inverse of the conductivity). If we assume that the major contribution to the overall conductivity is due to the ``filler" ions, $\text{Na}^+$ and $\text{Cl}^-$, this resistivity is given by \citep{Newman2021}:
\begin{equation}
\begin{split}
    \frac{1}{\rho} &= \frac{F^2 C_{\text{Cl}^-} (D^{\text{eff}}_{\text{Na}^+}+D^{\text{eff}}_{\text{Cl}^-})}{RT} \\ &= \frac{F^2 C_{\text{Cl}^-} \phi^{3/2}(D_{\text{Na}^+}+D_{\text{Cl}^-})}{RT} \left(\frac{S_\text{w}-S_\text{irr}}{1-S_\text{irr}}\right)^2 \\ & \approx 0.0197 C_{\text{Cl}^-} \phi^{3/2} (S_\text{w}-0.2)^2
    \end{split}\label{eq:Conductivity}
\end{equation}
This does not take into account the additional conductivity due to the concrete itself, as this is presumed much lower than the conductivity of the electrolyte contained within the concrete pores. As such, this assumes that the sole contribution to the conductivity of the (partially) saturated concrete is the diffusion of ions within the electrolyte. The corrosion products will also further enhance the local conductivity, but this is not taken into account within \cref{eq:Conductivity}. We note that this relation is not explicitly used within our model, as the electric potential field follows from \cref{eq:electroneutrality}, and is only used in this work to translate from porosity, saturation, and chlorine ion concentration to conductivity/resistivity during post-processing. As such, our simulations do take the local changes in resistivity due to the developed corrosion products into account, whereas any post-processing which translates our results into relations with the resistivity assumes that this resistivity is based on the initial conditions. 

\subsection{Solution method}
\cref{eq:massconserv,eq:current_conservation,eq:electroneutrality} are resolved using a finite element method, comparable to a scheme previously used for free-flowing electrolytes \citep{Hageman2023a}. The domain is discretised using quadratic tetrahedral elements, with a characteristic length of $1/10^\text{th}$ the size of the corrosion pit near the pit, $1/5^\text{th}$ the radius of the rebar near the rebar surface (excluding the pit area), and larger elements in the interior of the concrete. The temporal discretisation uses a backward Euler scheme, allowing for an unconditionally stable time discretisation. This scheme solves the concentrations and the electrolyte potential throughout the domain, and solves for a single metal potential. To retain stable and non-oscillatory results, and allow for large time increments, a lumped integration scheme is used for all reactions \citep{Hageman2023}. A more detailed description of the implementation is given in \ref{app:Implementation}. The implemented schemes have been verified to typical benchmark cases, e.g. one-dimensional diffusion/electro-migration, and it has been verified that obtained equilibrium concentrations match expected results. The code developed is made freely available at \url{https://mechmat.web.ox.ac.uk/codes} and \url{https://github.com/T-Hageman/ConcreteCorrosion}.

\section{Results}
\label{sec:Results}

\begin{table*}
    \centering
    \vspace*{-0.5cm}
    \hspace{-1.25cm}
    \begin{tabular}{|c l || r c|}
    \hline
    \hspace{1.0cm}Parameter \hspace{-1.0cm} & & Magnitude \hspace{-1.5cm} & \\
    \hline 
    \hline
         Diffusivity \citep{Lvov2015} & $D_{\text{H}^+}$ & $9.3\cdot10^{-9}$ & $\text{m}^2/\text{s}$  \\
         & $D_{\text{OH}^-}$ & $5.3\cdot10^{-9}$ & $\text{m}^2/\text{s}$  \\
         & $D_{\text{Na}^+}$ & $1.3\cdot10^{-9}$ & $\text{m}^2/\text{s}$  \\
         & $D_{\text{Cl}^-}$ & $2\cdot10^{-9}$ & $\text{m}^2/\text{s}$  \\
         & $D_{\text{Fe}^{2+}}$ & $1.4\cdot10^{-9}$ & $\text{m}^2/\text{s}$  \\
         & $D_{\text{FeOH}^+}$ & $1\cdot10^{-9}$ & $\text{m}^2/\text{s}$  \\
         & $D_{\text{O}_2}$ & $1\cdot10^{-9}$ & $\text{m}^2/\text{s}$ \\
         \hline
         Temperature & $T$ & $293.15$ & $\text{K}$ \\
         \hline
         Water auto-ionisation & $k_{\text{eq}}$ & $10^8$ & $\text{mol}/\text{m}^3/\text{s}$ \\
         constants & $K_{\text{w}}$ & $10^{-14}$ & $\;$ \\
         \hline
         Iron reaction constants & $k_{\text{fe}}$ & $10$ & $\text{mol}/\text{m}^3/\text{s}$ \\
         & $k_{\text{fe}}'$ & $10$ & $\text{mol}/\text{m}^3/\text{s}$ \\ 
         & $k_{\text{feoh}}$ & $10^{-2}$ & $\text{mol}/\text{m}^3/\text{s}$\\
         \hline
    \end{tabular}
        \begin{tabular}{|c c || c c|}
    \hline
    \hspace{0.7cm}Parameter\hspace{-0.7cm} & & Magnitude \citep{Holze2007} \hspace{-1.5cm} &  \\
    \hline 
    \hline
         Corrosion & $k_{\text{c}}$ & $0.5/F$ & $\text{mol}/\text{m}^2/\text{s}$  \\
         & $k_\text{c}'$ & $0.5/F$ & $\text{mol}/\text{m}^2/\text{s}$  \\
         & $E_{\text{eq},\text{c}}$ & $-0.4$ & $\text{V}_{\text{SHE}}$  \\
         & $\alpha_{\text{c}}$ & $0.5$ &   \\
         \hline
         Oxygen & $k_{\text{o}}$ & $2.5\cdot 10^{-5}/F$ & $\text{mol}/\text{m}^2/\text{s}$  \\
         & $k_\text{o}'$ & $2.5\cdot 10^{-5}/F$ & $\text{mol}/\text{m}^2/\text{s}$  \\
         & $E_{\text{eq},\text{o}}$ & $0.4$ & $\text{V}_{\text{SHE}}$  \\
         & $\alpha_{\text{o}}$ & $0.5$ &   \\
         \hline
         Hydrogen & $k_{\text{h}}$ & $5\cdot 10^{-3}/F$ & $\text{mol}/\text{m}^2/\text{s}$  \\
         & $k_\text{h}'$ & $0$ & $\text{mol}/\text{m}^2/\text{s}$  \\
         & $E_{\text{eq},\text{h}}$ & $0$ & $\text{V}_{\text{SHE}}$  \\
         & $\alpha_{\text{h}}$ & $0.5$ &   \\
         \hline
    \end{tabular} \hspace{-1cm}\\
    \caption{Properties, reaction constants, and equilibrium constants\textsuperscript{\ref{footnote:Eeq} } adopted for all the simulated cases. The values of the surface reaction constants are scaled by the Faraday constant $F$.}
    \label{tab:properites}
\end{table*}
\begin{figure}
    \centering
    \includegraphics{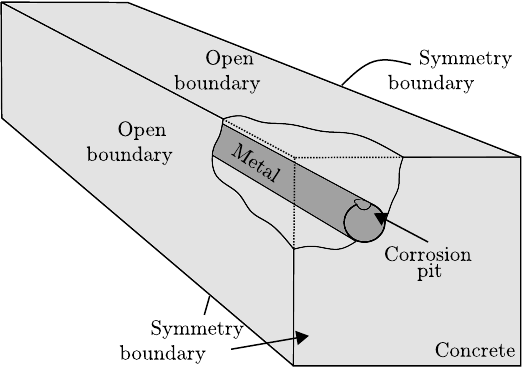}
    \caption{Overview of the simulated domain, including boundary conditions.}
    \label{fig:Domain}
\end{figure}

We consider a square concrete beam with rebar in each of its corners. Utilising symmetry, this reduces to the case shown in \cref{fig:Domain} composed of a single rebar in concrete, with symmetry planes on the front, bottom and right sides (modelling $1/8^\text{th}$ of the full beam). The dimensions of the modelled section use $10\;\text{cm}$ long concrete and rebar, a concrete cross section of $5\times5\;\text{cm}$ and a rebar with a diameter of $1\;\text{cm}$ (located $1\;\text{cm}$ deep within the concrete). On the centre of the rebar a corrosion pit is present with radius $0.4\;\text{mm}$ (a total area of $1\;\text{mm}^2$, or $0.5\;\text{mm}^2$ in our model due to symmetry) on which both the anodic and cathodic reactions are allowed, whereas on the exterior of the rebar, only the cathodic reactions are allowed to occur. Within the results, all total currents will refer to the case of \cref{fig:Domain}, such that our reported total currents correspond to the case of half a rebar. Thus to obtain the currents for a case with a pit in the centre of the rebar, the reported currents would need to be doubled, and for a full concrete beam, one would have to multiply the resulting currents by eight.

The exterior (the top and left sides) of the beam are exposed to the environment, prescribing constant ion concentrations $C_{\text{Fe}^{2+}}=C_{\text{FeOH}^{+}}=0\;\text{mol}/\text{m}^3$, $C_{\text{H}^+}=10^{-8}\;\text{mol}/\text{m}^3$ and $C_{\text{OH}^-}=1\;\text{mol}/\text{m}^3$ ($\text{pH}=11$). The $\text{Cl}^{-}$ concentration is altered between simulations to vary the conductivity, ranging from $10$ -- $500\;\text{mol}/\text{m}^3$ to cover the range from freshwater to seawater-saturated concrete, and the $\text{Na}^+$ concentration is adapted to fulfil the electroneutrality criterion (\cref{eq:electroneutrality}) at the boundaries. These concentrations are also imposed as initial conditions. Finally, simulations are performed with a prescribed oxygen concentration $C_{\text{O}_2}=1\;\text{mol}/\text{m}^3$ at the boundary, and with a no oxygen inflow boundary condition. These two conditions emulate the concrete either being exposed to an oxygen-containing environment, with this oxygen aiding the corrosion, or being isolated from it and having the corrosion solely able to use the initially present oxygen. For both these oxygen boundary conditions, the initial oxygen concentration is $C_{\text{O}_2}=1\;\text{mol}/\text{m}^3$.  
In our model, this concentration represents an average oxygen concentration, representing a weighted balance between oxygen dissolved in water and in air, with this corresponding to an assumption of instantaneous dissolution and desorption. This oxygen boundary condition is kept constant in all the cases considered here, spanning fully water-saturated and partially saturated concrete. This ensures that, in our results, variations in porosity remain the primary focus rather than changes in oxygen availability affecting the corrosion process. Similarly, the diffusivity of oxygen through the pores is assumed constant, rather than being dependent on changes in water saturation.
The reaction constants and properties are given in \cref{tab:properites}.

\begin{figure}
    \centering
    \begin{subfigure}{0.49\textwidth}
        \centering
        \includegraphics[clip=true, trim={0 0 0 0}]{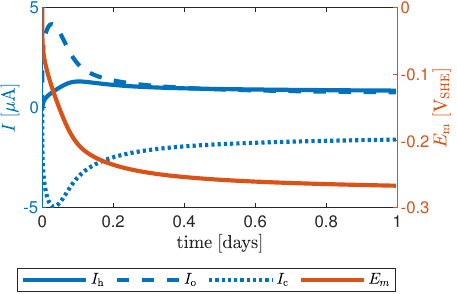}
        \caption{Reaction currents}
    \end{subfigure}
    \begin{subfigure}{0.49\textwidth}
        \centering
        \includegraphics[clip=true, trim={0 0 0 0}]{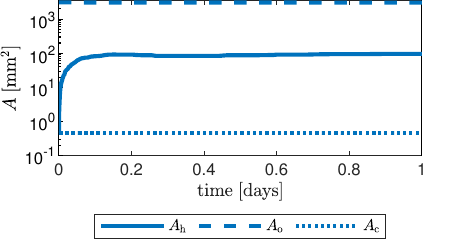}
        \caption{Reaction area}
    \end{subfigure}
    \caption{Reaction currents and area over which the reactions occur for the case with external oxygen, using $\phi=1\%$ and $C_\text{Cl}=500\;\mathrm{mol}/\mathrm{m}^3$. }
    \label{fig:Sw1_OX_Evolutions}
\end{figure}

\begin{figure}
    \centering
    \begin{subfigure}{0.49\textwidth}
        \includegraphics[clip=true, trim={0 00 0 0}]{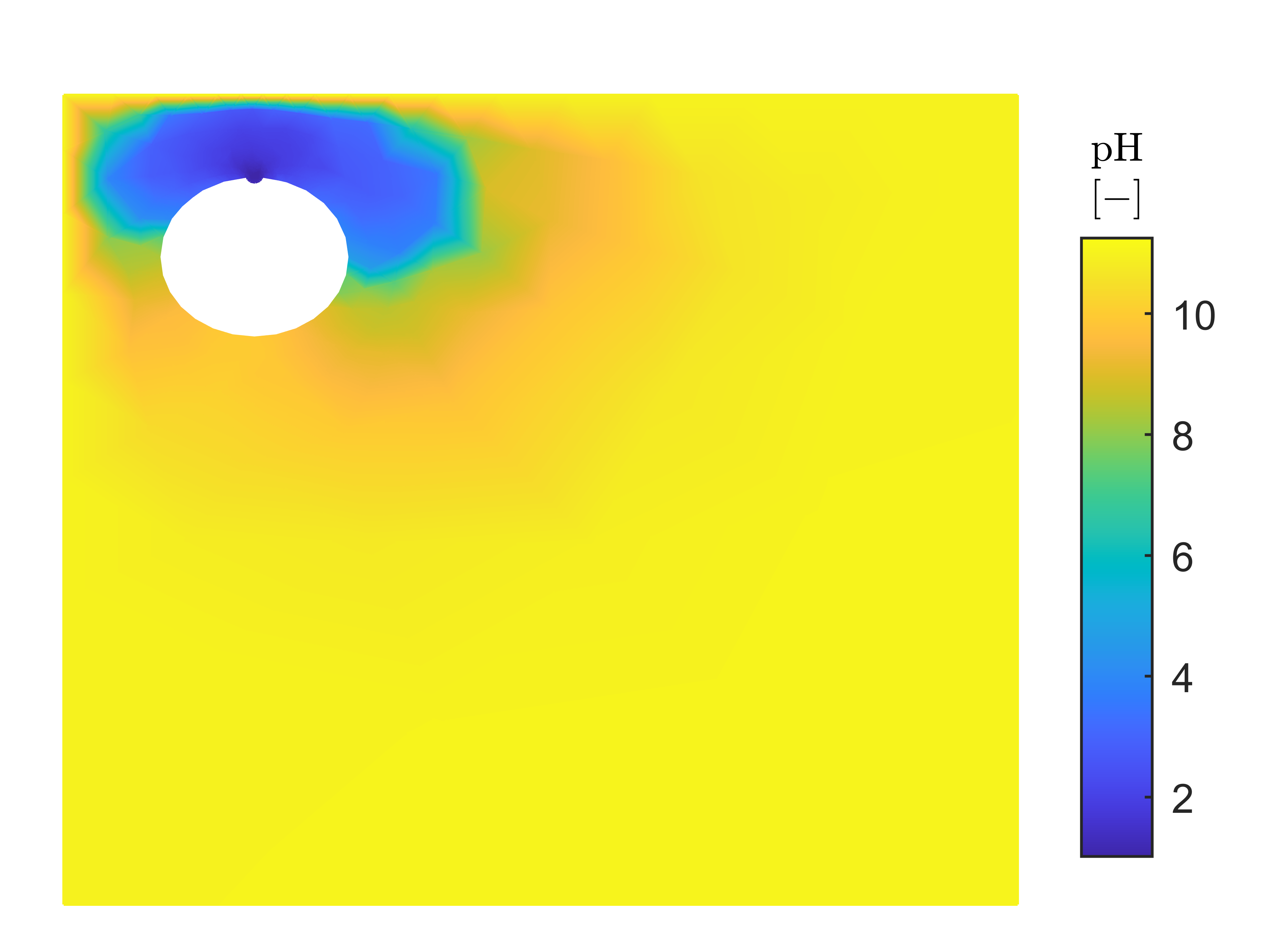}
        \caption{Electrolyte pH on the front symmetry surface}
    \end{subfigure}
    \begin{subfigure}{0.49\textwidth}
        \includegraphics[clip=true, trim={0 00 0 0}]{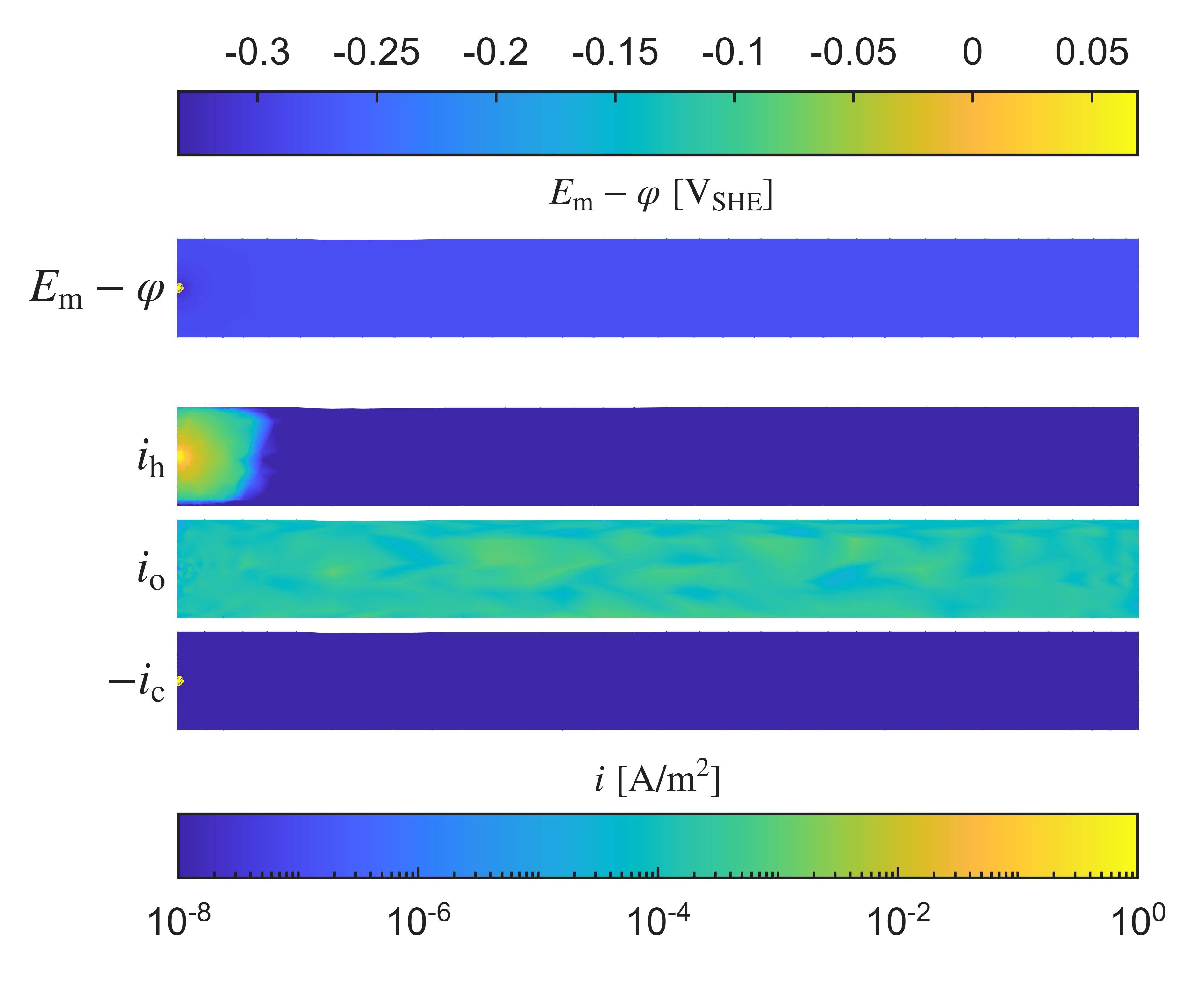}
        \caption{Reaction currents and electric overpotential on the top half of the rebar}
    \end{subfigure}
    \caption{pH and locations of reaction currents at steady-state ($28\;\text{days}$) for the case with external oxygen, using $\phi=1\%$ and $C_\text{Cl}=500\;\mathrm{mol}/\mathrm{m}^3$. }
    \label{fig:Sw1_OX_Surfs}
\end{figure}

\subsection{Localisation of reaction currents}
We first consider cases in which the concrete is fully water saturated, $S_\text{w}=1$, and consider a low porosity of $1\%$ and initial/boundary chlorine concentration of $C_\text{Cl}=500\;\text{mol}/\text{m}^3$. When external oxygen can enter the domain, the evolution of the reaction currents and reaction areas is shown in \cref{fig:Sw1_OX_Evolutions}. Initially, the corrosion reaction is fully supported by the oxygen evolution reaction. However, as the initially present oxygen depletes, the oxygen evolution reaction rate reduces. This is initially balanced by the changes in electric overpotential, allowing the oxygen evolution reaction rate to increase over the first couple of hours. After this initial time, the diffusion rate of oxygen from the boundary to the metal surface becomes limiting, resulting in a near-constant oxygen evolution reaction rate. In contrast, the hydrogen evolution reaction starts at a near-negligible rate. As corrosion occurs, and the corrosion products acidify the corrosion pit and surrounding region, the hydrogen evolution reaction accelerates. As this is dependent on the $\text{H}^+$ ions resulting from corrosion products, this reaction occurs over a smaller area compared to the oxygen reaction. After approximately a day, these reactions stabilise with the oxygen and hydrogen reaction rates having a similar order of magnitude, and the corrosion reaction having double the rate. Interestingly, this indicates that even though the corrosion reaction products are theoretically able to supply enough $\text{H}^+$ ions to sustain the corrosion reaction, only about half of these $\text{H}^+$ ions are used. This is likely due to the auto-ionisation reaction consuming $\text{H}^+$, the corrosion product hydrolysis not fully occurring, and part of the created $\text{H}^+$ ions diffusing out of the concrete. 

\begin{figure}
    \centering
    \begin{subfigure}{0.49\textwidth}
        \centering
        \includegraphics[clip=true, trim={0 0 0 0}]{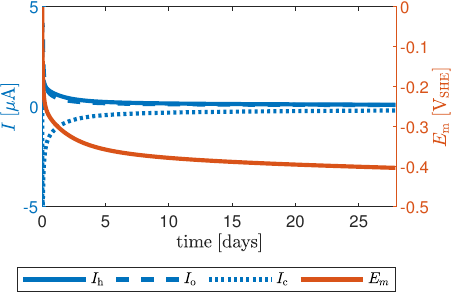}
        \caption{Reaction currents}
    \end{subfigure}
    \begin{subfigure}{0.49\textwidth}
        \centering
        \includegraphics[clip=true, trim={0 0 0 0}]{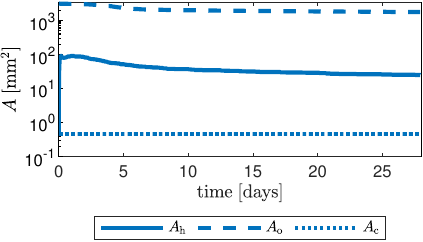}
        \caption{Reaction area}
    \end{subfigure}
    \caption{Reaction currents and area over which the reactions occur for the case without external oxygen, using $\phi=1\%$ and $C_\text{Cl}=500\;\mathrm{mol}/\mathrm{m}^3$. }
    \label{fig:Sw1_NoOX_Evolutions}
\end{figure}

\begin{figure*}
    \centering
    \begin{subfigure}{0.3\textwidth}
        \includegraphics[width=\textwidth]{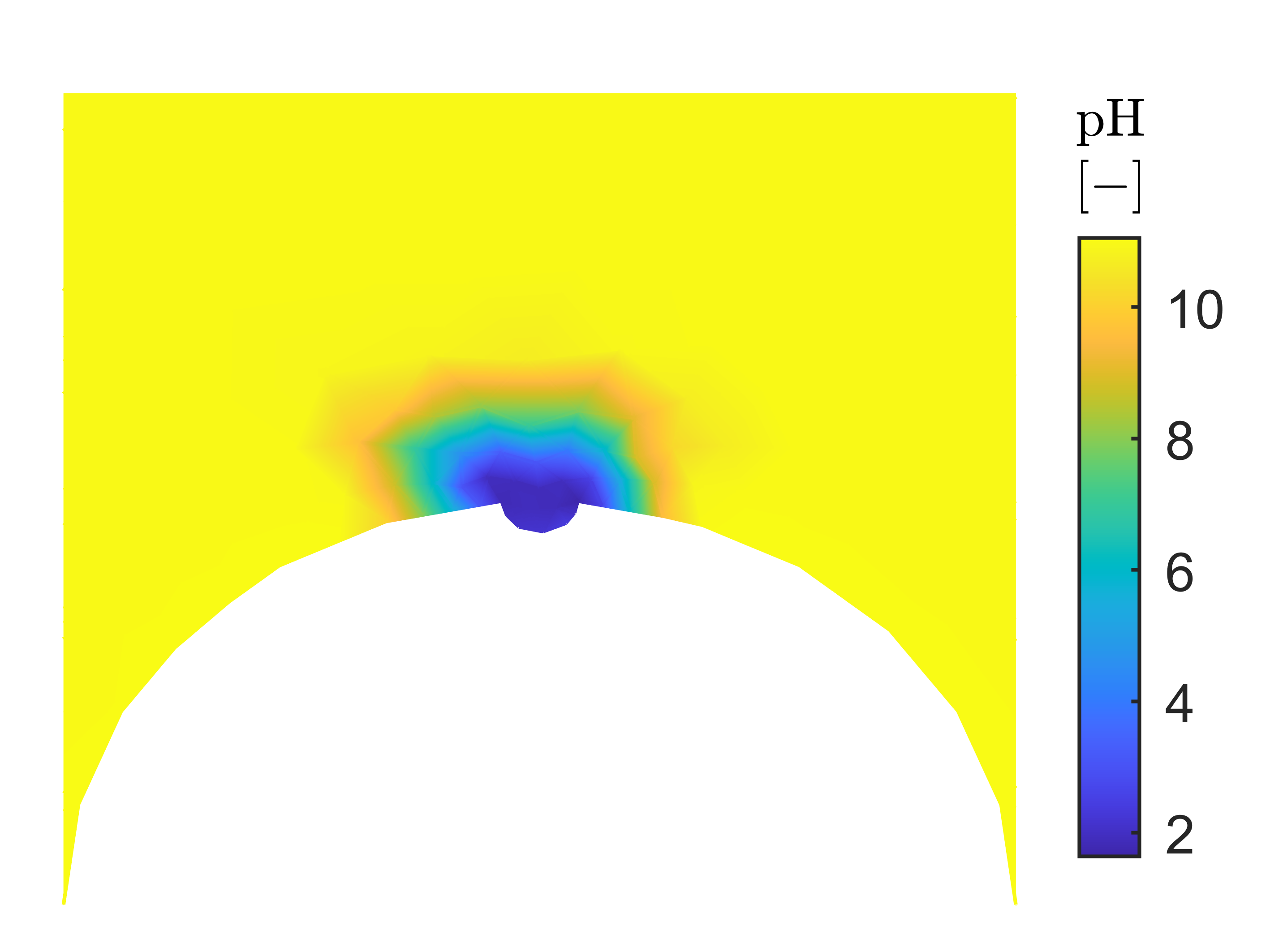}
        \caption{$t=30\;\text{s}$}
    \end{subfigure}
    \begin{subfigure}{0.3\textwidth}
        \includegraphics[width=\textwidth]{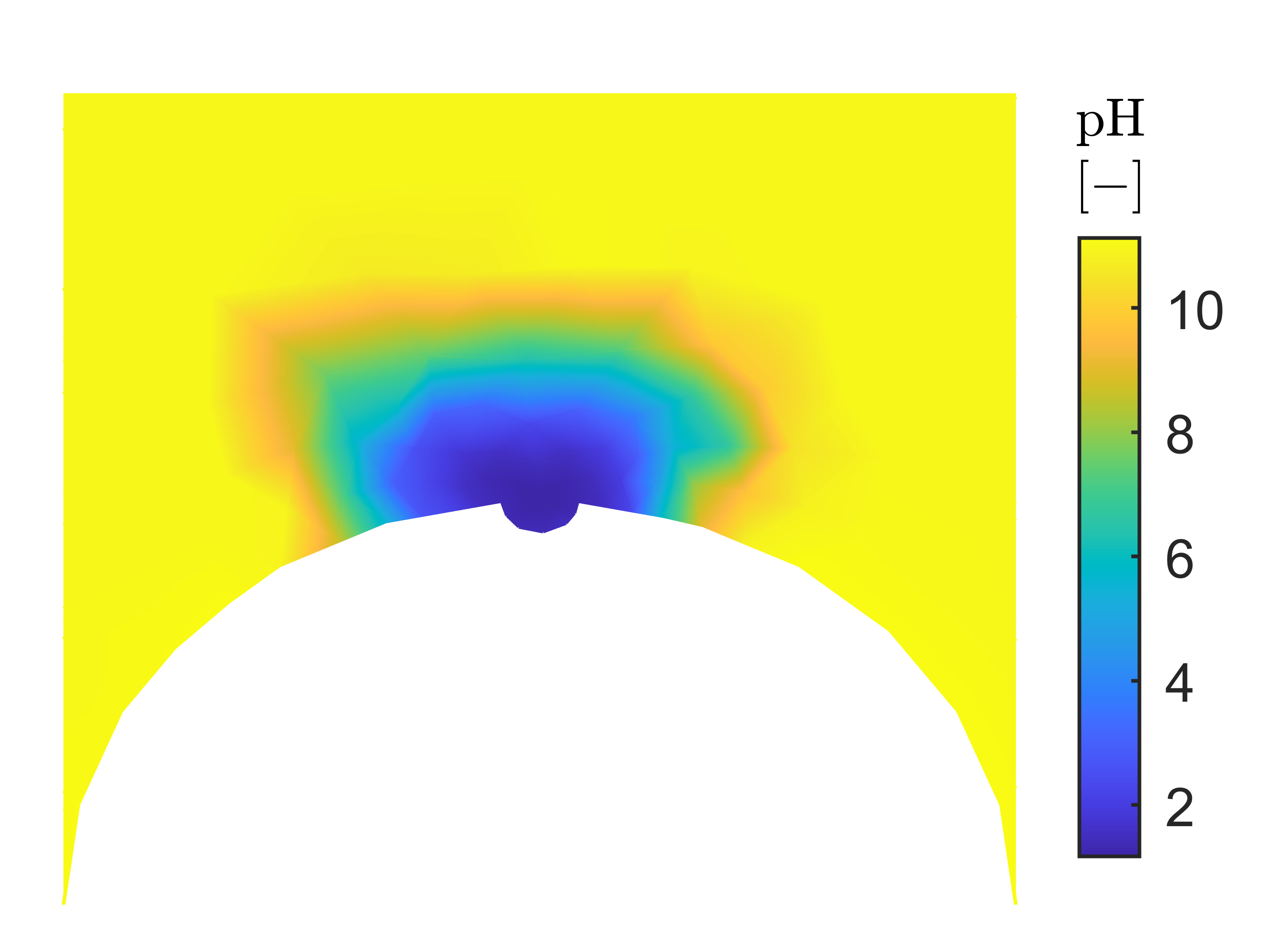}
        \caption{$t=2\;\text{min}$}
    \end{subfigure}
    \begin{subfigure}{0.3\textwidth}
        \includegraphics[width=\textwidth]{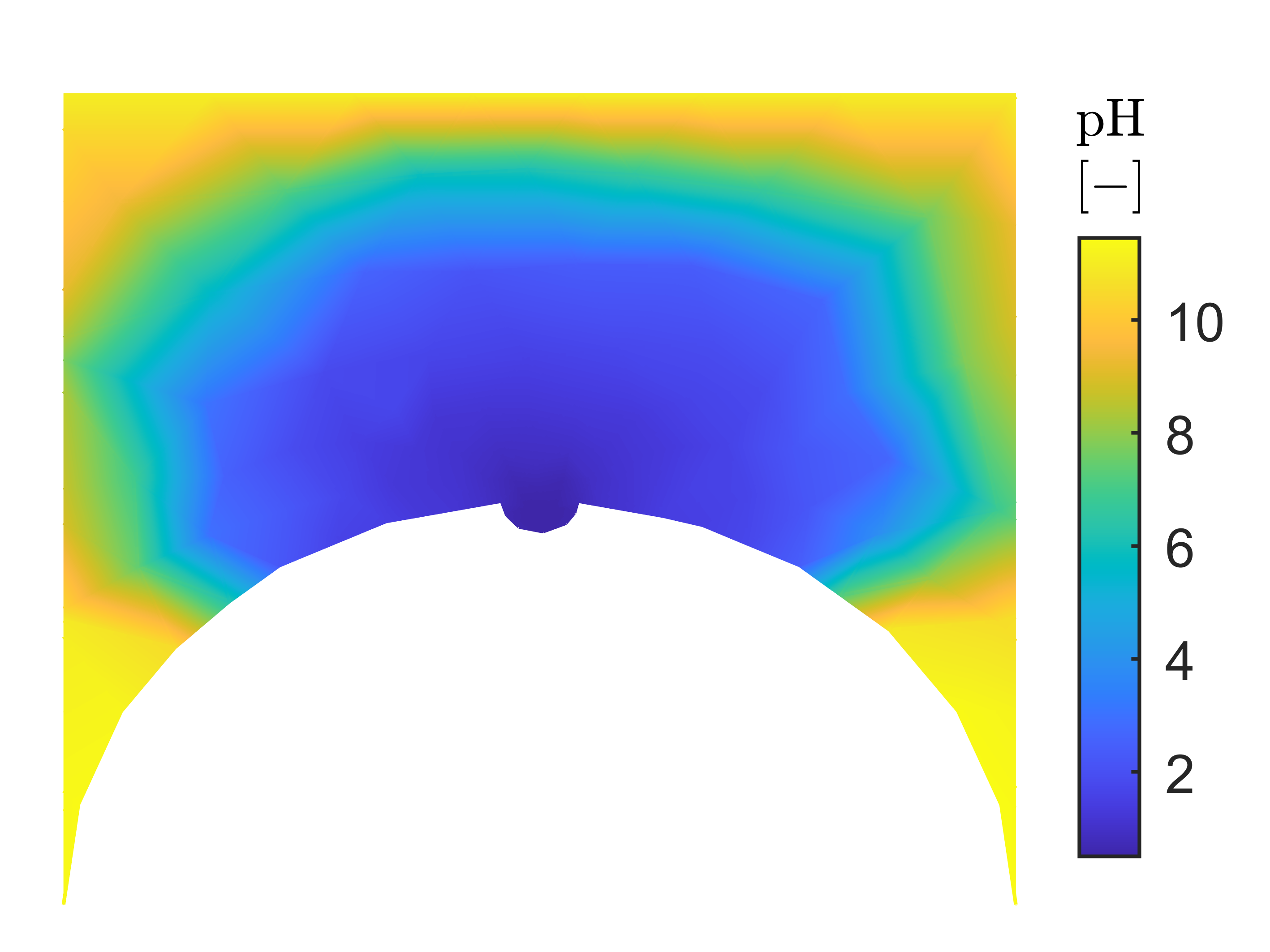}
        \caption{$t=15\;\text{min}$}
    \end{subfigure}
    \begin{subfigure}{0.3\textwidth}
        \includegraphics[width=\textwidth]{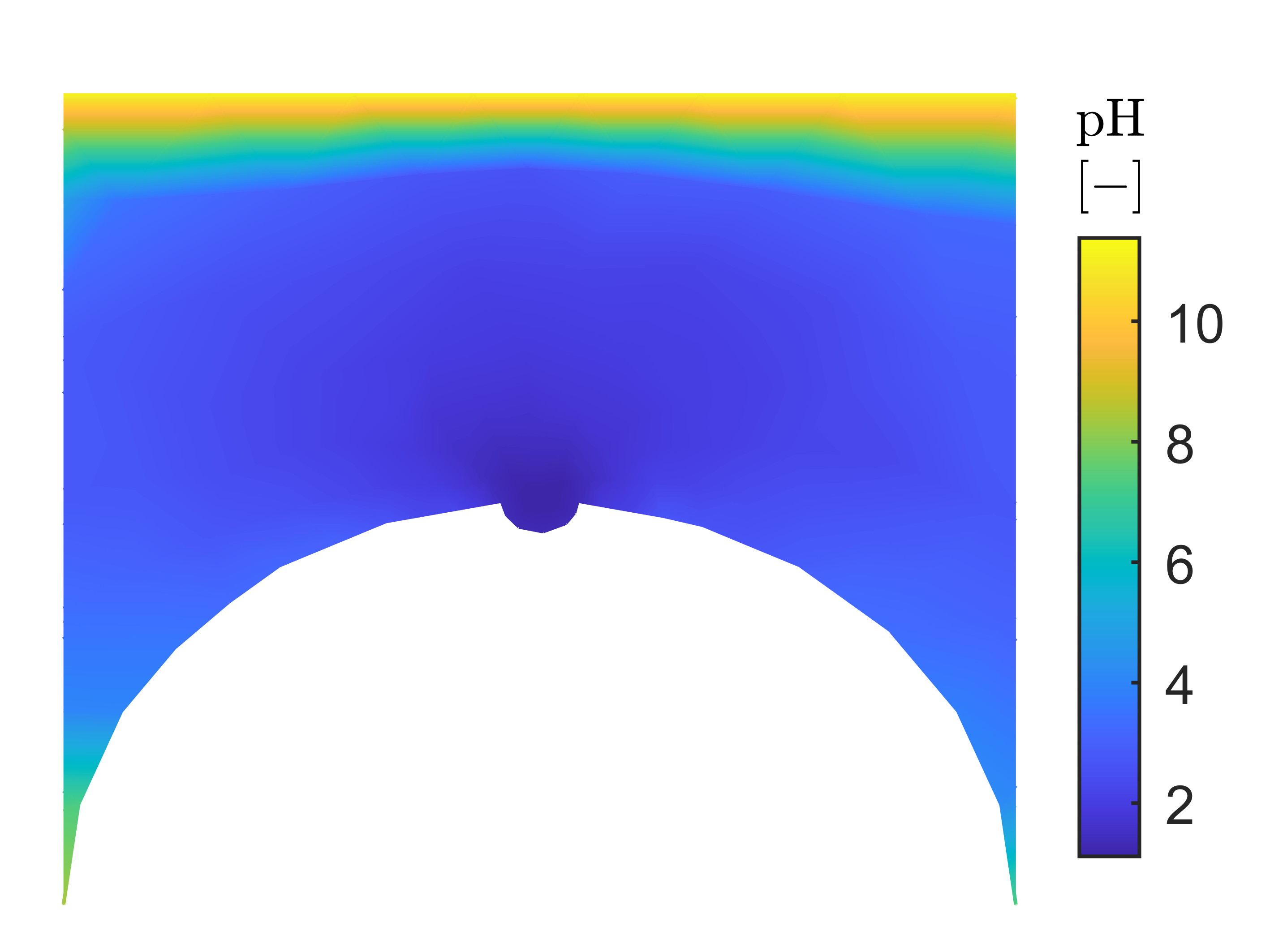}
        \caption{$t=1\;\text{days}$}
    \end{subfigure}
    \begin{subfigure}{0.3\textwidth}
        \includegraphics[width=\textwidth]{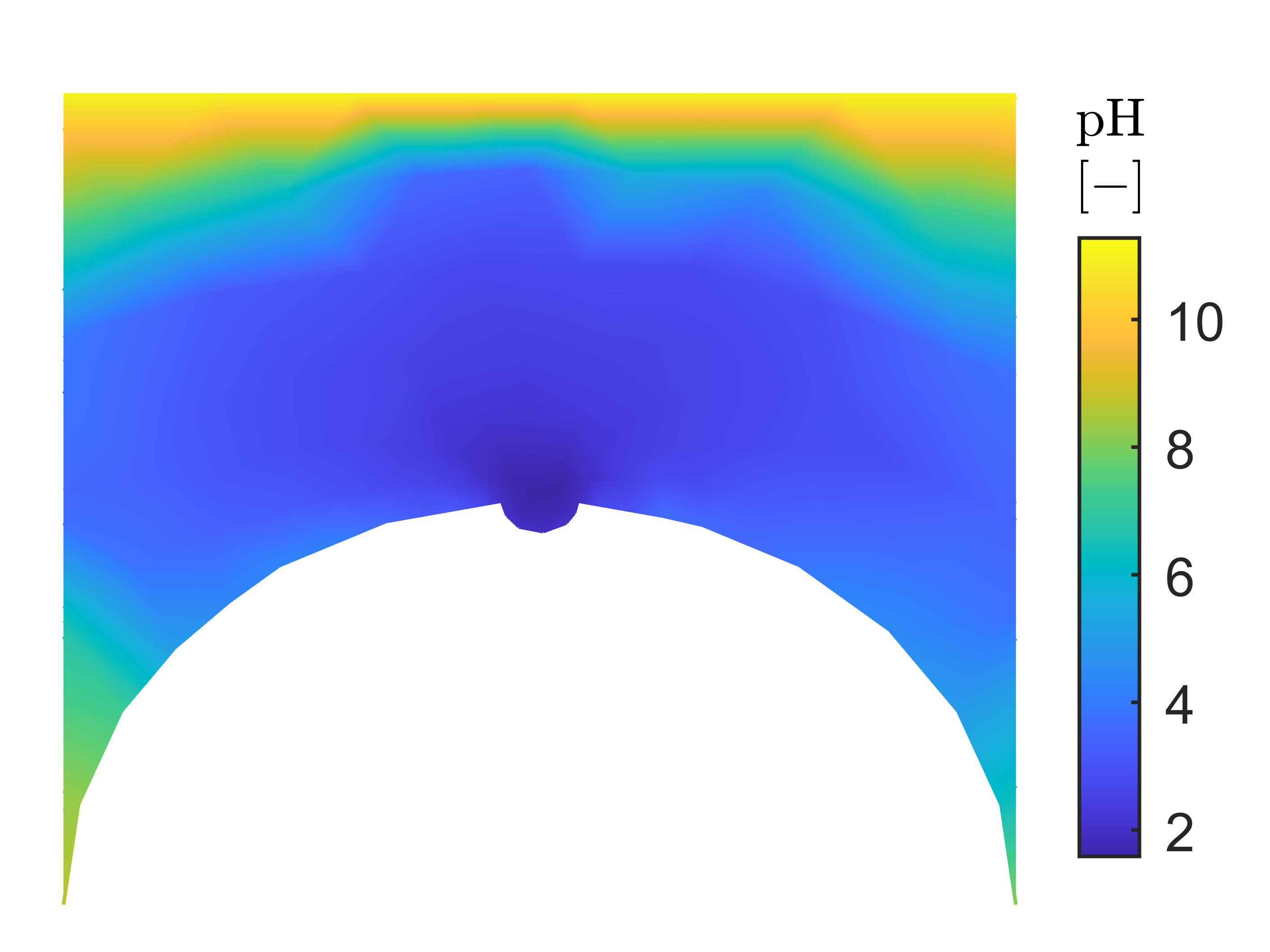}
        \caption{$t=4\;\text{days}$}
    \end{subfigure}
    \begin{subfigure}{0.3\textwidth}
        \includegraphics[width=\textwidth]{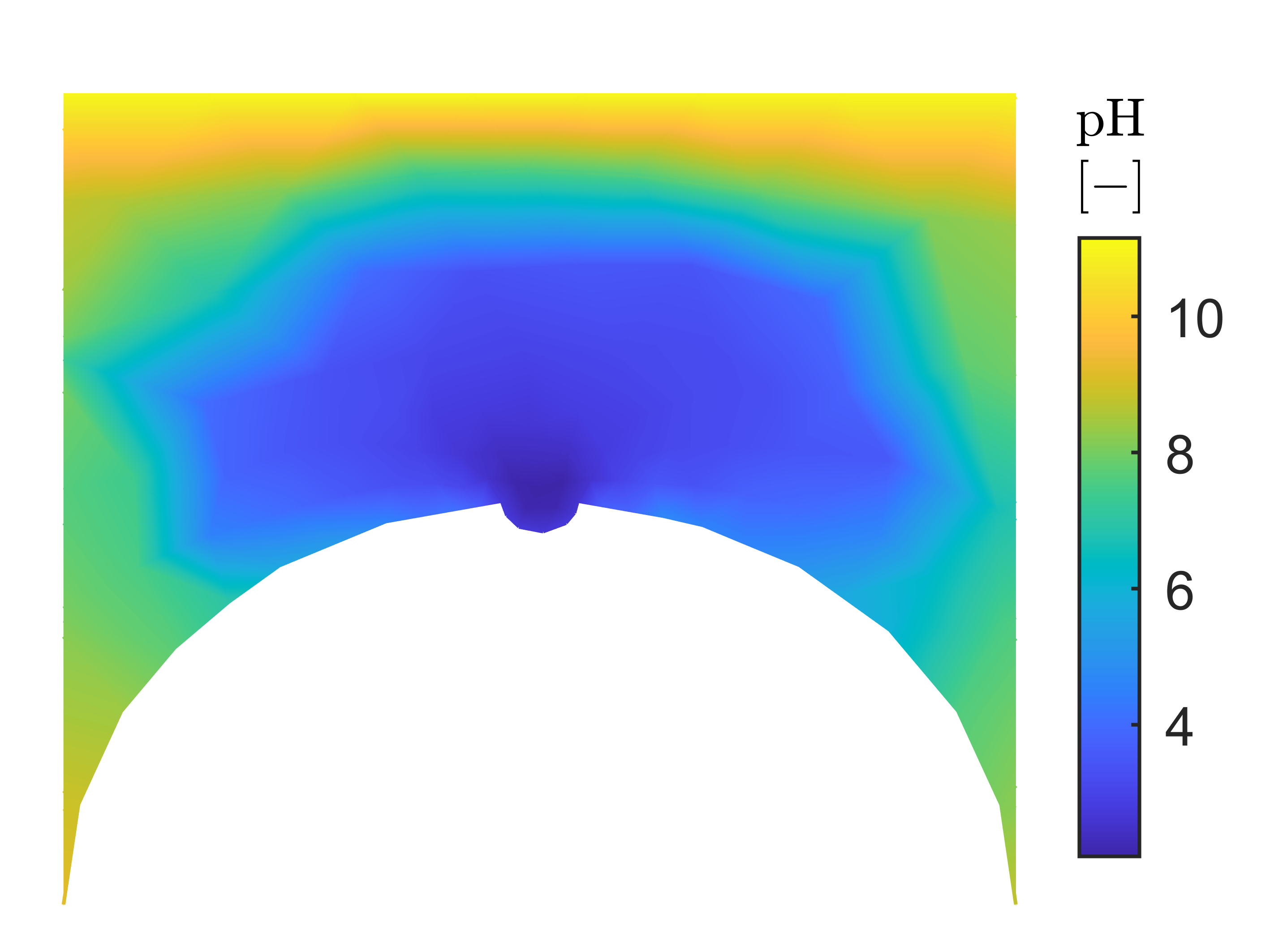}
        \caption{$t=28\;\text{days}$}
    \end{subfigure}
    \caption{Evolution of pH for $\phi=1\%$ and $C_\text{Cl}=500\;\text{mol}/\text{m}^3$, for the case without external oxygen. }
    \label{fig:Evolution_pH_over_time}
\end{figure*}

The resulting pH in the electrolyte and the location of the reactions are shown in \cref{fig:Sw1_OX_Surfs}. The corrosion pit and the area surrounding it become highly acidic, with a pH around 2. Farther from the corrosion pit, the original pH is maintained by the $\text{OH}^-$ ions resulting from the oxygen evolution reactions. As a result, the hydrogen evolution reaction only occurs in the area surrounding the pit, as shown in \cref{fig:Sw1_OX_Surfs}b. In contrast, as the oxygen reaction rate is dictated by the diffusion of oxygen to the metal surface, this reaction occurs at a similar rate everywhere on the metal surface. It should be noted, however, that the hydrogen evolution reaction current reaches $i_\text{h}\approx 1\;\mu\text{A}/\text{mm}^2$ inside the corrosion pit, whereas the oxygen evolution current is orders of magnitude lower, $i_\text{o}\approx 10^{-3}\;\mu\text{A}/\text{mm}^2$. As a result, the oxygen evolution reaction current is likely to be strongly dependent on the length of the rebar (see also the discussion in \ref{App:Geo}), whereas the hydrogen evolution reaction only interacts with the $1\text{--} 2\;\text{cm}$ of rebar surrounding the corrosion pit, and only with the side where the pit is located (an area of $\approx 100 \;\mathrm{mm}^2$). Looking at the electric overpotential, it is clear that the corrosion reaction rate is limited by the potential within the corrosion pit, with this potential accelerating the hydrogen reaction, whereas the oxygen reaction is impacted by the near-constant overpotential on the outside of the rebar. 

\begin{figure*}
    \centering
    \begin{subfigure}{0.3\textwidth}
        \includegraphics[width=\textwidth]{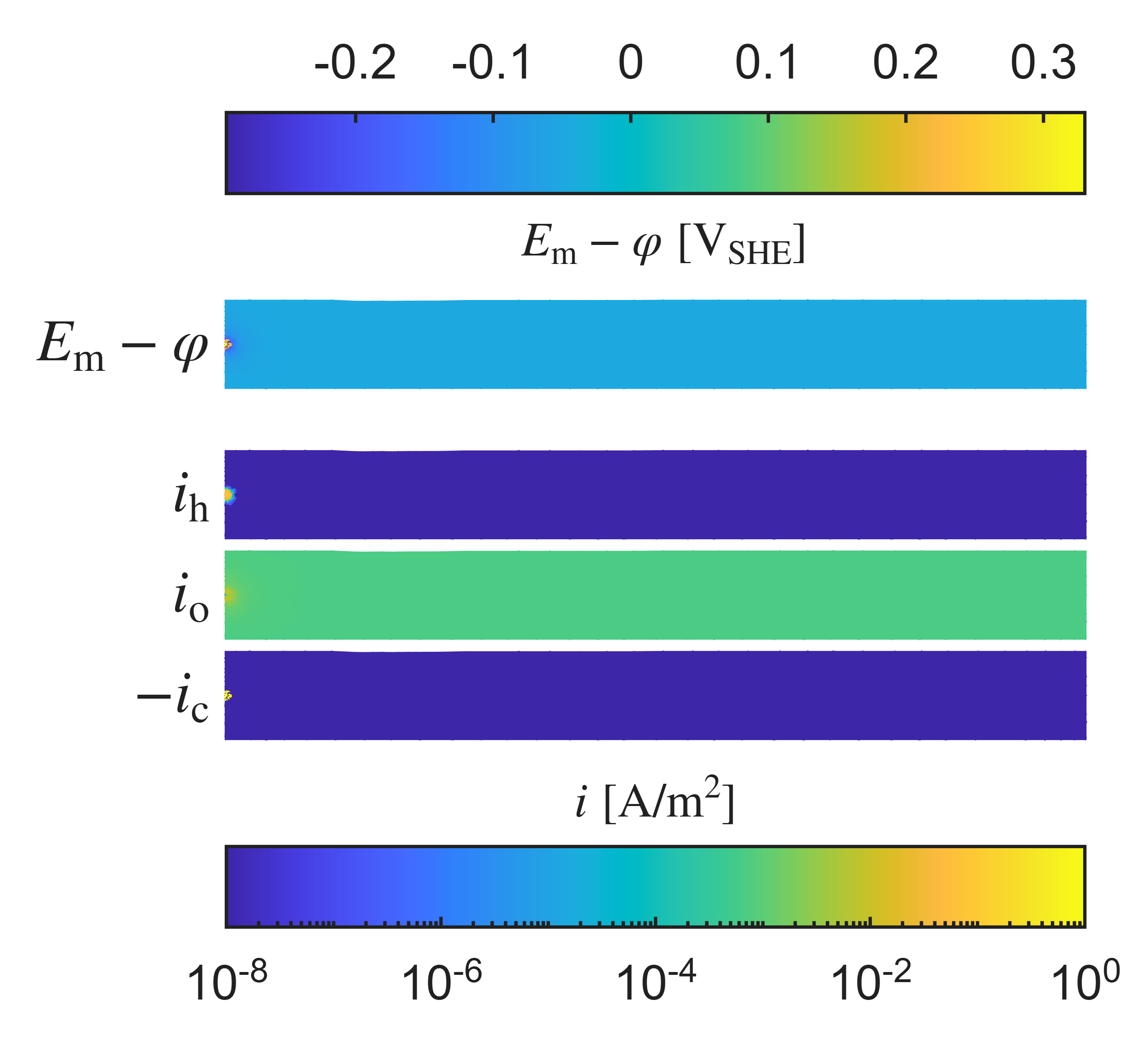}
        \caption{$t=30\;\text{s}$}
    \end{subfigure}
    \begin{subfigure}{0.3\textwidth}
        \includegraphics[width=\textwidth]{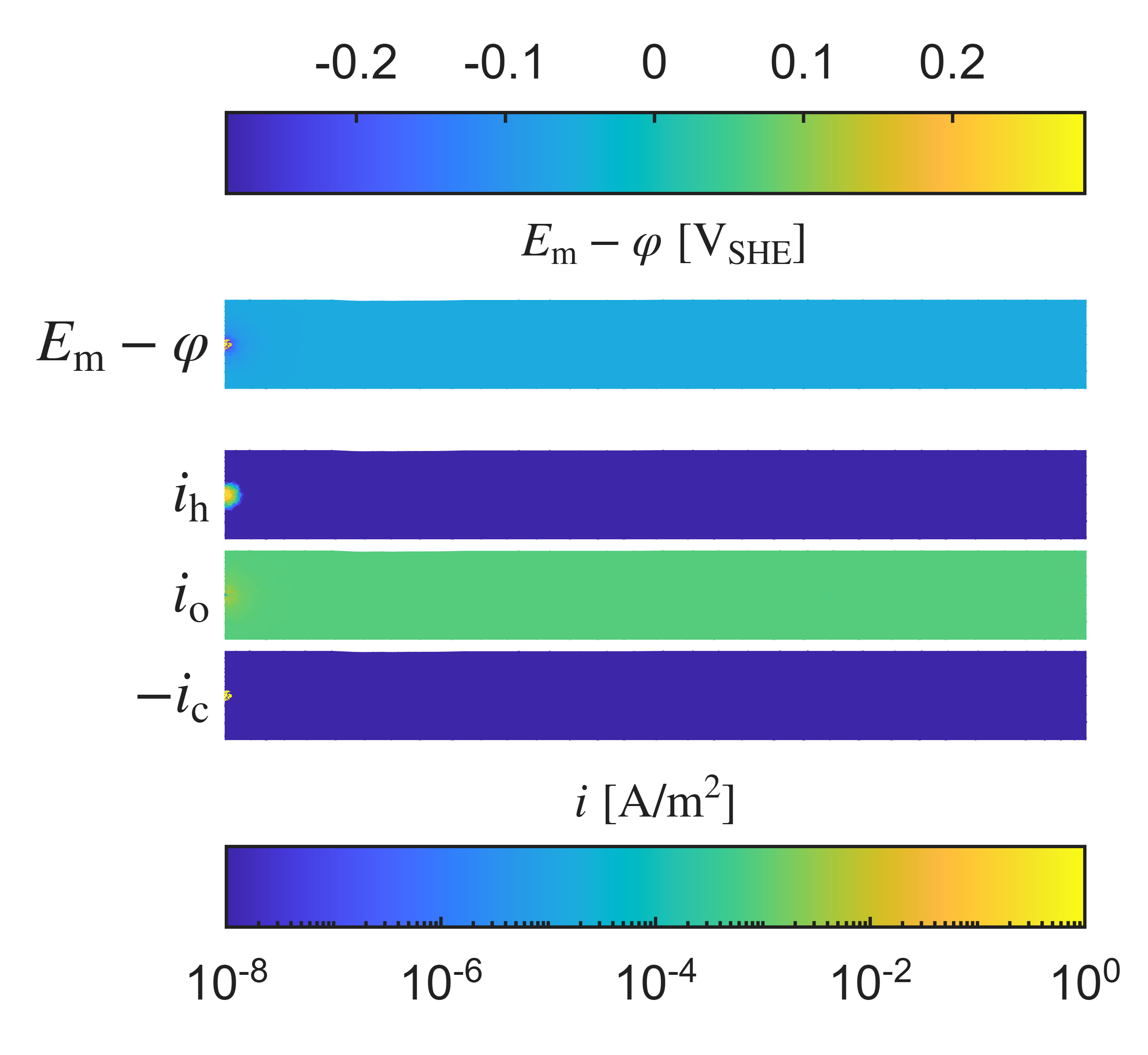}
        \caption{$t=2\;\text{min}$}
    \end{subfigure}
    \begin{subfigure}{0.3\textwidth}
        \includegraphics[width=\textwidth]{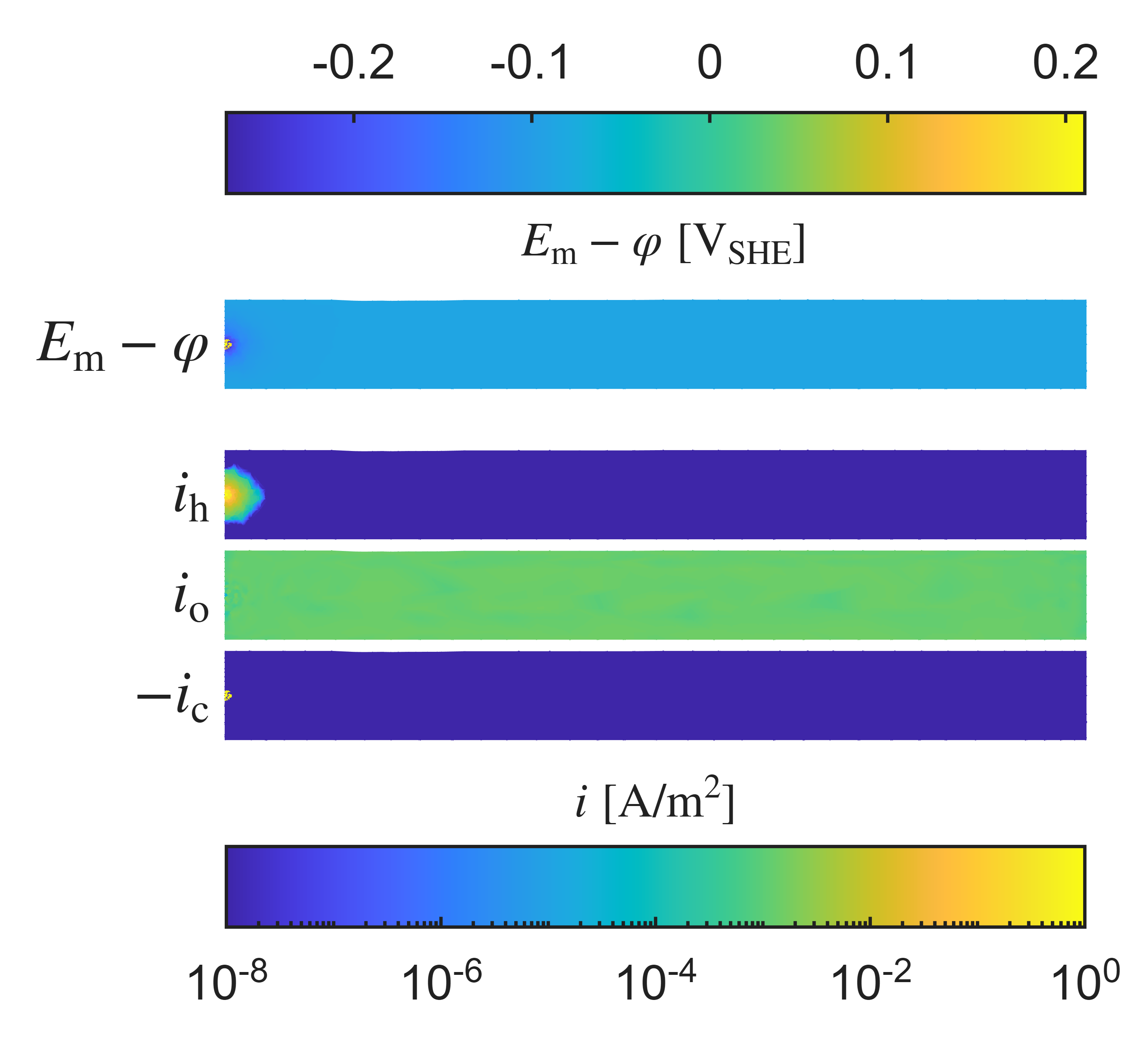}
        \caption{$t=15\;\text{min}$}
    \end{subfigure}
    \begin{subfigure}{0.3\textwidth}
        \includegraphics[width=\textwidth]{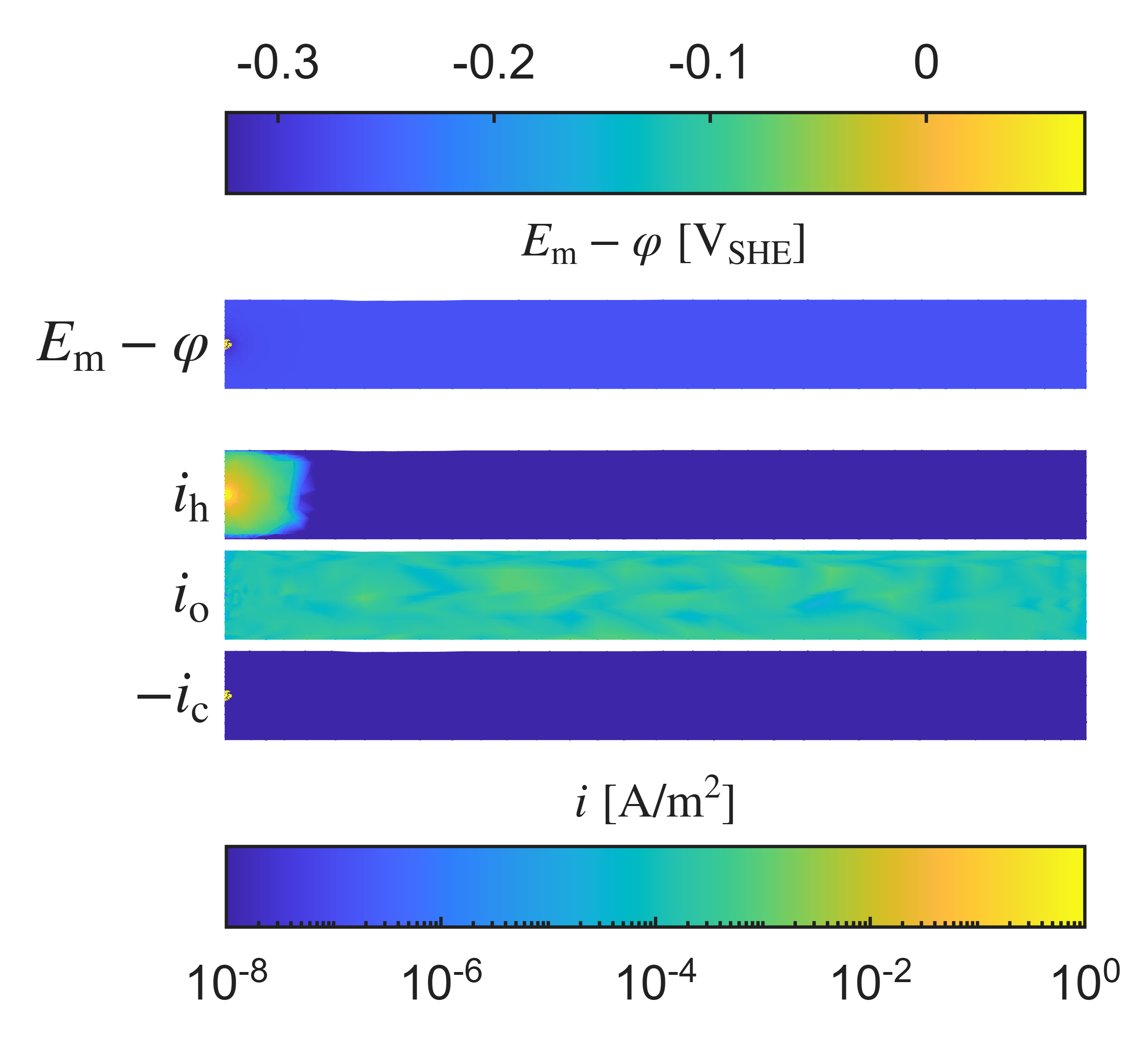}
        \caption{$t=1\;\text{days}$}
    \end{subfigure}
    \begin{subfigure}{0.3\textwidth}
        \includegraphics[width=\textwidth]{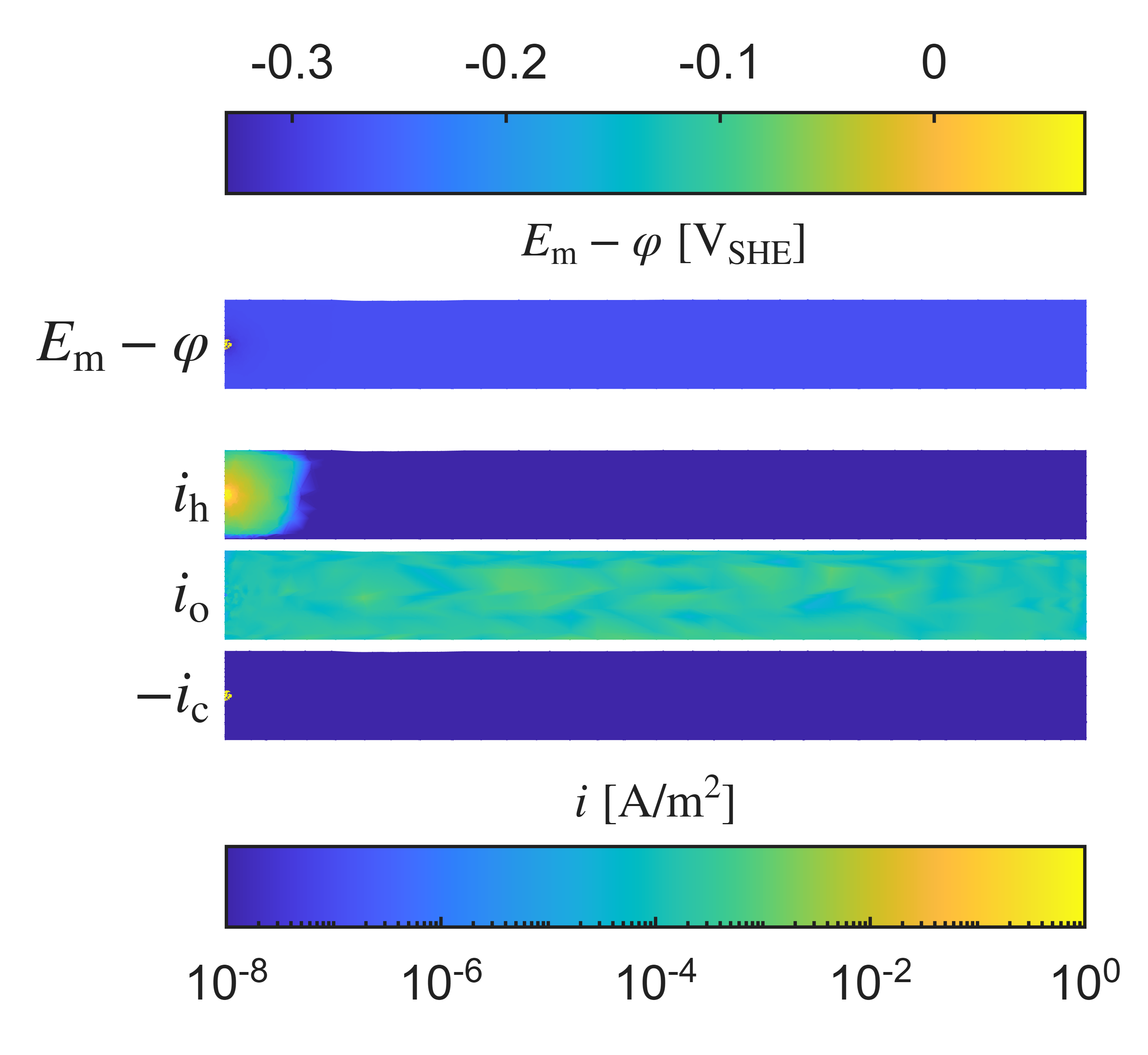}
        \caption{$t=4\;\text{days}$}
    \end{subfigure}
    \begin{subfigure}{0.3\textwidth}
        \includegraphics[width=\textwidth]{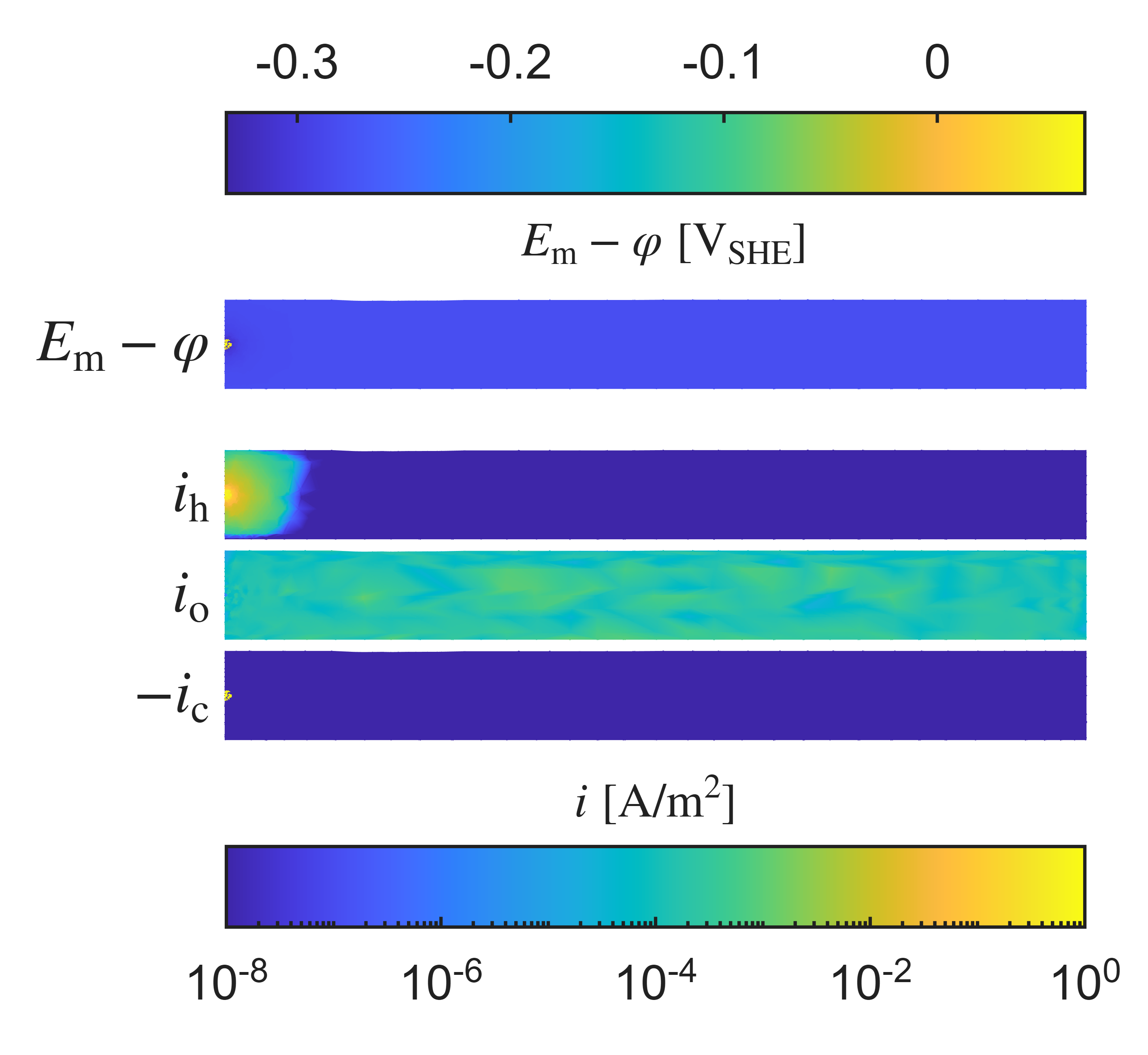}
        \caption{$t=28\;\text{days}$}
    \end{subfigure}
    \caption{Evolution of surface reaction rates for $\phi=1\%$ and $C_\text{Cl}=500\;\text{mol}/\text{m}^3$, for the case without external oxygen. }
    \label{fig:Evolution_rates_over_time}
\end{figure*}

If we instead consider a situation where oxygen is initially present, but no additional oxygen can enter from the external boundaries, the reaction currents and areas from \cref{fig:Sw1_NoOX_Evolutions} are obtained. In the first hours, the reaction currents, electric potential, and ion concentrations are  identical to that of cases with external oxygen (see \cref{fig:Sw1_OX_Evolutions}), as this is driven by the oxygen present from the onset of the simulations. However, as the oxygen depletes the oxygen reaction current is decreased. Over time, this lowers the corrosion current and, as fewer corrosion products are available to acidify the corrosion pit, also lowers the hydrogen evolution reaction rate. 

As the corrosion rate decreases, so does the size of the acidic region surrounding the corrosion pit. This is shown in \cref{fig:Evolution_pH_over_time}. During the initial minutes-hours, the corrosion pit goes from being basic to being highly acidic, driven by the high rate of corrosion due to the presence of oxygen. As the oxygen depletes, and the reaction rates are reduced, the acidic region shrinks in size. The impact of this on the reaction rates is shown in \cref{fig:Evolution_rates_over_time}: the region over which the hydrogen evolution reaction occurs initially increases, peaks after approximately one day, and then slowly shrinks in size. While the oxygen initially present is sufficient to start the corrosion process, and initialise the acidic pit, it is expected that as time passes, the corrosion rate will keep decreasing and, without additional oxygen entering the concrete, will eventually stop. 

While we do not include detailed results for all the cases studied, they all follow a similar trend to that described in the above section. To more quantitatively analyse the differences between the cases with regards to the reaction rates, aggregate data will be presented in the next sections.

\begin{figure*}
    \centering
    \begin{subfigure}{8cm}
         \centering
         \includegraphics{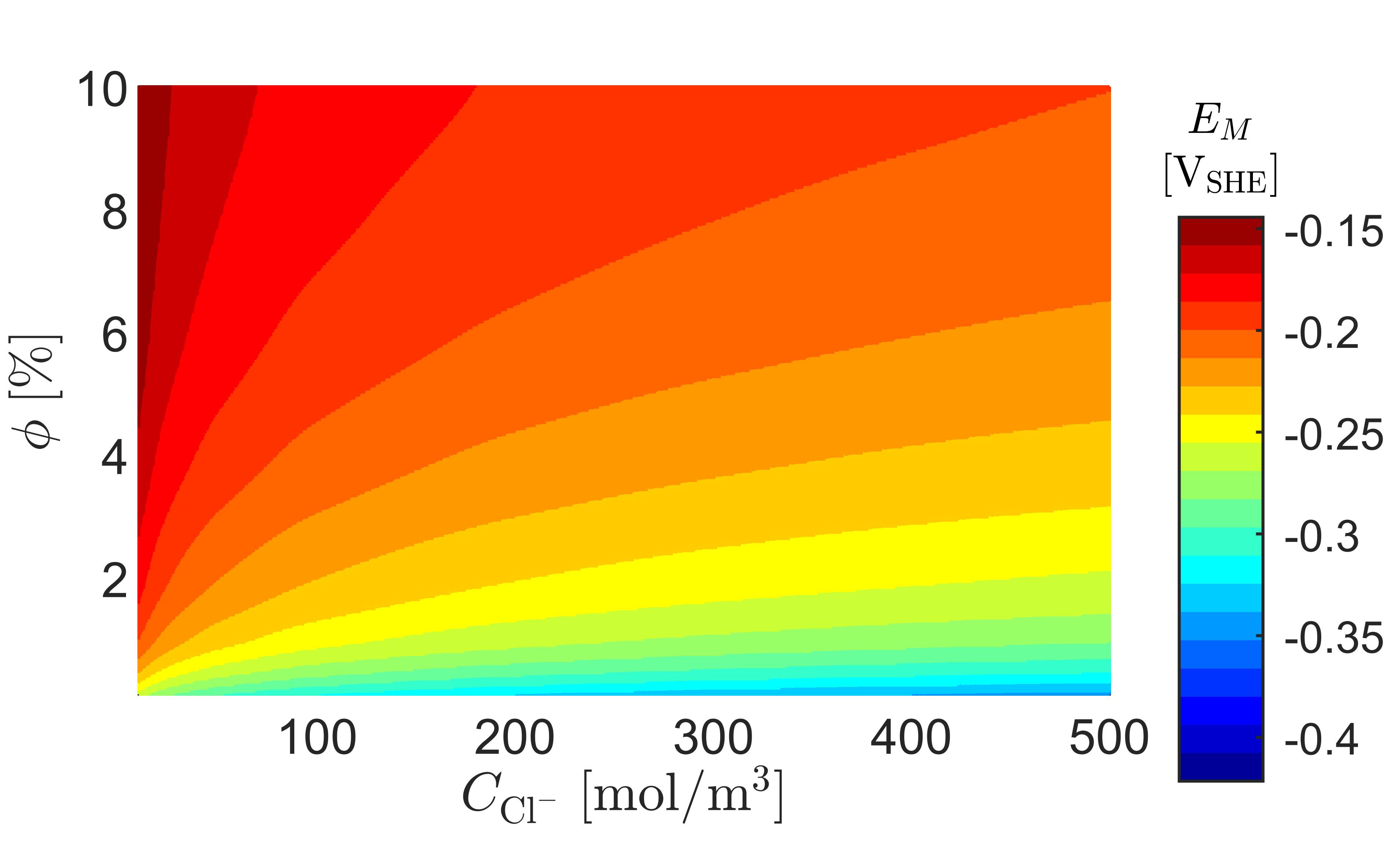}
         \caption{ }
         \label{fig:surfs_Em_ox}
    \end{subfigure}
    \begin{subfigure}{8cm}
         \centering
         \includegraphics{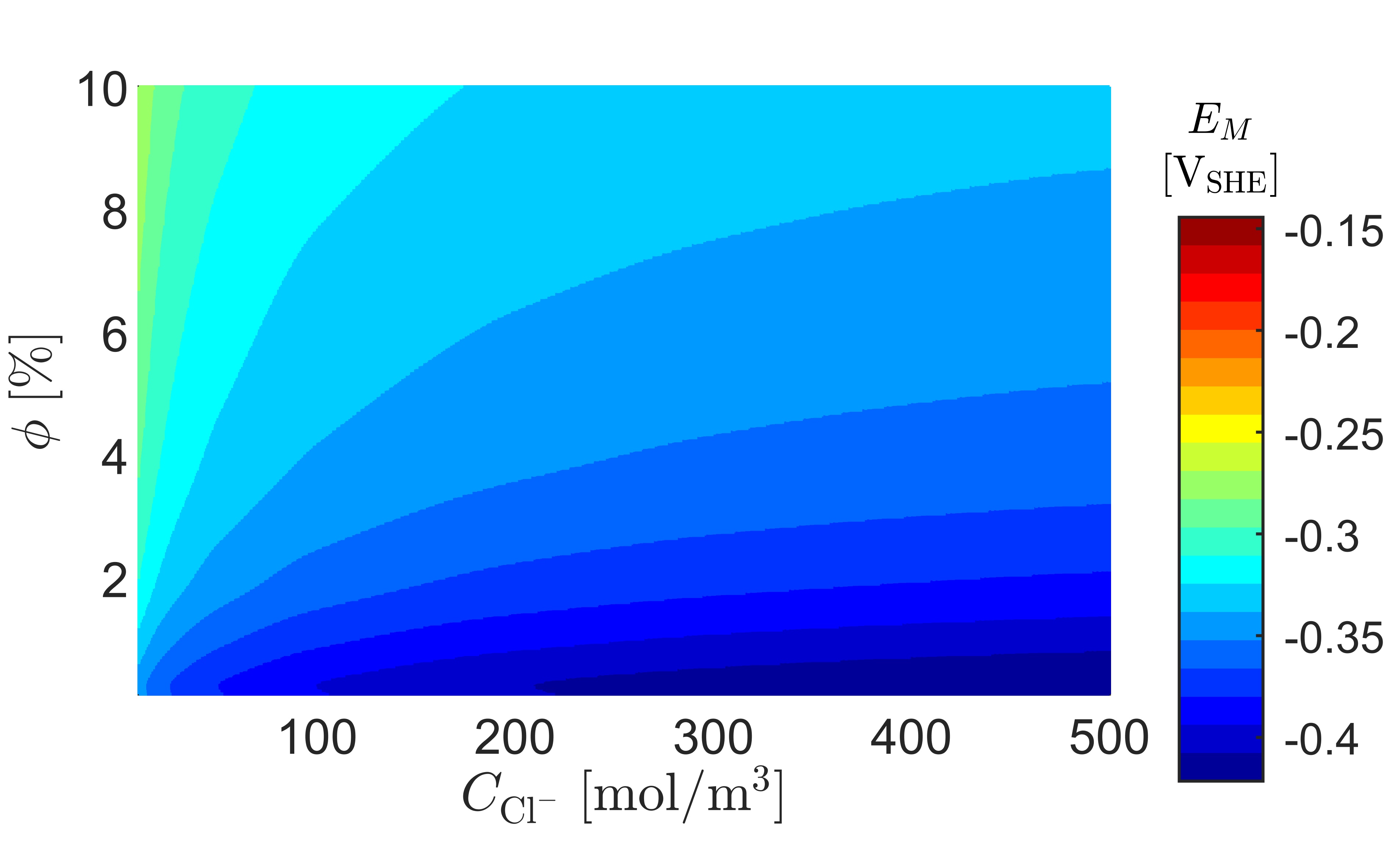}
         \caption{ }
         \label{fig:surfs_Em_noox}
     \end{subfigure}
    \begin{subfigure}{8cm}
         \centering
         \includegraphics{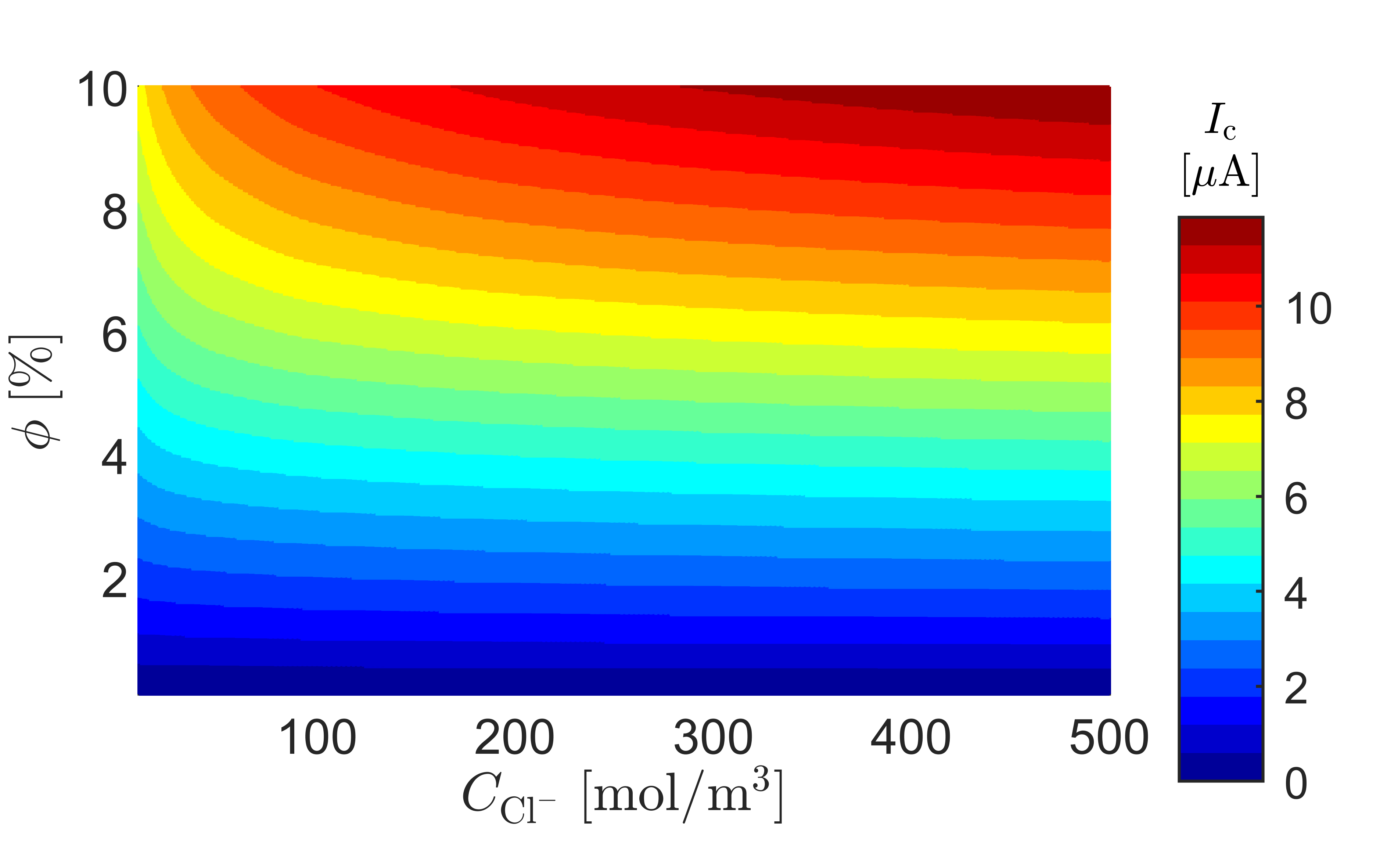}
         \caption{ }
         \label{fig:surfs_Fe_ox}
    \end{subfigure}
    \begin{subfigure}{8cm}
         \centering
         \includegraphics{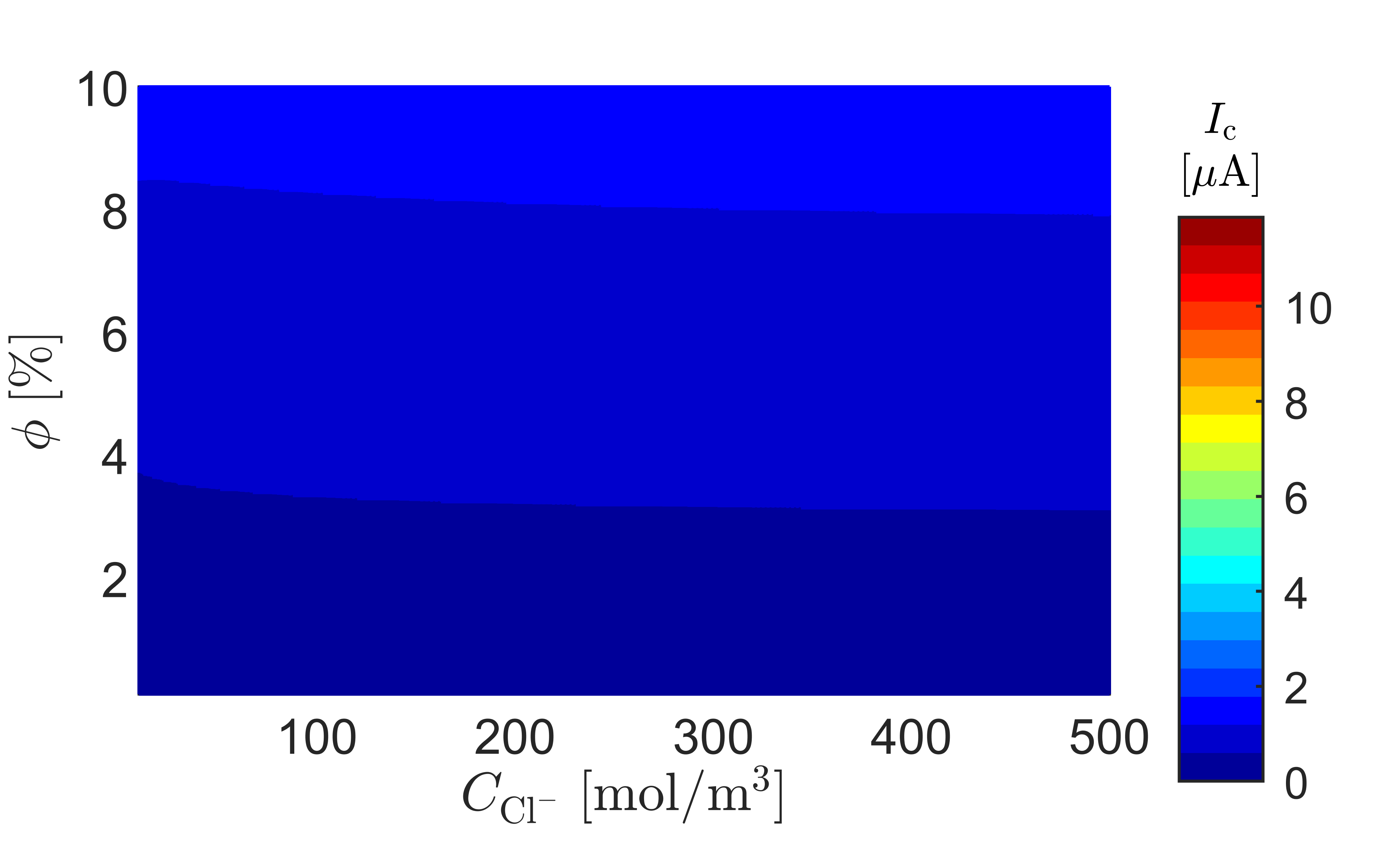}
         \caption{ }
         \label{fig:surfs_Fe_noox}
     \end{subfigure}
    \begin{subfigure}{8cm}
         \centering
         \includegraphics{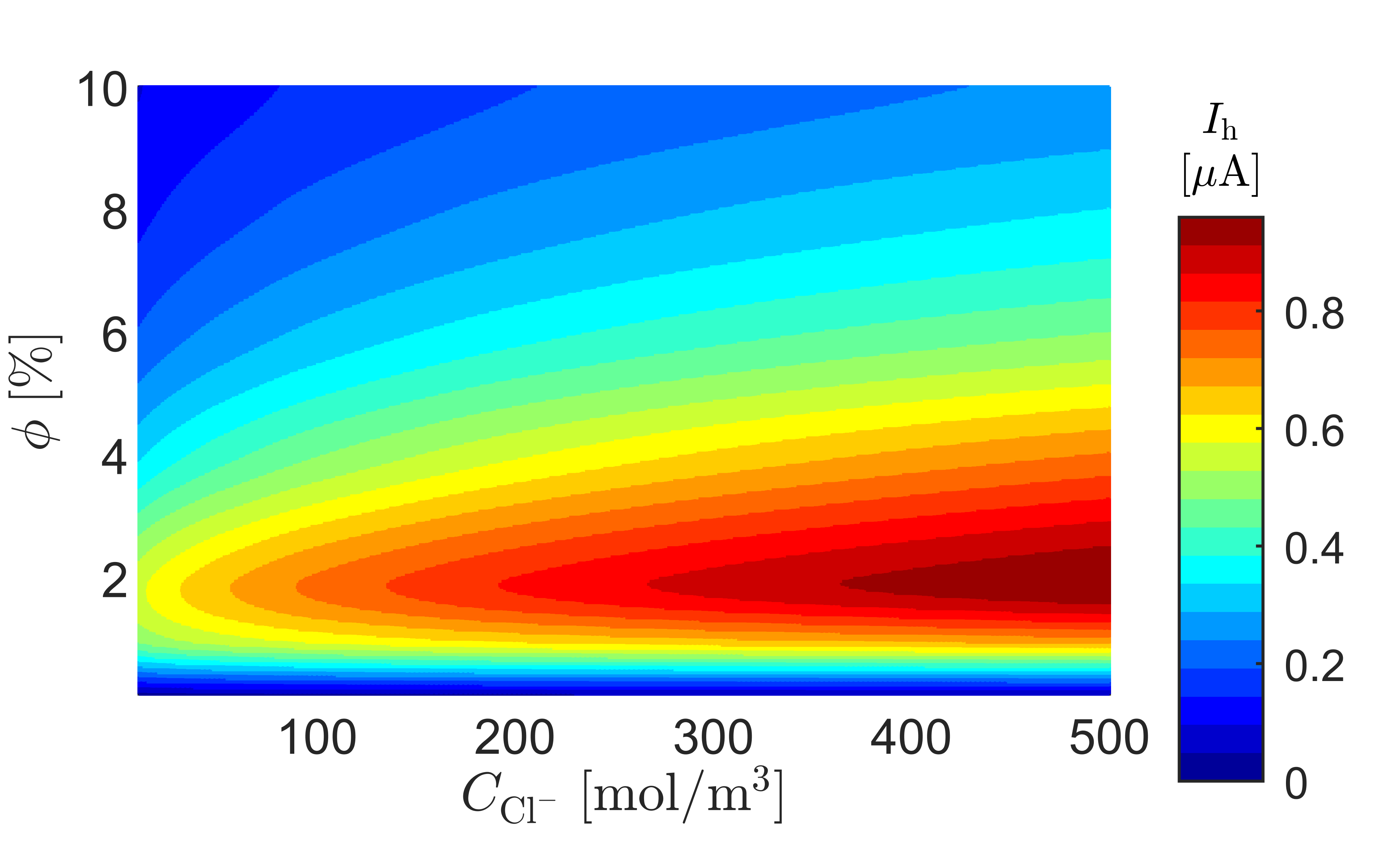}
         \caption{ }
         \label{fig:surfs_H_ox}
    \end{subfigure}
    \begin{subfigure}{8cm}
         \centering
         \includegraphics{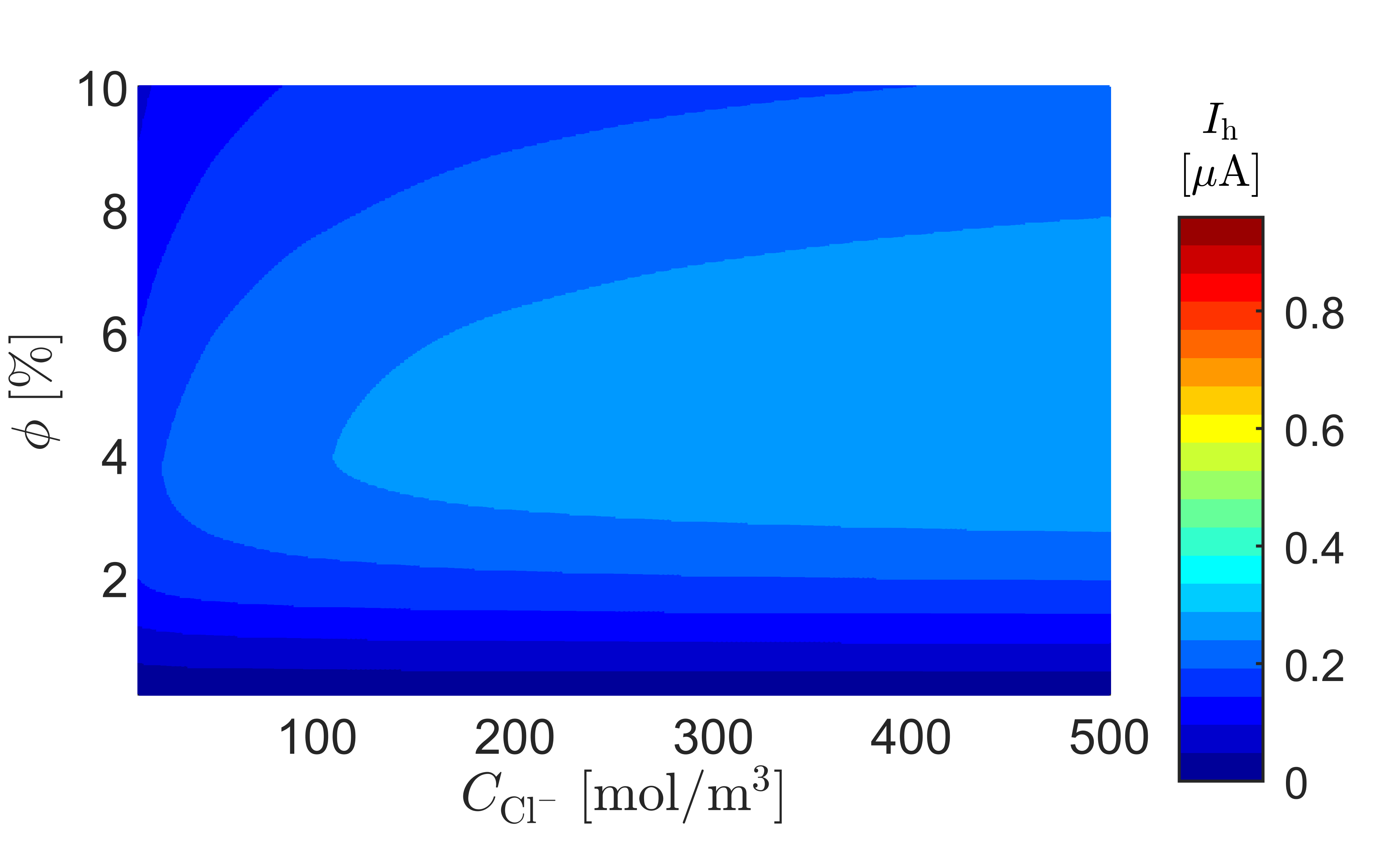}
         \caption{ }
         \label{fig:surfs_H_noox}
     \end{subfigure}
    \begin{subfigure}{8cm}
         \centering
         \includegraphics{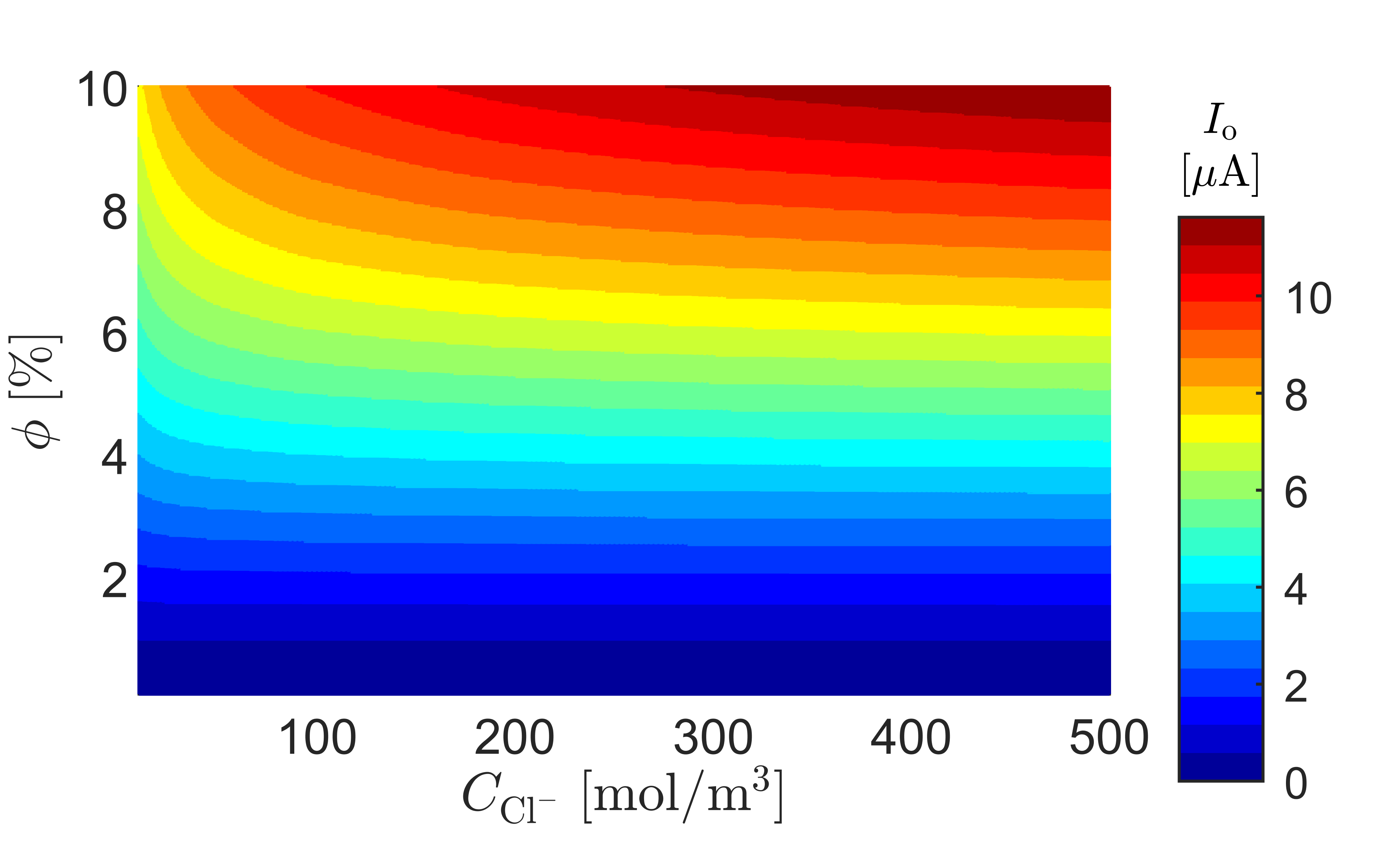}
         \caption{ }
         \label{fig:surfs_O_ox}
    \end{subfigure}
    \begin{subfigure}{8cm}
         \centering
         \includegraphics{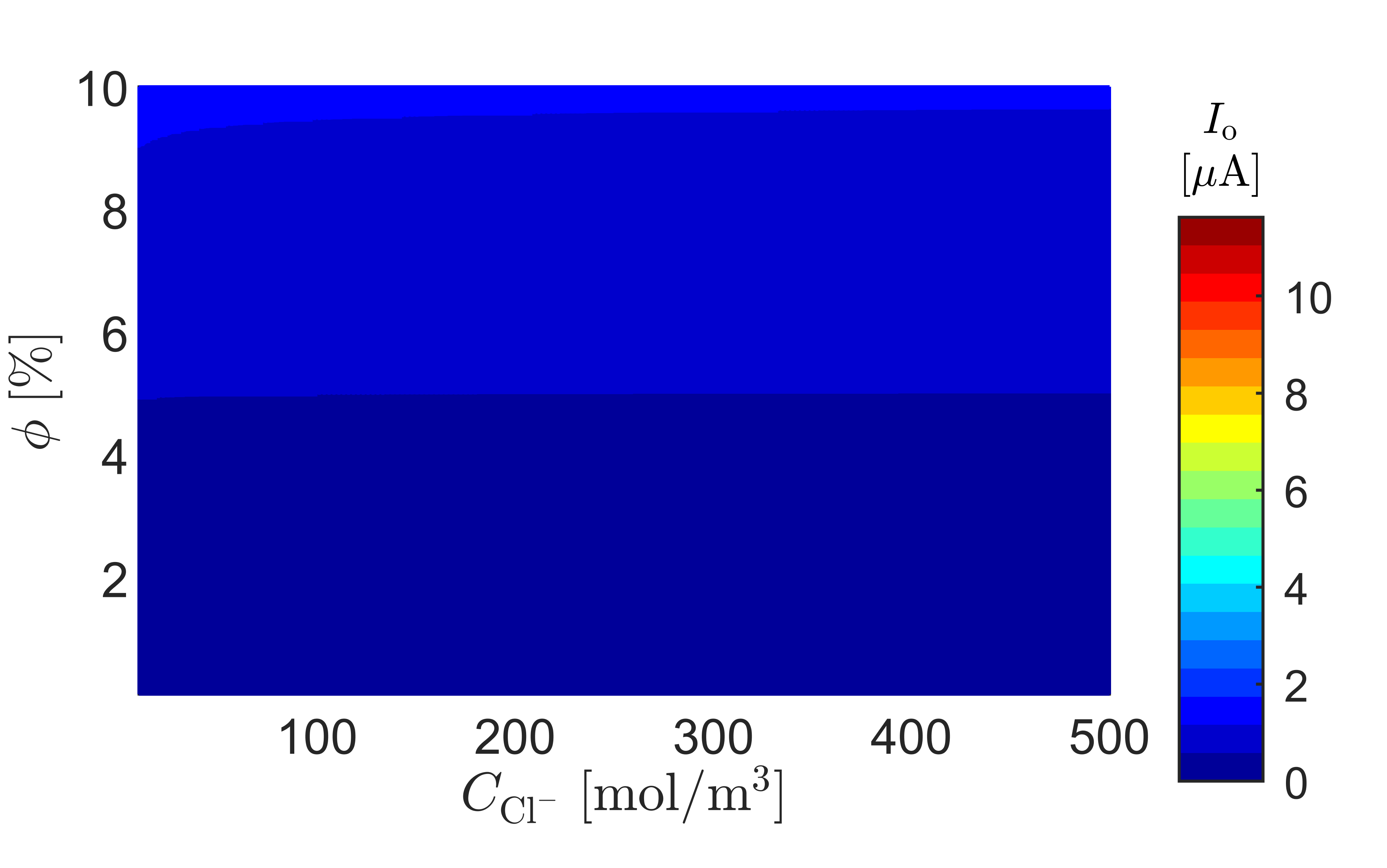}
         \caption{ }
         \label{fig:surfs_O_noox}
     \end{subfigure}
    \caption{Metal potentials (a \& b), corrosion currents (c \& d), hydrogen evolution reaction currents (e \& f), and oxygen evolution current (g \& h) for cases with external oxygen entering the domain (left column) and without external oxygen (right column) after $28\;\text{days}$, considering fully saturated concrete. }
    \label{fig:rateSurfaces}
\end{figure*}

\begin{figure*}
    \centering
    \begin{subfigure}{8cm}
         \centering
         \includegraphics{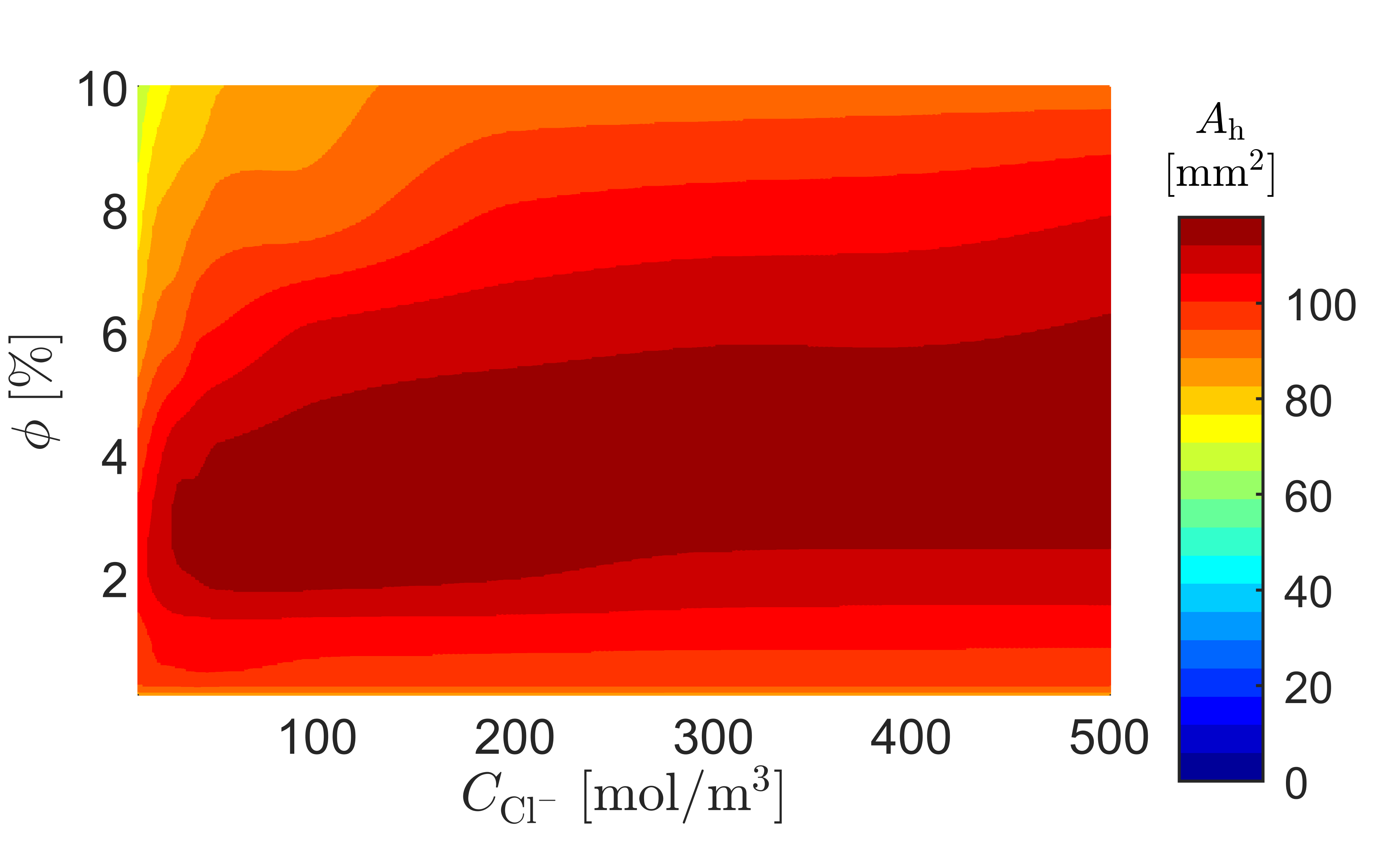}
         \caption{ }
         \label{fig:Area_H_ox}
    \end{subfigure}
    \begin{subfigure}{8cm}
         \centering
         \includegraphics{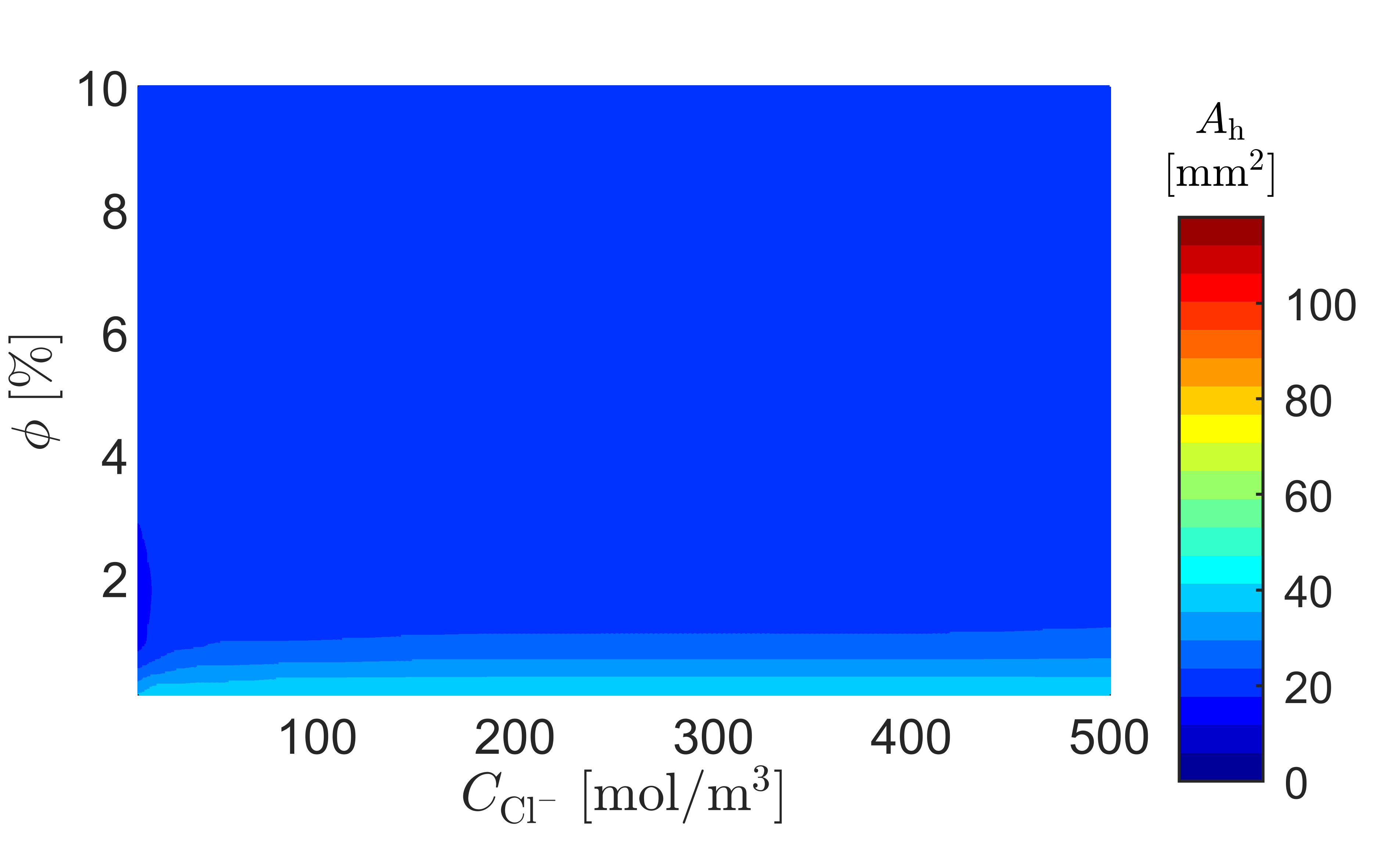}
         \caption{ }
         \label{fig:Area_H_noox}
     \end{subfigure}
    \begin{subfigure}{8cm}
         \centering
         \includegraphics{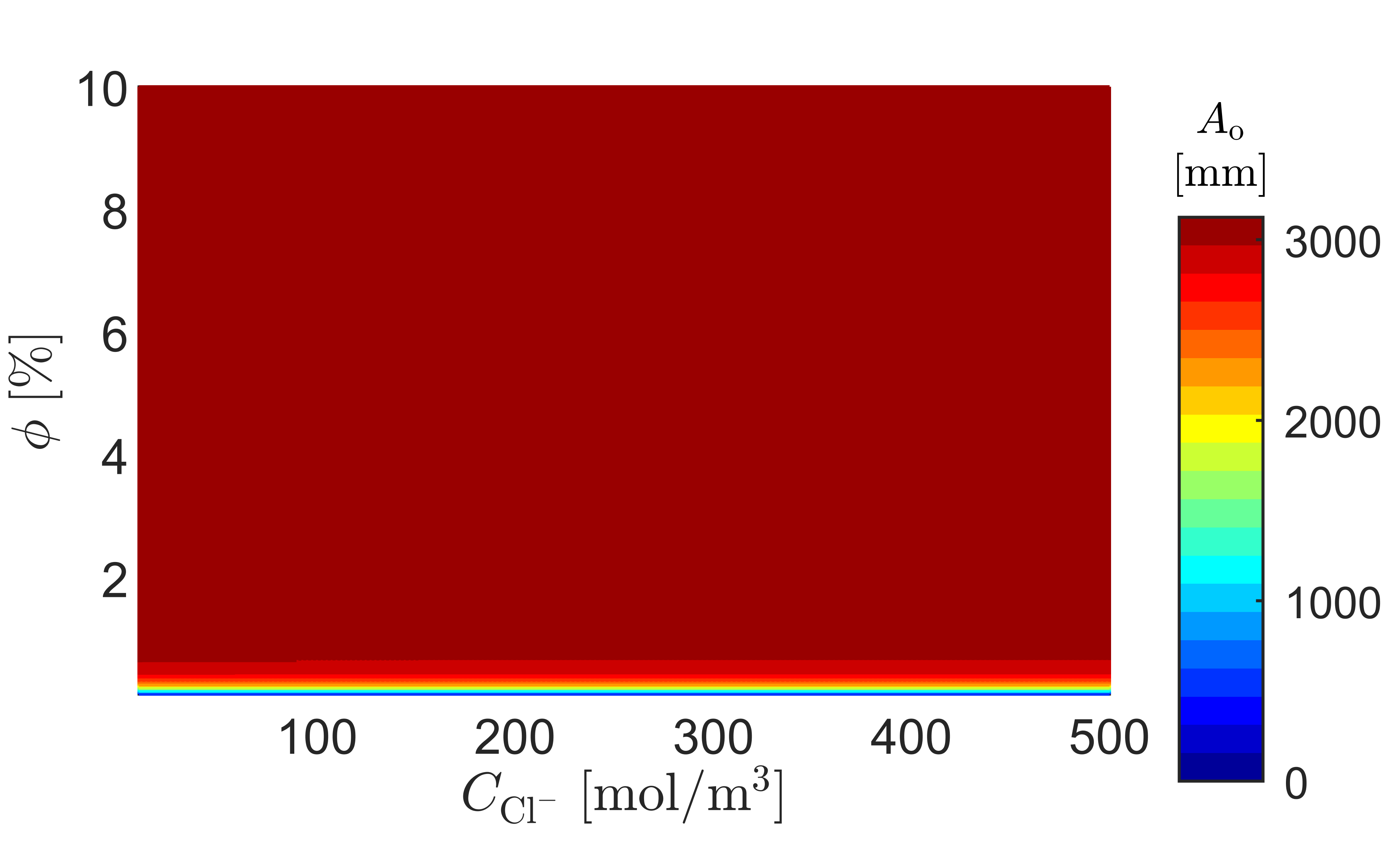}
         \caption{ }
         \label{fig:Area_O_ox}
    \end{subfigure}
    \begin{subfigure}{8cm}
         \centering
         \includegraphics{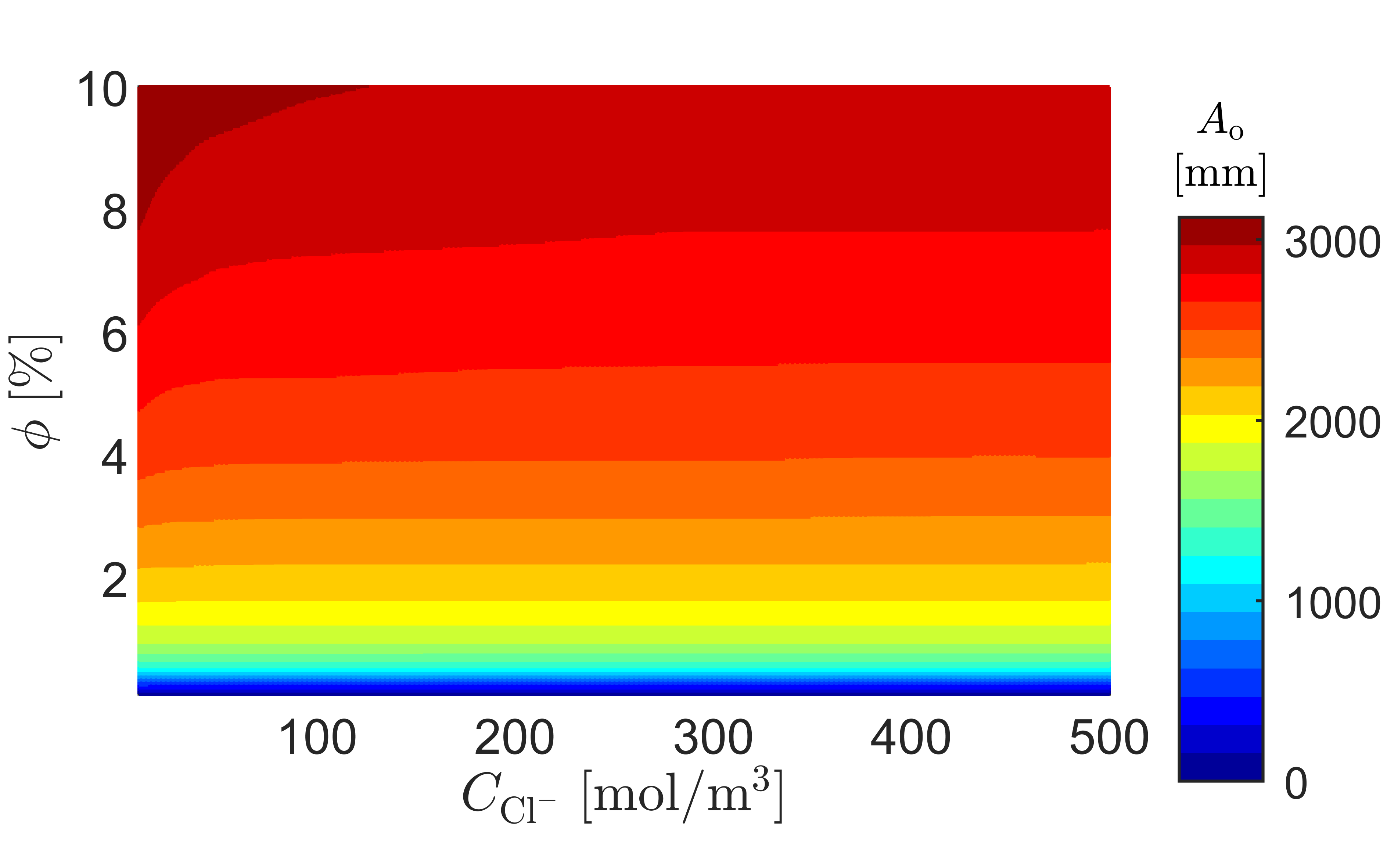}
         \caption{ }
         \label{fig:Area_O_noox}
     \end{subfigure}
    \caption{Surface area on which the Hydrogen evolution reaction (a \& b), and oxygen evolution reaction (c \& d) occur with a rate of $i>10^{-5}\;\mu\text{A}/\text{mm}^2$ for cases with external oxygen entering the domain (left column) and without external oxygen (right column) after $28\;\text{days}$. }
    \label{fig:Area_rateSurfaces}
\end{figure*}

\subsection{Influence of porosity and conductivity}
Repeating the above analysis for a range of porosity and chlorine concentrations results in the reaction rates shown in \cref{fig:rateSurfaces}. When oxygen is able to enter the concrete, the main factor that influences the oxygen evolution reaction rate is the porosity of the concrete, as indicated by \cref{fig:rateSurfaces}g. A minor contribution due to the chlorine concentration (and thus the conductivity of the electrolyte) is also observed, with this stemming from the changes in metal potential observed (\cref{fig:rateSurfaces}a): if the conductivity of the electrolyte is lower (a low concentration of $\text{Cl}^-$ ions), higher metal potentials are observed, which in turn slow the oxygen reaction rate. For higher porosities, this oxygen evolution reaction is the main cathodic reaction supporting corrosion, such that the corrosion rate closely follows the oxygen evolution rate. When the porosity is lower, the hydrogen and oxygen evolution reactions contribute about equally, as was also discussed in the previous section. While the oxygen reaction is hindered by the lower porosity, as less oxygen is able to diffuse toward the rebar, the hydrogen evolution reaction is aided by this lower porosity, ensuring that more corrosion products are retained within the concrete and thereby acidifying the corrosion pit. Looking at the areas associated with the oxygen and hydrogen evolution reactions, Figs. \ref{fig:Area_rateSurfaces}a and \ref{fig:Area_rateSurfaces}c, confirms this: the oxygen reaction covers the full rebar for all but the lowest values of concrete porosity, whereas the area over which the hydrogen evolution reaction occurs initially increases with porosity (as higher porosities result in higher corrosion rates, hence more corrosion products) up to about 4\% porosity, after which the effect of $\text{H}^+$, $\text{Fe}^{2+}$, and $\text{FeOH}^+$ diffusing out of the domain becomes dominant, reducing the area over which hydrogen evolution occurs. 

The results without external oxygen being able to enter the domain show much lower corrosion rates compared to the cases with external oxygen, \cref{fig:rateSurfaces}b. These results no longer indicate the steady-state reaction rates, as was the case for the results with external oxygen allowed, rather detailing the rates after $28\;\text{days}$. The main dependence of the oxygen reaction rate is still the porosity of the concrete, now determining the amount of total oxygen that was initially present. As the oxygen evolution reaction rates are much lower, so are the rates for the corrosion reaction. As a result, less $\text{H}^+$ is generated through the hydrolysis of corrosion products, leading to the hydrogen evolution rate also being lower for all cases considered without external oxygen. Interestingly, the region over which the oxygen evolution reaction occurs shrinks as the porosity is reduced, as shown in \cref{fig:Area_rateSurfaces}d. This reduction in area is due to parts of the rebar dropping below the threshold set to consider an area as ``active'', $i_\text{o}<10^{-5}\;\mu\text{A}/\text{mm}^2$. In contrast, the hydrogen reaction area follows a similar trend as for the case with external oxygen, where increased corrosion rates and decreased porosity both aid in sustaining this reaction, causing the maximum reaction area to be attained for porosities of $5\%$. 

\begin{figure*}
    \includegraphics{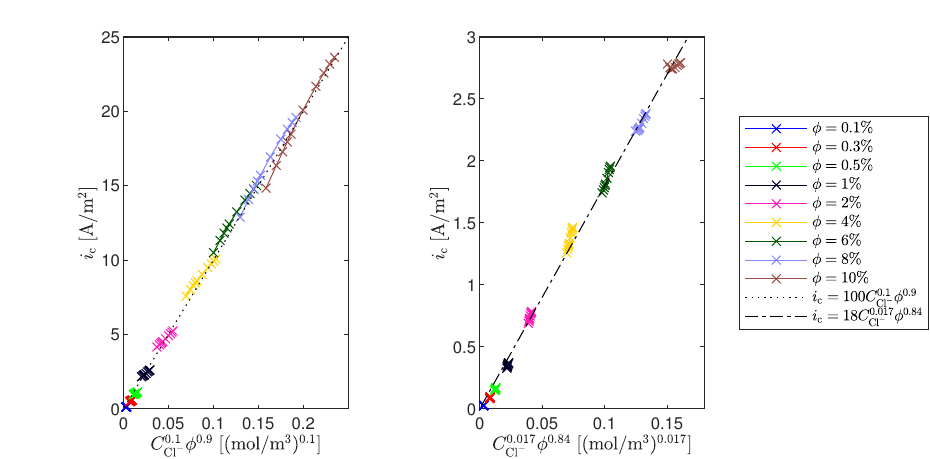}
    \begin{subfigure}{6cm}
        \centering
        \caption{External $\text{O}_2$}
    \end{subfigure}
    \begin{subfigure}{6cm}
        \centering
        \caption{No external $\text{O}_2$}
    \end{subfigure}
    \caption{Comparison between simulation results with (left) and without (right) external oxygen inflow, and the curve-fits from \cref{tab:fitParams,eq:fit}}
    \label{fig:LineFit}
\end{figure*}

To analyse the results more quantitatively, we define an estimate of the corrosion current density, relative to the surface area of the corrosion pit, as:
\begin{equation}
    i_\text{c} = a C_{\text{Cl}^-}^b \phi^c = \frac{a}{0.01264^{2c/3}} C_{\text{Cl}^-}^{b-2c/3} \rho^{-2c/3}\label{eq:fit}
\end{equation}
with $a$, $b$, and $c$ fitting parameters, indicating the magnitude of the corrosion current ($a$), and the impact of the chlorine concentration and porosity on this current ($b$ and $c$ respectively). The last expression, in terms of the resistivity of the concrete and chlorine concentration, uses $S_\text{w}=1$ and the definition of the concrete resistivity $\rho$ from \cref{eq:Conductivity}, which neglects the local variations in conductivity due to the iron ions. We note, however, that this does take into account variations in resistivity due to the used concentration of $\text{Cl}^-$ ions, such that the expression from \cref{eq:fit} using resistivity/Chlorine concentrations has its dependence on $C_{\text{Cl}^-}$ both explicitly included in $C_{\text{Cl}^-}$ as well as implicitly in $1/\rho$. We use the $aC_\text{Cl}^b\phi^c$ expression, dependent on the chlorine concentration ($C_{\text{Cl}^-}=10\text{--}500 \;\text{mol}/\text{m}^3)$) and porosity ($\phi = 0.001\text{--}0.1$), and fit this function to our simulation results using the \texttt{MATLAB} function \texttt{fit()} \citep{MATLAB}. The resulting coefficients are provided in \cref{tab:fitParams}, and visually shown in \cref{fig:LineFit}. As we have only fitted our simulation results for this one specific geometry and boundary conditions, these coefficients should not be interpreted as widely applicable, but rather to give an indication of the importance of each term. The applicability of this relation is inherently limited by geometric sensitivity, as variations in factors such as concrete cover thickness, the presence of multiple corrosion pits, or complex boundary conditions (e.g., temporal changes in salinity or saturation) can significantly alter the corrosion process. Additionally, further interactions between corrosion products and the surrounding concrete could influence the reaction mechanisms in ways not captured by this simplified expression. Due to these complexities, a general formulation that accounts for all possible variations remains infeasible within a simple empirical relation. 

For both cases, this simple function is able to accurately capture the corrosion current for a wide range of porosity and chlorine concentration. Notably, the corrosion current scales using an exponent $0.9$ with the porosity when oxygen is allowed to enter the domain, while it scales with an exponent $0.8$ when no external oxygen is present. The chlorine concentration has much less of an influence, scaling with an exponent of $0.01$ and $0.017$ for both cases. It should be noted that this does not take the interactions of chlorine with passivation mechanisms into account. As such, a high chlorine concentration could be relevant for initialising pitting corrosion (which is not captured within our simulation scheme), while once the corrosion process is started it becomes less relevant for sustained natural corrosion. 

\begin{table}
    \centering
    \begin{subtable}{0.55\textwidth}
    \centering
    \begin{tabular}{|c||c|c||c|c|}
    \hline
         & \multicolumn{2}{c||}{$S_\text{w}=1$, Oxygen} & \multicolumn{2}{c|}{$S_\text{w}=1$, No oxygen} \\ \hline
         &  Value & 95\% interval & Value & 95\% interval \\ \hline 
         $a$ & 100 & 95.2-105.4     & 17.9 & 16.94-18.86\\
         $b$ & 0.10 & 0.0968-0.1078 & 0.017 & 0.0110-0.0232\\
         $c$ & 0.90 & 0.8835-0.9169   & 0.84 & 0.8219-0.8568\\ \hline
         $R^2$ & \multicolumn{2}{c||}{0.998} & \multicolumn{2}{c|}{0.997}  \\ \hline
    \end{tabular}
    \caption{}
    \end{subtable}
    \begin{subtable}{0.55\textwidth}
    \centering
    \begin{tabular}{|c||c|c||c|c|}
    \hline
         & \multicolumn{2}{c||}{$S_\text{w}=0.25$, Oxygen} & \multicolumn{2}{c|}{$S_\text{w}=0.25$, No oxygen} \\ \hline
         &  Value & 95\% interval & Value & 95\% interval \\ \hline 
         $a$ & 20 & 17.36-22.39     & 24 & 21.11-26.73\\
         $b$ & 0.25 & 0.2326-0.2654 & 0.13 & 0.1139-0.1438\\
         $c$ & 0.71 & 0.6767-0.7453   & 0.69 & 0.652-0.7196\\ \hline
         $R^2$ & \multicolumn{2}{c||}{0.985} & \multicolumn{2}{c|}{0.982}  \\ \hline
    \end{tabular}
    \caption{}
    \end{subtable}
    \caption{Coefficients for \cref{eq:fit}, 95\% confidence interval, and $R^2$ coefficient indicating the quality of the fit, considering fully saturated (a), and partially saturated (b) concrete.}
    \label{tab:fitParams}
\end{table}

Evaluating the expression from \cref{eq:fit} in terms of conductivity for the case in which external oxygen is allowed provides the relation as:
\begin{equation}
    i_\text{c} = 100 C_{\text{Cl}^-}^{0.1} \phi^{0.9} \approx \frac{1.4\cdot10^3}{\rho^{0.6} C_{\text{Cl}^-}^{0.5} } 
    \label{eq:ic_Sw1}
\end{equation}
using all SI units and producing the corrosion current in $\text{A}/\text{m}^2$. Using the units of $\mu \text{A}/\text{cm}^2$, $K\Omega \text{cm}$, and $\text{mol}/\text{dm}^3$ (more commonly used in experimental studies), the numerical factor becomes approximately $138$. Comparing this to experimentally obtained relations, for instance $i_\text{c}=26/\rho$ \citep{Lambert1991, Andrade2018} (at a constant but unknown concentration of chlorine ions within the electrolyte), shows that similar simple relations have been previously observed, albeit with a different exponent for the resistivity. As the main contribution to the corrosion current within our simulations has been observed to be driven by the oxygen diffusion, one potential source of this different exponent could be the tortuosity relation used in \cref{eq:Deff}. While the exponent of $3/2$ used is fairly common for concrete diffusion, different exponents in the tortuosity relation could produce different corrosion current relations, which could match closer to reality. Furthermore, we assume fully saturated concrete, whereas the experimental setup is likely to be under a partially saturated state. Even though this difference exists, it is interesting to note the role of chlorine ions when the corrosion current is related to the resistivity. If the resistivity is kept constant (e.g. by altering both the porosity and chlorine concentration) increasing the salinity of the pore solution decreases the corrosion current density, whereas following from the porosity-chlorine expression in \cref{eq:ic_Sw1}, it is clear that the corrosion current increases with increased chlorine concentrations. This indicates that while resistivity is a good measure for comparing the corrosion performance of concretes under a constant salinity, it should not be used when both the concrete porosity and the pore solution are varied. Instead, these two variations should be separated, for instance by directly using the pore solution conductivity/chlorine concentration, and the concrete porosity, among other parameters.

\subsection{Impact of saturation}

\begin{figure}
    \centering
    \begin{subfigure}{0.49\textwidth}
        \centering
        \includegraphics[clip=true, trim={0 0 0 0}]{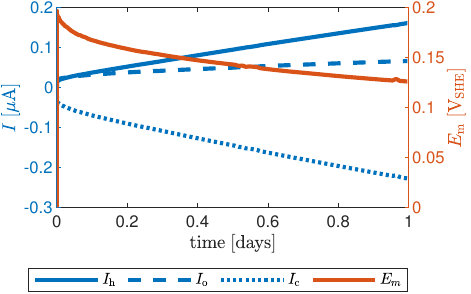}
        \caption{Reaction currents}
    \end{subfigure}
    \begin{subfigure}{0.49\textwidth}
        \centering
        \includegraphics[clip=true, trim={0 0 0 0}]{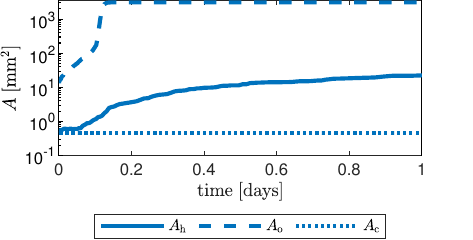}
        \caption{Reaction area}
    \end{subfigure}
    \caption{Reaction currents and area over which the reactions occur for the case with external oxygen, using $\phi=1\%$, $C_\text{Cl}=500\;\mathrm{mol}/\mathrm{m}^3$, and $S_\text{w}=0.25$. }
    \label{fig:Sw025_OX_Evolutions}
\end{figure}

\begin{figure*}
    \centering
    \begin{subfigure}{8cm}
         \centering
         \includegraphics{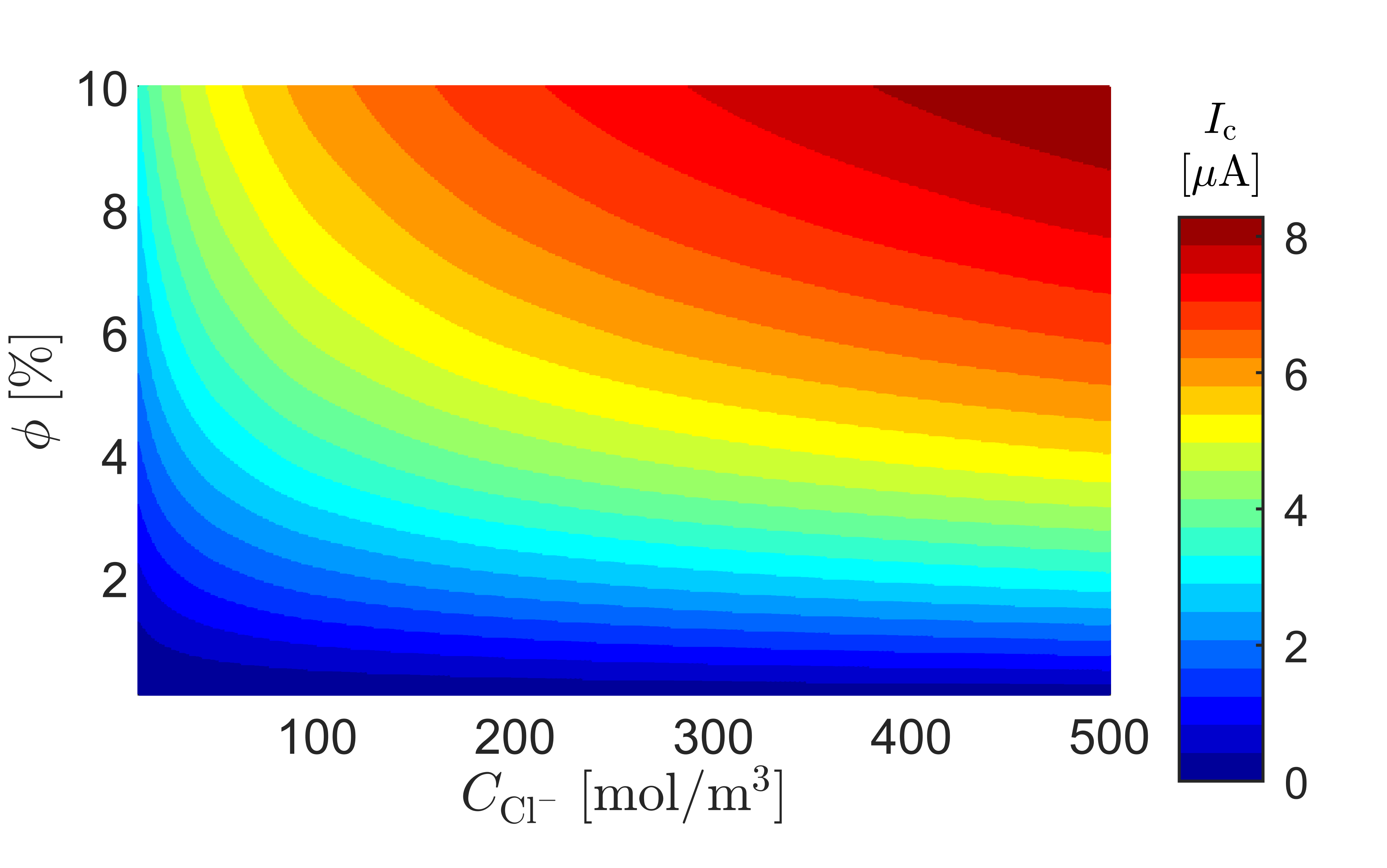}
         \caption{ }
    \end{subfigure}
    \begin{subfigure}{8cm}
         \centering
         \includegraphics{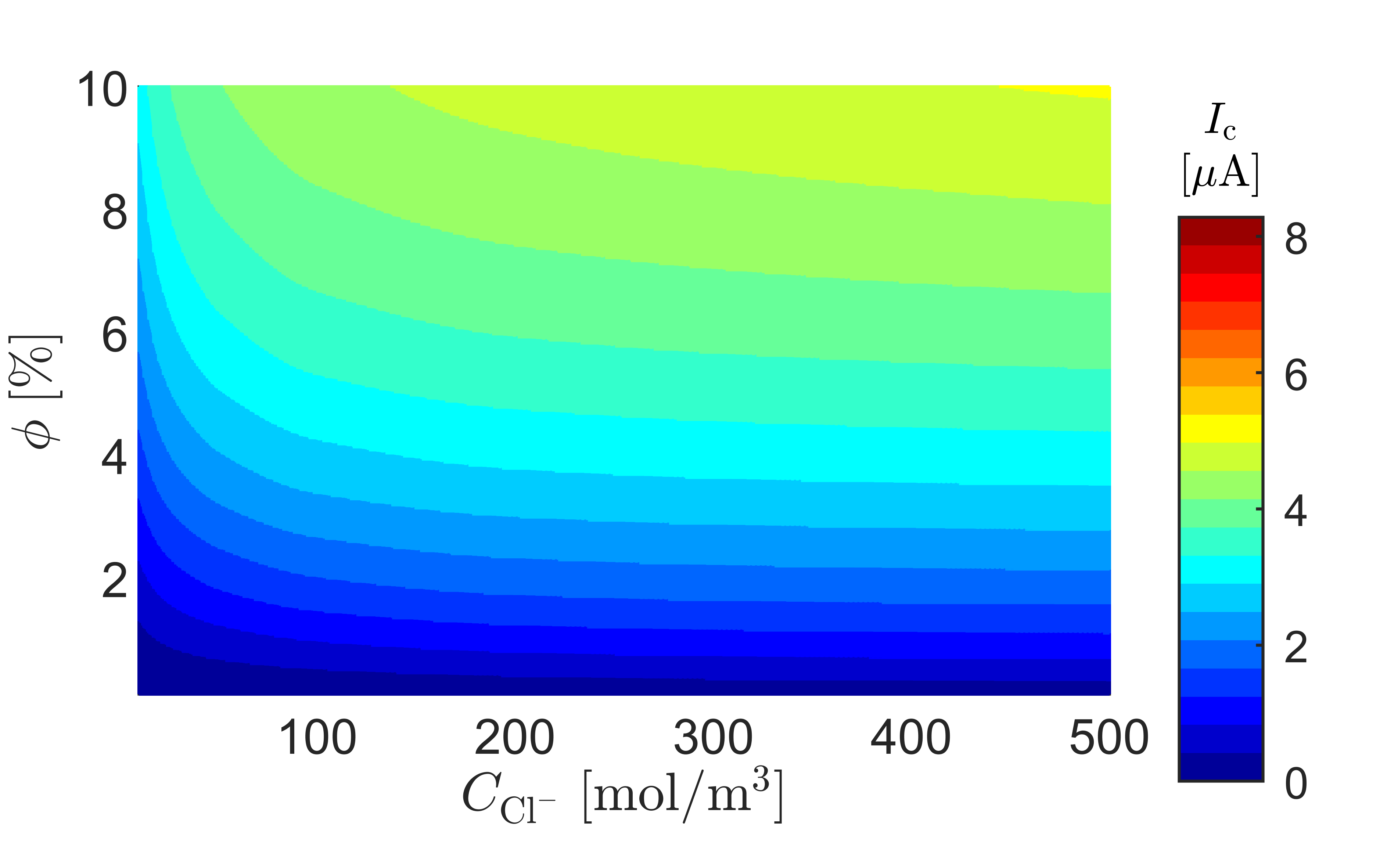}
         \caption{ }
    \end{subfigure}
    \begin{subfigure}{8cm}
         \centering
         \includegraphics{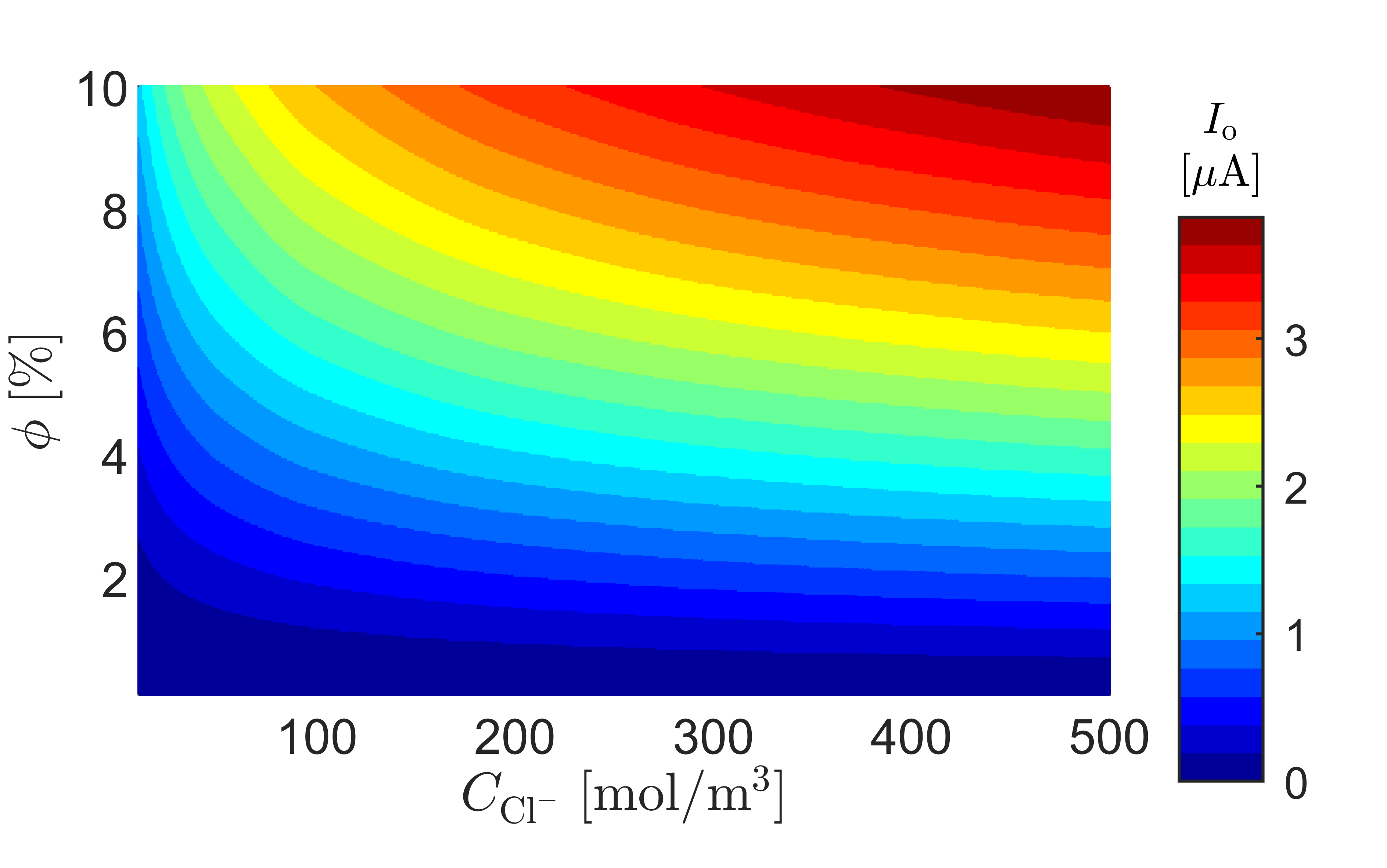}
         \caption{ }
     \end{subfigure}
    \begin{subfigure}{8cm}
         \centering
         \includegraphics{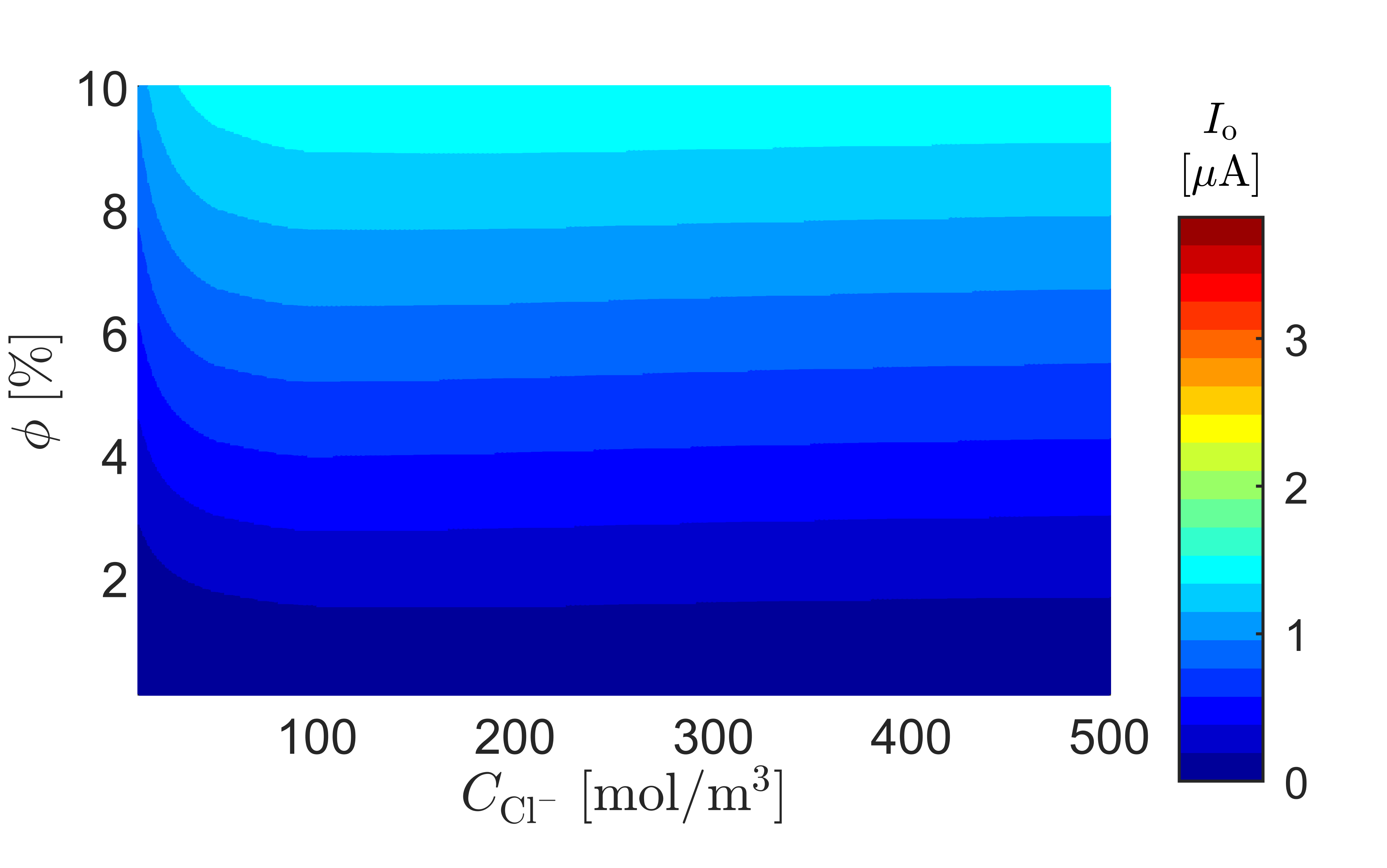}
         \caption{ }
     \end{subfigure}
    \begin{subfigure}{8cm}
         \centering
         \includegraphics{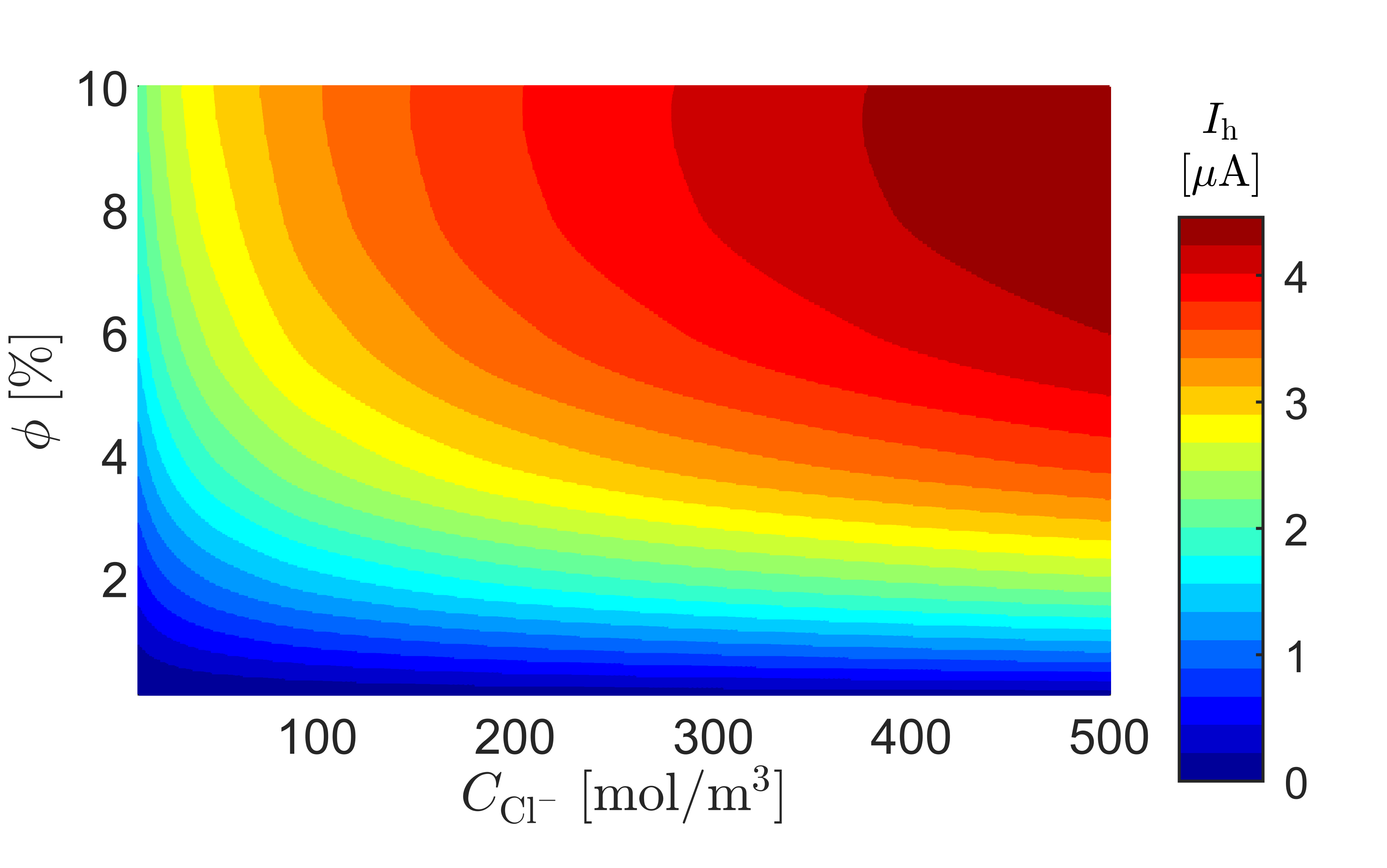}
         \caption{ }
    \end{subfigure}
    \begin{subfigure}{8cm}
         \centering
         \includegraphics{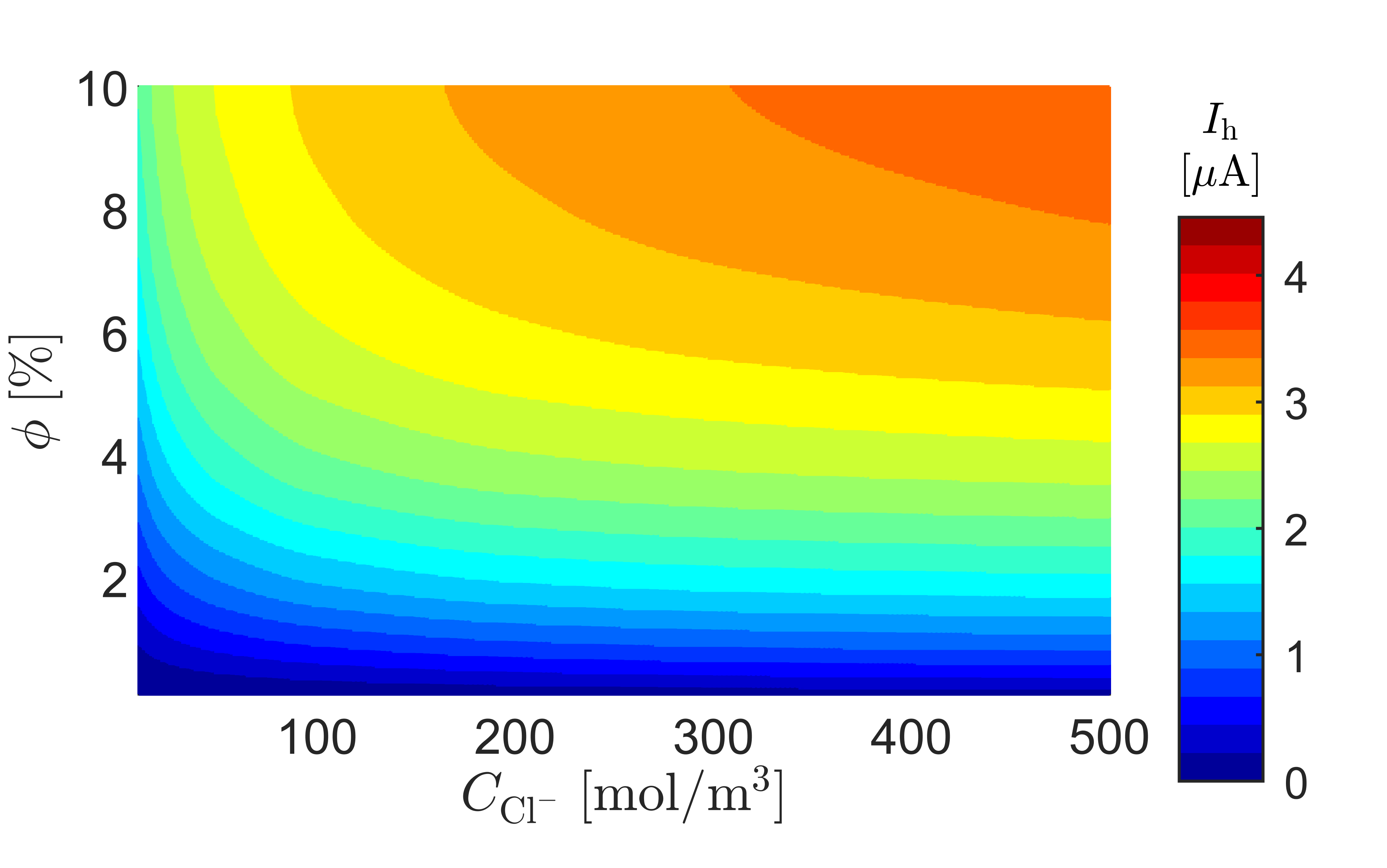}
         \caption{ }
    \end{subfigure}
    \begin{subfigure}{8cm}
         \centering
         \includegraphics{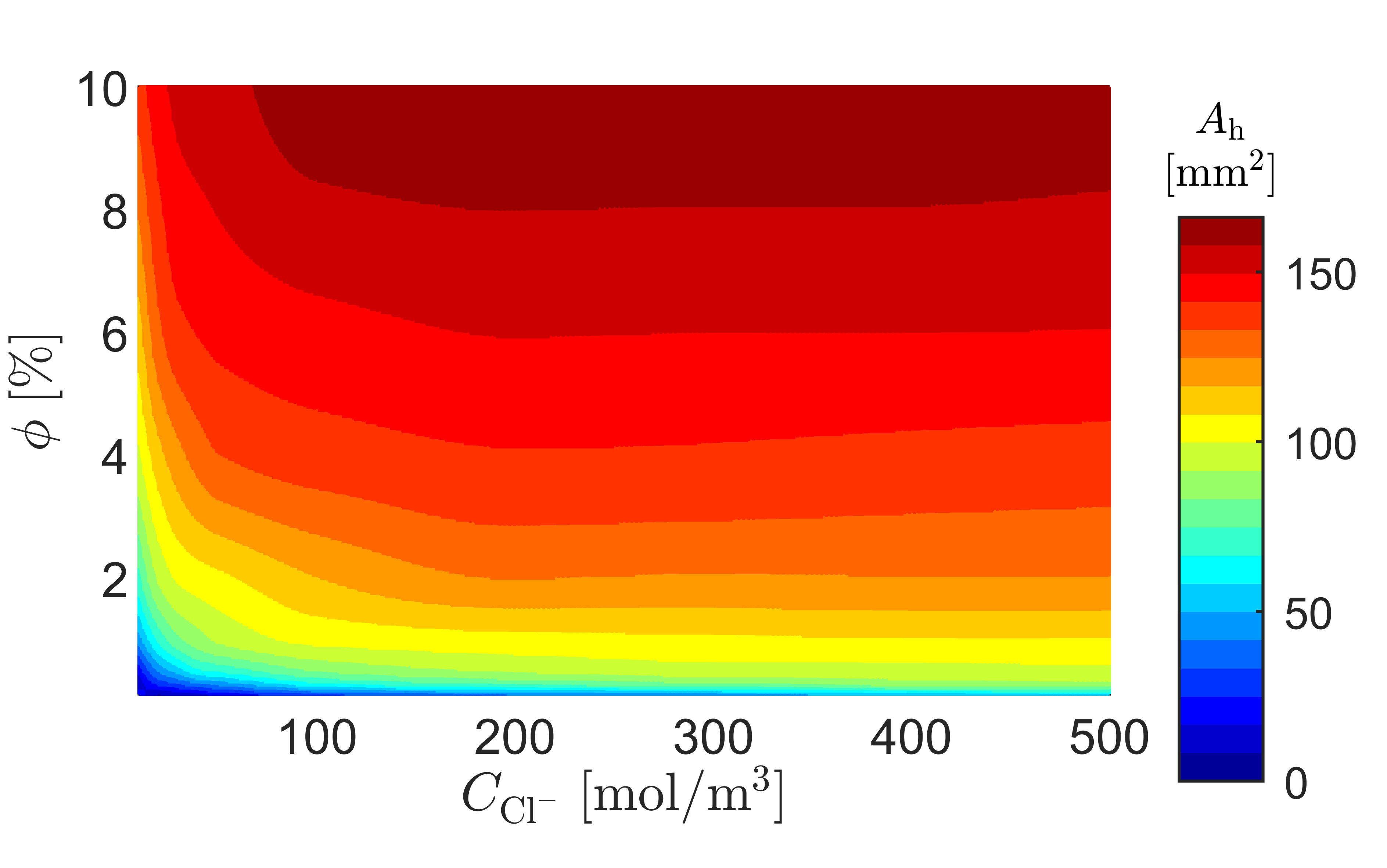}
         \caption{ }
     \end{subfigure}
    \begin{subfigure}{8cm}
         \centering
         \includegraphics{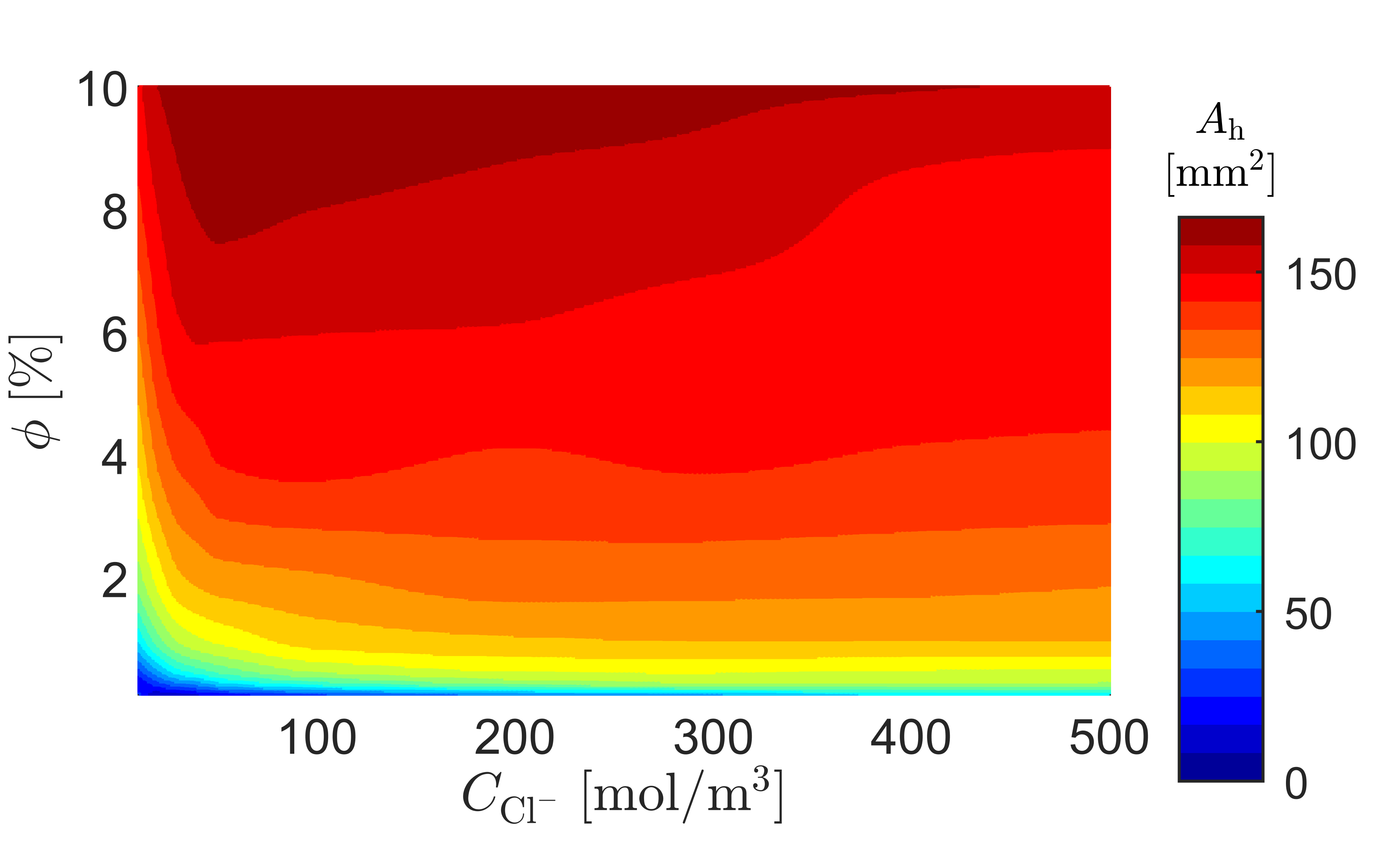}
         \caption{ }
     \end{subfigure}
    \caption{Corrosion, oxygen, and hydrogen reaction currents (a -- f), and area involved in the hydrogen reaction (g-h) for $S_\text{w}=0.25$ and with (left column) and without (right column) external oxygen entering the domain after $28\;\text{days}$. }
    \label{fig:SW025_Surfs}
\end{figure*}

Finally, we consider the case in which the concrete is not fully water-saturated, using $S_\text{w}=0.25$. Following \cref{eq:Deff}, the main effect of this is that the diffusivity of all ionic species (but not oxygen) gets reduced by $D_\text{eff}(S_\text{w}=0.25)=0.0039D_\text{eff}(S_\text{w}=1.0)$, a $99.6\%$ reduction in diffusivity, while the capacity terms and volume reaction rates are reduced by $75\%$. The effect of this reduced diffusivity is observed in \cref{fig:Sw025_OX_Evolutions}: while under fully saturated conditions the oxygen rate was equal or higher compared to the hydrogen evolution reaction rate, if we consider $S_\text{w}=0.25$ the hydrogen rate is significantly increased. Furthermore, both the hydrogen and the corrosion reaction rates keep increasing for the full $28\;\text{days}$ that are simulated, while the oxygen evolution reaction rate stabilises quickly. This indicates that, due to the significantly lower diffusivity for ionic species, the corrosion products are retained within the porous material, where they are close to fully converted through hydrolysis to make the corrosion reaction self-sufficient: for every $\text{Fe}^{2+}$ ion produced via corrosion, about $1.8\text{H}^+$ ions are produced which partake in the hydrogen evolution reaction. While the area partaking in the hydrogen reaction remains constant, see \cref{fig:Sw025_OX_Evolutions}b, the reduction in metal potential required to balance the corrosion and cathodic rates is sufficient to accelerate the hydrogen reaction. 

Similar to what is presented for the full saturation cases, results for a range of porosities and $\text{Cl}^-$ concentrations are shown in \cref{fig:SW025_Surfs} for $S_\text{w}=0.25$. While the oxygen reaction still follows similar trends, being strongly dependent on porosity and weakly dependent on the metal potential, the hydrogen reaction increases with both porosity and conductivity. As the diffusivity is reduced due to the partial saturation, fewer hydrogen and iron ions diffuse out of the concrete even for the highest porosity. However, slight increases in diffusivity (via an increased porosity) do aid in increasing the area of the acidic pit, and thus increasing the area over which the hydrogen evolution reaction occurs. As additional increases in porosity allow for the corrosion reaction to occur faster, thus enhancing the amount of available $\text{H}^+$ ions even further, the hydrogen reaction scales very strongly with porosity. More importantly, the hydrogen reaction rate exceeds that of the oxygen reaction for almost all combinations of porosity and $\text{Cl}^-$ concentration, indicating that under partially saturated conditions, the oxygen rate is no longer a direct indicator of the corrosion rate. Even though a large saturation difference exists between these and the previous cases, the maximum corrosion rates observed are fairly similar ($11\;\mu\text{A}$ for the fully saturated case, vs $8\;\mu\text{A}$ for $S_\text{w}=0.25$). However, the cathodic reaction consuming the electrons produced by this corrosion switches under these low saturation conditions from oxygen-reaction dominated to being driven by both the oxygen and hydrogen reactions. 

\begin{figure}
    \centering
    \includegraphics{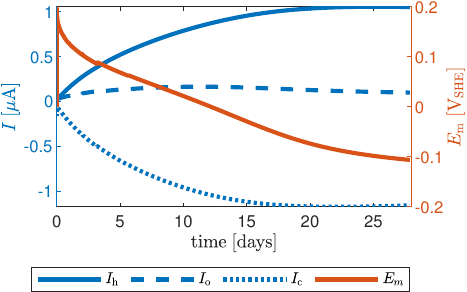}
    \caption{Corrosion, hydrogen and oxygen reaction currents over time for a partially saturated case, $S_\text{w}=0.25$, in the absence of external oxygen using $C_\text{Cl}=500\;\text{mol}/\text{m}^3$ and $\phi=1\%$.}
    \label{fig:Sw025_noox_timeseries}
\end{figure}

When no external oxygen is able to enter the concrete, the corrosion reaction is much more able to sustain itself under these partially saturated conditions. As the corrosion reaction is close to self-sufficiency, even after $28\;\text{days}$ there is still oxygen remaining within the concrete, and the hydrogen reaction current is only reduced to $3\;\mu\text{A}$. Furthermore, the area over which this hydrogen reaction occurs does not significantly differ from the case where oxygen is able to enter. Consequently, under these partially saturated conditions, the corrosion rate observed is higher compared to when the concrete is fully saturated. While similar to the fully saturated conditions, this corrosion will eventually stop, only a slight reduction in corrosion rate is observed, see \cref{fig:Sw025_noox_timeseries}. 

\begin{figure*}
    \includegraphics{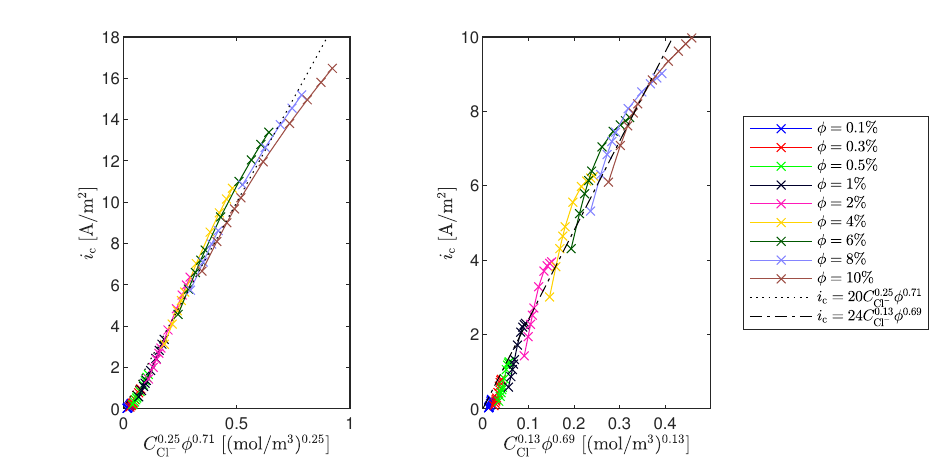}
    \begin{subfigure}{6cm}
        \centering
        \caption{External $\text{O}_2$}
    \end{subfigure}
    \begin{subfigure}{6cm}
        \centering
        \caption{No external $\text{O}_2$}
    \end{subfigure}
    \caption{Comparison between simulation results for $S_\text{w}=0.25$ with (left) and without (right) external oxygen inflow, and the curve-fits from \cref{tab:fitParams,eq:fit}}
    \label{fig:Sw025_LineFit}
\end{figure*}

Finally, we can fit \cref{eq:fit} to the results for partially saturated conditions, producing the coefficients shown in \cref{tab:fitParams}b and \cref{fig:Sw025_LineFit}. Compared to the saturated conditions, the role of the porosity is slightly reduced (coefficients $c=0.71$/$c=0.69$ for cases with and without external oxygen under partially saturated conditions, whereas $c=0.9$/$c=0.84$ when fully saturated), whereas the role of the $\text{Cl}^-$ concentration is significantly increased. For the external oxygen results, the corrosion current density is then given by:
\begin{equation}
    i_\text{c} =20C_\text{Cl}^{0.25}\phi^{0.71} \approx \frac{2.2\cdot10^3}{\rho^{0.47}C_\text{Cl}^{0.22}}
    \label{eq:ic_Sw025}
\end{equation}
Comparing \cref{eq:ic_Sw1,eq:ic_Sw025} shows that for unsaturated conditions, the corrosion current is less dependent on both the resistivity and the $\text{Cl}^-$ concentration. However, it should be noted that this is an effect of the $\text{Cl}^-$ concentration also influencing the resistivity. This matches the conclusion obtained for saturated conditions that solely using the conductivity of concrete to estimate the susceptibility of rebar to pitting corrosion is insufficient, and that instead porosity and $\text{Cl}^-$ concentration need to be considered as two different variables. Additionally, the effects of saturation are not only included in the resistivity, but also directly influence the exponents of the resistivity and concentration, and the constant factor included. 

\subsection{Limitations}
While this work has aimed to include the natural corrosion process in a realistic manner within the computational model, there are intrinsic limitations and assumptions. The first set of assumptions is related to the (electro-) chemical reactions included: while we include the dominant reactions at the metal surface, reactions between the concrete and water (such as carbonation) are not included. Furthermore, corrosion reactions are only included within the pit, while it is feasible that the reverse corrosion reaction (depositing iron ions) could occur on the passivated surface, especially away from the corrosion pit. These assumptions are not expected to influence the qualitative conclusions regarding the regions partaking in the corrosion process, extending beyond just the corrosion pit by providing the supporting current for the corrosion process. However, neglecting the plethora of additional reactions taking place between the concrete and electrolyte will have influenced the quantitative results: e.g. the carbonation process typically try to retain an pH close to 7, whereas the buffer capacity provided by the cement aims to retain a basic pH, both preventing the spread of the acidic region surrounding the corrosion pit. Similarly, by neglecting porosity changes due to corrosion product precipitates and reactions between chloride ions and Friedel salt, the diffusivity in the area surrounding the corrosion pit is likely to be lower lower in reality, as are the ion concentrations in the pit, impacting the development of acidic regions surrounding the pit. 

A second source of uncertainty is due to the used boundary conditions: we assume ions to be able to enter freely (and thus currents can flow into/out of the concrete), whereas actual conditions are more complex, e.g. due to the chlorine ions entering the porous concrete due to fog or sea spray. While this is occasionally imitated within experiments by spraying the concrete with a saline solution at regular intervals, rather than having it submerged or fully dry, this would impose additional complexities on the data analysis by preventing a steady-state from being attained. The exact composition of the electrolyte within the concrete pores is also more complex than modelled: While we assume the $\text{Na}^+$ and $\text{Cl}^-$ ions to be the predominant source of the conductivity/resistivity of the concrete, experimental observations have indicated that this resistivity might be close to independent of these concentrations, likely due to the other constituents in the electrolyte providing a contribution to the overall conductivity of the concrete. Similarly, the relations between the relative diffusivity and the porosity/saturation used in this work, \cref{eq:Deff}, are only one among many different relations from literature \citep{Tartakovsky2019, Ghanbarian2013, Bentz1991, Liu2022, Patel2016, Zheng2024, Ahmad2013}, with the impact of these relations on the overall corrosion rate warranting further research.

Finally, we have only analysed a single corrosion pit on a rebar with a pre-determined size. As such, the relations stated previously, \cref{eq:fit,eq:ic_Sw1,eq:ic_Sw025}, should not be interpreted as being correct for corrosion under general circumstances, but instead indicate how corrosion of a single pit occurs under known and idealised conditions. Furthermore, interactions between multiple corrosion pits and geometric variations such as the effect of cover thickness are not included, which all are likely to alter the corrosion rate: While multiple corrosion pits can each develop acidic regions surrounding the pits (and thus scale the hydrogen evolution reaction current with the amount of pits present), the oxygen evolution reaction current is already occurring over the complete metal surface, and thus if multiple pits were present this current would be split between supporting these pits, thereby limiting the corrosion rate.  However, despite these limitations of the developed computational models, the detailed look into the development of corrosion currents, and the localization of reactions in the area surrounding the corrosion pit, are expected to be similar if these limitations are not present.

\section{Conclusions}
\label{Sec:Conclusions}

We present a novel computational model capable of simulating natural corrosion of reinforcements within concrete. Mimicking natural corrosion, this model assumes the rebar to be isolated from any current sources, except for the electro-chemical reactions occurring on its own surface. As a result, the corrosion rate is directly coupled to the rates of the hydrogen and oxygen evolution reactions, requiring the reaction currents across reactions to be conserved. This allows the electric potential of the metal to change, dynamically adapting to enforcing the conservation of currents. Additionally, no prior assumptions are made regarding the area in which cathodic reactions occur, instead these reaction areas follow automatically from the simulations. The model thus allows us to gain unprecedented insight into concrete corrosion; The main findings include:
\begin{itemize}
    \item The hydrogen reaction is highly localised around the corrosion pit, where an acidic region is created due to the corrosion products. In contrast, the oxygen reaction occurs on the complete surface of the rebar, but with local reaction currents orders of magnitude lower compared to the hydrogen evolution reaction. 
    \item While the corrosion current is much higher when external oxygen is allowed to diffuse into the concrete, just the pre-existing oxygen within the concrete is sufficient to initialise and sustain the corrosion reaction for over $28$ days. However, without external oxygen the corrosion process will eventually stop.
    \item Under these natural corrosion conditions, the electric potential of the metal depends on the conditions of the concrete and the conductivity of the electrolyte. The metal potential varies over time as reactions stabilise. 
    \item The porosity of the concrete has a large influence on the corrosion rate, for our cases scaling with an exponent $0.9$ (or with an exponent $0.84$ without external oxygen), with this large influence being due to the oxygen evolution reaction.
    \item The conductivity of the electrolyte (excluding effects related to porosity changes) has a much smaller influence, with this influence being related to the hydrogen evolution reaction rate increasing for higher conductivity. 
    \item Within the range considered in this work ($S_\text{w}=0.25$ \& $S_\text{w}=1$), a decreased saturation slows the oxygen evolution reaction, but also prevents hydrogen ions from diffusing out of the concrete thereby accelerating the hydrogen reaction. These two effects in isolation are strongly dependent on the saturation levels, but act to decrease/increase the corrosion currents respectively, cancelling out part of each other their effect on the corrosion.
    \item The combined resistivity of the concrete (e.g. the combined effects of porosity, saturation, and ion concentrations) is not a good estimate of its corrosion potential: While the chlorine ion concentration is generally considered to be the controlling factor for depasivation, effects of ion concentrations (such as chlorine), concrete properties (such as porosity), and environmental factors (such as saturation and boundary conditions) should be separately considered when considering the corrosion rate for existing corrosion pits, instead of being lumped into a single resistivity.
\end{itemize}

\section*{Data availability}
\noindent The finite element code used within this study, together with documentation and example inputs and instructions on how to compile the C++ code, is available at \url{https://github.com/T-Hageman/ConcreteCorrosion}. Example inputs and files for post-processing, allowing the results from \cref{fig:Evolution_pH_over_time,fig:Evolution_rates_over_time} to be re-produced are also included. [Repository currently set to private, will be set to public nearer to publication]

\section*{Declaration of competing interest}
\noindent The authors declare that they have no known competing financial interests or personal relationships that could have appeared to influence the work reported in this paper.

\section*{Acknowledgements}
\noindent Tim Hageman acknowledges support through the research fellowship scheme of the Royal Commission for the Exhibition of 1851, and Emilio Mart\'{\i}nez-Pa\~neda acknowledges financial support from UKRI's Future Leaders Fellowship programme [grant MR/V024124/1]. The authors also acknowledge computational resources and support provided by the Imperial College London Research Computing Service (\url{http://doi.org/10.14469/hpc/2232}), and the University of Oxford Advanced Research Computing Service (\url{http://dx.doi.org/10.5281/zenodo.22558}).

\section*{CRediT authorship contribution statement}
\textbf{Tim Hageman}: Methodology, Software, Validation, Formal analysis, Investigation, Data curation, Visualization, Writing – original draft. \textbf{Carmen Andrade}: Conceptualization, Writing – review \& editing. \textbf{Emilio Martínez-Pañeda}: Conceptualization, Methodology, Project administration, Writing – review \& editing

\FloatBarrier
\appendix
\section{Numerical implementation}
\label{app:Implementation}
For the finite element discretisation, concentrations and the electrolyte potential are discretised throughout the concrete as:
\begin{equation}
    C_\pi = \sum_{el} \mathbf{N}^{el} \mathbf{C}_\pi^{el} \qquad \qquad \varphi = \sum_{el} \mathbf{N}^{el} \mathbf{\upvarphi}^{el}
\end{equation}
using quadratic Bernstein shape functions $\mathbf{N}^{el}$ for the tetrahedral elements. In addition to these $7+1$ nodal degrees of freedom for the concentrations and electrolyte potential, the metal potential is added as a singular degree of freedom $E_\text{m}$.

Using this discretisation, the discretised weak form of the Nernst-Planck equation from \cref{eq:massconserv} is given as:
\begin{equation} \begin{split}
    \mathbf{f}_{\text{c}\pi} = &\int_{\Omega} \frac{\phi S^*}{\Delta t}\mathbf{N}^T \mathbf{N} \left(\mathbf{C}_\pi^{t+\Delta t} - \mathbf{C}_\pi^t\right)\;\text{d}\Omega \\ &+ \int_{\Omega} D_\pi^\text{eff} \left(\bm{\nabla}\mathbf{N}\right)^T\bm{\nabla}\mathbf{N}\mathbf{C}_\pi^{t+\Delta t} \; \text{d}\Omega \\
    &+ \int_{\Omega} \frac{D_\pi^\text{eff} z_\pi F}{RT} \left(\bm{\nabla}\mathbf{N}\right)^T \left(\mathbf{N} \mathbf{C}_\pi^{t+\Delta t}\right) \bm{\nabla}\mathbf{N} \bm{\upvarphi}^{t+\Delta t}\;\text{d}\Omega \\ & - \phi S^* \mathbf{R}_{\pi} + \mathbf{\upnu}_{\pi} - \int_{\Gamma} \mathbf{N} j_\pi \; \text{d}\Gamma = \mathbf{0}
\end{split} \label{eq:f_c} \end{equation} 
where the boundary flux $j_\pi$ and the reaction flux $\bm{\nu}_\pi$ interact with the full boundary (in contrast to these being corrected for the porosity). The volume reactions are integrated using a lumped integration scheme \citep{Schellekens1993, Hageman2020a, Hageman2023}, using lumped integration weights $W^{nd}$ associated with each node, providing the reaction terms as:
\begin{subequations}
\begin{align}
    \begin{split}
    \mathbf{R}_{\text{H}^+} &= \sum_{nd} W^{nd} \bm{i}^{nd} \Big(k_{\text{eq}}\left(K_\text{w}-\frac{\mathbf{C}_{\text{H}^+}^{t+\Delta t}\mathbf{C}_{\text{OH}^-}^{t+\Delta t}}{C_\text{ref}^2}\right) \\ &+ \left( k_{\text{fe}}\frac{\mathbf{C}_{\text{Fe}^{2+}}^{t+\Delta t}}{C_\text{ref}}-k'_{\text{fe}}\frac{\mathbf{C}_{\text{FeOH}^+}^{t+\Delta t}\mathbf{C}_{\text{H}^+}^{t+\Delta t}}{C_\text{ref}^2} + k_{\text{feoh}} \frac{\mathbf{C}_{\text{FeOH}^+}}{C_\text{ref}} \right) \Big) 
    \end{split}\label{eq:RH} \\
    \mathbf{R}_{\text{OH}^-} &= \sum_{nd} W^{nd} \bm{i}^{nd} k_{\text{eq}} \left(K_\text{w}-\frac{\mathbf{C}_{\text{H}^+}^{t+\Delta t}\mathbf{C}_{\text{OH}^-}^{t+\Delta t}}{C_\text{ref}^2}\right) \label{eq:ROH} \\
    \mathbf{R}_{\text{Fe}^{2+}} &= \sum_{nd} W^{nd} \bm{i}^{nd} \left(-k_{\text{fe}} \frac{\mathbf{C}_{\text{Fe}^{2+}}}{C_\text{ref}} + k_{\text{fe}}' \frac{\mathbf{C}_{\text{FeOH}^+}^{t+\Delta t}\mathbf{C}_{\text{H}^+}^{t+\Delta t}}{C_\text{ref}^2}\right) \label{eq:RFE} \\
    \begin{split}
    \mathbf{R}_{\text{FeOH}^+} &=\sum_{nd} W^{nd} \bm{i}^{nd} \Big( k_{\text{fe}}\frac{\mathbf{C}_{\text{Fe}^{2+}}^{t+\Delta t}}{C_\text{ref}} \\ &-k'_{\text{fe}} \frac{\mathbf{C}_{\text{FeOH}^+}^{t+\Delta t}\mathbf{C}_{\text{H}^+}^{t+\Delta t}}{C_\text{ref}^2} - k_{\text{feoh}} \frac{\mathbf{C}_{\text{FeOH}^+}}{C_\text{ref}}  \Big) \end{split}  \label{eq:RFEOH} \\
    \mathbf{R}_{\text{O}_2}&=\mathbf{R}_{\text{Na}^+} = \mathbf{R}_{\text{Cl}^-} = \mathbf{0} \label{eq:RNACL}
\end{align}
\end{subequations}
where the used concentrations correspond to the nodal values. The allocation vectors $\mathbf{i}^{nd}$ are used to allocate the contributions to the corresponding nodes. The weight vectors for the bulk and surface reactions are determined in a standard manner as:
\begin{equation}
    \mathbf{W} = [W^{nd_1}\; W^{nd_2} \; ... ] = \int_{\Omega} \mathbf{N}\;\text{d}\Omega \qquad \mathbf{W}^{\text{surf}} = \int_{\Gamma} \mathbf{N}\;\text{d}\Gamma
\end{equation}
and the surface reaction rates are integrated in a similar manner, 
\begin{equation}
    \mathbf{\upnu}_{\pi} = \sum_{nd} W^{\text{surf},nd} \mathbf{i}^{nd} \nu_\pi (C^{t+\Delta t})
\end{equation}

The discretised weak form of the electroneutrality condition, \cref{eq:electroneutrality}, is given by:
\begin{equation}
    \bm{f}_\varphi = \int_{\Omega} \sum_\pi z_\pi  \mathbf{N}_\varphi^T\mathbf{N}_\text{C}\mathbf{C}_\pi^{t+\Delta t} \; \text{d}\Omega = \mathbf{0} \label{eq:f_phi}
\end{equation}

Finally, the current conservation condition imposed to emulate natural corrosion, \cref{eq:current_conservation}, is discretised and integrated using a lumped scheme as:
\begin{equation}
    f_{E_\text{m}} = \sum_{nd} W^{\text{surf},nd} \left(2F\nu_\text{c}+ 2F\nu_\text{h} + 4F\nu_\text{o}  \right) = 0 
\end{equation}

These three discretised equations are solved in a monolithic manner using an iterative scheme:
\begin{equation}
    \begin{bmatrix} \frac{\partial \mathbf{f}_{\text{c}\pi}}{\partial \mathbf{C}_{\pi}}  &  \frac{\partial \mathbf{f}_{\text{c}\pi}}{\partial E_\text{m}} & \frac{\partial \mathbf{f}_{\text{c}\pi}}{\partial \mathbf{\upvarphi}} \\ 
    \frac{\partial f_{E_\text{m}}}{\partial \mathbf{C}_{\pi}} & \frac{\partial f_{E_\text{m}}}{\partial E_\text{m}} & \frac{\partial f_{E_\text{m}}}{\partial \mathbf{\upvarphi}} \\ 
    \frac{\partial \mathbf{f}_{\varphi}}{\partial \mathbf{C}_{\pi}} & 0 & 0\end{bmatrix}_i
    \begin{bmatrix} \text{d} \mathbf{C}_{\pi}  \\ \text{d} E_\text{m} \\ \text{d} \bm{\upvarphi} \end{bmatrix}_{i+1} 
    = - \begin{bmatrix} \mathbf{f}_{\text{c}\pi}  \\ f_{E_\text{m}} \\ \mathbf{f}_\varphi \end{bmatrix}_{i}
\end{equation}
which is iterated until a solution is obtained that satisfies the convergence criteria:
\begin{equation}
\begin{split}
     &\frac{\epsilon}{\epsilon_0}<10^{-12} \qquad \text{or} \qquad \epsilon_i<10^{-15} \qquad \\ &\text{where} \qquad \epsilon_i = \left| \begin{bmatrix} \text{d} \mathbf{C}_{\pi}  & \text{d} E_\text{m} & \text{d} \bm{\upvarphi} \end{bmatrix}_{i+1} \cdot \begin{bmatrix} \mathbf{f}_{c\pi}  & f_{E_\text{m}} & \mathbf{f}_\varphi \end{bmatrix}_{i+1} \right|
\end{split}
\end{equation}

\section{Effect of beam length}
\label{App:Geo}
Throughout this work, we have focused on variations in the saturation, conductivity, and porosity of the concrete, considering a single corrosion pit on a concrete beam with a set length. One of the findings of this research is that the oxygen reaction occurs over the complete surface of the rebar, while the hydrogen reaction is more localised. In this appendix, we present results for a range of beam lengths (ranging from $0.05-0.3\;\text{m}$), confirming that this conclusion holds for other geometries as well. In these studies, the porosity of the concrete is kept constant at $5\%$, the $\text{Cl}^\text{-}$ concentration at $C_{\text{Cl}^-}=10\;\text{mol}/\text{m}^3$, and the concrete is considered to be fully water-saturated. 

\begin{figure}
    \centering
    \begin{subfigure}{0.49\textwidth}
        \centering
        \includegraphics[]{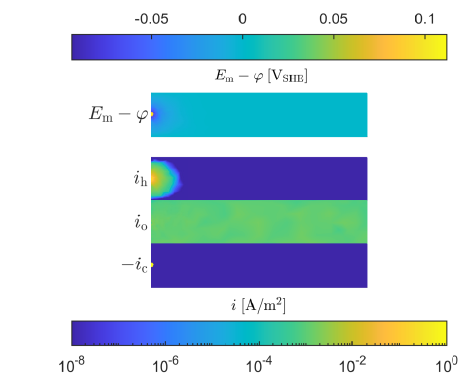}
        \caption{$L=0.05\;\text{m}$}
    \end{subfigure}
    \begin{subfigure}{0.49\textwidth}
        \centering
        \includegraphics[]{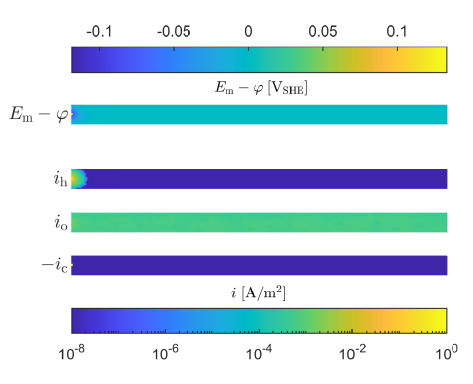}
        \caption{$L=0.2\;\text{m}$}
    \end{subfigure}
    \caption{Electric overpotential and reaction currents for cases with varying rebar lengths.}
    \label{fig:LStudy}
\end{figure}

\begin{figure}
    \centering
    \begin{subfigure}{0.49\textwidth}
        \centering
        \includegraphics[]{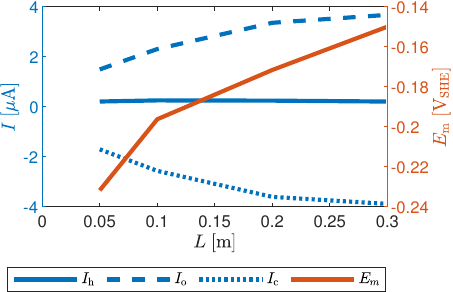}
        \caption{}
    \end{subfigure}
    \begin{subfigure}{0.49\textwidth}
        \centering
        \includegraphics[]{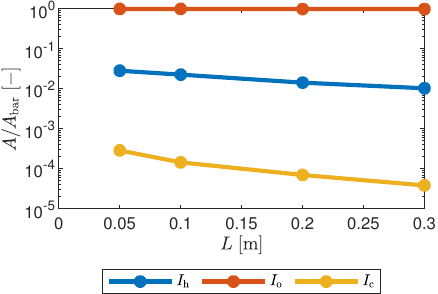}
        \caption{}
    \end{subfigure}
    \caption{Total reaction currents and electric potential (a) and reaction areas normalised by the rebar surface area (b) for varying rebar lengths.}
    \label{fig:LStudyLines}
\end{figure}

The reaction currents after 28 days are shown in \cref{fig:LStudy} for modelled domain lengths of $L=0.05\;\text{m}$ and $L=0.2\;\text{m}$, half and double the length used for our main results. As expected, for both these cases, the oxygen reaction occurs over the complete rebar length, whereas the hydrogen reaction remains localised around the corrosion pit. As the corrosion reaction current is higher for the longer rebar (due to the increased oxygen reaction current), the acidic region around the corrosion pit increases in size with increased rebar length, thereby also increasing the region over which the hydrogen evolution reaction occurs.

More quantitative results are shown in \cref{fig:LStudyLines}. Even though the oxygen reaction occurs over the complete surface of the rebar, doubling the rebar length does not double the reaction currents. Instead, as the area over which the oxygen reaction occurs increases, the metal potential also increases, slowing the oxygen evolution reaction while accelerating the corrosion reaction to balance the reaction currents. This increased metal potential also causes the hydrogen reaction rate to decrease, with the total reaction current remaining close to constant even though the area over which this reaction occurs increases with rebar length. While these results still only consider a single corrosion pit in an otherwise passivated rebar, they demonstrate that the conclusions regarding the localisation of reaction areas obtained in this work still hold for different rebar lengths.

\end{document}